\documentstyle[]{article}
\begin{document}
\overfullrule 0 mm
\language 0
\rightline {{\it Dedicated to Irene Z.}}
\vskip 0.5 cm
\centerline { \bf{ CLASSICAL
TUNNELING }} \centerline { \bf{AS THE RESULT OF RETARDATION }}
\centerline { \bf {IN CLASSICAL ELECTRODYNAMICS:}}
\centerline { \bf {NONRELATIVISTIC CASE}}
\vskip 0.5 cm

\centerline {\bf{ Alexander A.  Vlasov}}
\vskip 0.3 cm
\centerline {{  High Energy and Quantum Theory}}
\centerline {{Department of Physics}}
\centerline {{ Moscow State University}}
\centerline {{  Moscow, 119899}}
\centerline {{ Russia}}
\vskip 0.3cm
{\it In nonrelativistic approximation   one-dimensional motion of
Sommerfeld sphere in the case of potential barrier  is numerically
investigated.  The effect of classical tunneling is confirmed once
 more - Sommerfeld sphere overcomes the barrier and finds itself in
the forbidden, from classical point of view, area}

03.50.De
\vskip 0.3 cm
The problem of radiation reaction in classical electrodynamics is
still discussed in the literature (for ex. see [1]). This problem can
be formulated in the following way: it is known that classical
charged body moving with acceleration must radiate. Thus there is
back reaction of outgoing electromagnetic waves. But what quantity
feels this back reaction - pure mechanical mass of a charged body
or an effective mass, constructed from the mechanical mass and energy
of self electromagnetic field? Is this effective mass constant on the
trajectory of a moving body  or a function of time? In another words,
what is the dynamics of a charged body due to radiation reaction?

To answer  these questions long time ago [2] was proposed by
Sommerfeld model of sphere with uniform surface charge $Q$ and
mechanical mass $m$.  In nonrelativistic approximation such sphere
obeys the equation (see also [3,4,5,6]):
$$m\dot{\vec v} =\vec F_{ext}+  \eta\left[\vec v(t-2a/c) - \vec
v(t)\right] \eqno(1)$$ here $a$ - radius of the sphere,
 $\eta= {Q^2 \over 3  c a^2},\ \ \vec v= d \vec R /dt,\ \ \vec R$ -
coordinate of the center of the shell.

One can find in the literature the opinion [1], that the equation (1)
has no unphysical solutions and "free of the problems that have
plagued the theory for most of this century".

But the fact is that (as was shown in [7]) equation of
motion for Sommerfeld model possesses some strange solution which can
be interpreted as "classical tunneling" (see also [8,9] ). The
physics of this effect is simple: due to retardation the body
"understands" that there is the potential barrier "too late" and thus
can fall through the barrier.

Here we consider  one-dimensional motion of the shell in more
simple, then in [7], case - in nonrelativistic  case for
potential barrier, produced by homogeneous static electric field
$E_z$, stretched in $z$ - direction for $0<z<L$ (like in plane
condenser):  $$ E_z= \left\{\matrix{ 0,& z<0; \cr E,& 0<z<L; \cr 0,&
L<z; \cr }\right.$$

For dimensionless variables $y= R/L,\ \ x=ct/L,\ \ a^{*}=2a/L$,
 taking for simplicity $a^{*}=1$, the  equation of motion of
Sommerfeld sphere in nonrelativistic approximation (1) with external
force produced by $E_z$
$$ F_{ext} =\int d \vec r \rho \cdot E_z = EQ\cdot f,$$
where
$$\rho = Q\delta( |\vec r - \vec R| -a)/4\pi a^2,$$
$$f=\left\{\matrix{
0,& y<-1/2; \cr
(2y +1)/2,& -1/2<y<1/2; \cr
(-2y +3)/2,& 1/2<y<3/2;\cr
0,& 3/2<y; \cr
}\right.$$
reads

$${d^2 y \over dx^2} =
 k \cdot \left[ {d y(x-1) \over dx} -{d y(x) \over dx}\right]
+ \lambda f \eqno(2)$$
here $k={2Q^2 \over 3 m c^2 a},\ \  \lambda={L Q E \over m c^2}$,

It is useful to compare solutions of (1) with  classical point
charge motion in the same field, governed by the following
nonrelativistic equation without radiation force:
$${d^2 y \over dx^2} = F_E \eqno(3)$$
here
$$F_E=\lambda \left\{\matrix{
0,& y<0; \cr
1,& 0<y<1; \cr
0,& 1<y; \cr
}\right.$$

\vskip 0.5 cm {\bf A.}
Dividing $x$-axis into unit intervals, one can find solutions of (2)
on each interval in elementary functions (exponents) and then sew
them together with appropriate boundary conditions (position of the
center of the shell and its velocity must be continuous) thus
constructing the solution of (2) on the whole $x$-axis. But for our
goal it will be more effective to obtain solutions of (2) through
numerical calculations.

Numerical calculations of eq. (2) show that there is the
effect of classical tunneling for Sommerfeld sphere.

Indeed, classical point particle motion, governed by eq. (3), is
simple:
$$v^2 = 2\lambda +v_{0}^2,\ \ \ 0<y<1 $$
here ${d y \over dx}=v$,  $v_{0} $ - initial velocity.

Thus for given initial velocity for $2|\lambda|>v_{0}^2 $ there is
the turning point - i.e. classical particle cannot overcome the
potential barrier.

But for Sommerfeld sphere the result is different.

Numerical results  are on fig. (A.1-A.3) (vertical axis is velocity
$dy/dx$, horizontal axis is coordinate $y,\ \ -1/2<y<3/2$ - i.e. 
inside the barrier).

On {\bf fig. A.1} we can see the effect of tunneling for the
following values of $k$ and $\lambda$:

$k=1,\ \  \lambda=-0.5$.

Velocities of the shell are

$v=0.4,\ \ v=0.6,\ \ v=0.7$ (- all give rebounce); $\ \ v=0.8$ (and
here is tunneling)

and all of them are from the "forbidden area" $v\leq
\sqrt{2|\lambda|}= 1.0$.

On {\bf fig. A.2} we can see the effect of tunneling for the
following values of $k$ and $\lambda$:

$k=1,\ \  \lambda=-0.1$.

Velocities of the shell are:

$v=0.12,\ \ v=0.3$ (rebounce); $\ \ v=0.4$ (tunneling)

and all of them are from the "forbidden area" $v\leq
\sqrt{2|\lambda|}= 0.4472...$.

Comparing {\bf fig. A.3} with {\bf fig. A.2}, we can see that the
more greater the value of $k$ ("more" retardation), the more stronger
becomes the effect of tunneling:

on {\bf fig. A.3}:
$k=10,\ \  \lambda=-0.1$;

velocities of the shell are the same as for fig. A.2:

$v=0.12$ (rebounce);$\ \ v=0.3,\ \ v=0.4$ (tunneling)

Thus we see that the effect of classical tunneling exists not only 
for point-like particles, governed by Lorentz-Dirac equation [8], but 
also exists for charged bodies of finite size.

 \vskip 0.5 cm

 \centerline {\bf{REFERENCES}}

  \begin{enumerate}
\item  F.Rohrlich, Am.J.Phys., 65(11), 1051(1997).
\item A.Sommerfeld, Gottingen Nachrichten, 29 (1904), 363 (1904), 201
  (1905).
\item L.Page, Phys.Rev., 11, 377 (1918)
\item T.Erber, Fortschr. Phys., 9, 343 (1961)
\item P.Pearle in "Electromagnetism",ed. D.Tepliz, (Plenum, N.Y.,
1982), p.211.
\item A.Yaghjian, "Relativistic Dynamics of a Charged Sphere".
  Lecture Notes in Physics, 11 (Springer-Verlag, Berlin, 1992).
\item Alexander A.Vlasov, physics/9905050.
\item F.Denef et al, Phys.Rev. E56, 3624 (1997); hep-th/9602066.
\item Alexander A.Vlasov, Theoretical and
Mathematical Physics, 109, n.3, 1608(1996).

\end{enumerate}

\newcount\numpoint
\newcount\numpointo
\numpoint=1 \numpointo=1
\def\emmoveto#1#2{\offinterlineskip
\hbox to 0 true cm{\vbox to 0
true cm{\vskip - #2 true mm
\hskip #1 true mm \special{em:point
\the\numpoint}\vss}\hss}
\numpointo=\numpoint
\global\advance \numpoint by 1}
\def\emlineto#1#2{\offinterlineskip
\hbox to 0 true cm{\vbox to 0
true cm{\vskip - #2 true mm
\hskip #1 true mm \special{em:point
\the\numpoint}\vss}\hss}
\special{em:line
\the\numpointo,\the\numpoint}
\numpointo=\numpoint
\global\advance \numpoint by 1}
\def\emshow#1#2#3{\offinterlineskip
\hbox to 0 true cm{\vbox to 0
true cm{\vskip - #2 true mm
\hskip #1 true mm \vbox to 0
true cm{\vss\hbox{#3\hss
}}\vss}\hss}}
\special{em:linewidth 0.8pt}

\vrule width 0 mm height                0 mm depth 90.000 true mm

\special{em:linewidth 0.8pt}
\emmoveto{130.000}{10.000}
\emlineto{12.000}{10.000}
\emlineto{12.000}{80.000}
\emmoveto{71.000}{10.000}
\emlineto{71.000}{80.000}
\emmoveto{12.000}{45.000}
\emlineto{130.000}{45.000}
\emmoveto{130.000}{10.000}
\emlineto{130.000}{80.000}
\emlineto{12.000}{80.000}
\emlineto{12.000}{10.000}
\emlineto{130.000}{10.000}
\special{em:linewidth 0.4pt}
\emmoveto{12.000}{17.000}
\emlineto{130.000}{17.000}
\emmoveto{12.000}{24.000}
\emlineto{130.000}{24.000}
\emmoveto{12.000}{31.000}
\emlineto{130.000}{31.000}
\emmoveto{12.000}{38.000}
\emlineto{130.000}{38.000}
\emmoveto{12.000}{45.000}
\emlineto{130.000}{45.000}
\emmoveto{12.000}{52.000}
\emlineto{130.000}{52.000}
\emmoveto{12.000}{59.000}
\emlineto{130.000}{59.000}
\emmoveto{12.000}{66.000}
\emlineto{130.000}{66.000}
\emmoveto{12.000}{73.000}
\emlineto{130.000}{73.000}
\emmoveto{23.800}{10.000}
\emlineto{23.800}{80.000}
\emmoveto{35.600}{10.000}
\emlineto{35.600}{80.000}
\emmoveto{47.400}{10.000}
\emlineto{47.400}{80.000}
\emmoveto{59.200}{10.000}
\emlineto{59.200}{80.000}
\emmoveto{71.000}{10.000}
\emlineto{71.000}{80.000}
\emmoveto{82.800}{10.000}
\emlineto{82.800}{80.000}
\emmoveto{94.600}{10.000}
\emlineto{94.600}{80.000}
\emmoveto{106.400}{10.000}
\emlineto{106.400}{80.000}
\emmoveto{118.200}{10.000}
\emlineto{118.200}{80.000}
\special{em:linewidth 0.8pt}
\emmoveto{12.000}{60.282}
\emlineto{12.472}{60.290}
\emmoveto{12.472}{60.280}
\emlineto{12.944}{60.284}
\emmoveto{12.944}{60.274}
\emlineto{13.416}{60.274}
\emmoveto{13.416}{60.264}
\emlineto{13.887}{60.261}
\emmoveto{13.887}{60.251}
\emlineto{14.358}{60.244}
\emmoveto{14.358}{60.234}
\emlineto{14.829}{60.224}
\emmoveto{14.829}{60.214}
\emlineto{15.299}{60.200}
\emmoveto{15.299}{60.190}
\emlineto{15.768}{60.172}
\emmoveto{15.768}{60.162}
\emlineto{16.237}{60.141}
\emmoveto{16.237}{60.131}
\emlineto{16.705}{60.107}
\emmoveto{16.705}{60.097}
\emlineto{17.172}{60.070}
\emmoveto{17.172}{60.060}
\emlineto{17.638}{60.030}
\emmoveto{17.638}{60.020}
\emlineto{18.104}{59.986}
\emmoveto{18.104}{59.976}
\emlineto{18.568}{59.940}
\emmoveto{18.568}{59.930}
\emlineto{19.031}{59.891}
\emmoveto{19.031}{59.881}
\emlineto{19.493}{59.839}
\emmoveto{19.493}{59.829}
\emlineto{19.953}{59.784}
\emmoveto{19.953}{59.774}
\emlineto{20.412}{59.726}
\emmoveto{20.412}{59.716}
\emlineto{20.870}{59.666}
\emmoveto{20.870}{59.656}
\emlineto{21.326}{59.603}
\emmoveto{21.326}{59.593}
\emlineto{21.781}{59.537}
\emmoveto{21.781}{59.527}
\emlineto{22.234}{59.470}
\emmoveto{22.234}{59.460}
\emlineto{22.686}{59.399}
\emmoveto{22.686}{59.389}
\emlineto{23.135}{59.327}
\emmoveto{23.135}{59.317}
\emlineto{23.583}{59.252}
\emmoveto{23.583}{59.242}
\emlineto{24.030}{59.175}
\emmoveto{24.030}{59.165}
\emlineto{24.474}{59.096}
\emmoveto{24.474}{59.086}
\emlineto{24.916}{59.014}
\emmoveto{24.916}{59.004}
\emlineto{25.357}{58.931}
\emmoveto{25.357}{58.921}
\emlineto{25.795}{58.845}
\emmoveto{25.795}{58.835}
\emlineto{26.231}{58.758}
\emmoveto{26.231}{58.748}
\emlineto{26.666}{58.669}
\emmoveto{26.666}{58.659}
\emlineto{27.098}{58.578}
\emmoveto{27.098}{58.568}
\emlineto{27.528}{58.485}
\emmoveto{27.528}{58.475}
\emlineto{27.955}{58.391}
\emmoveto{27.955}{58.381}
\emlineto{28.381}{58.295}
\emmoveto{28.381}{58.285}
\emlineto{28.804}{58.197}
\emmoveto{28.804}{58.187}
\emlineto{29.224}{58.098}
\emmoveto{29.224}{58.088}
\emlineto{29.643}{57.997}
\emmoveto{29.643}{57.987}
\emlineto{30.058}{57.895}
\emmoveto{30.058}{57.885}
\emlineto{30.472}{57.791}
\emmoveto{30.472}{57.781}
\emlineto{30.883}{57.686}
\emmoveto{30.883}{57.676}
\emlineto{31.291}{57.580}
\emmoveto{31.291}{57.570}
\emlineto{31.697}{57.472}
\emmoveto{31.697}{57.462}
\emlineto{32.100}{57.363}
\emmoveto{32.100}{57.353}
\emlineto{32.501}{57.254}
\emmoveto{32.501}{57.244}
\emlineto{32.899}{57.142}
\emmoveto{32.899}{57.132}
\emlineto{33.294}{57.030}
\emmoveto{33.294}{57.020}
\emlineto{33.687}{56.917}
\emmoveto{33.687}{56.907}
\emlineto{34.076}{56.803}
\emmoveto{34.076}{56.793}
\emlineto{34.463}{56.688}
\emmoveto{34.463}{56.678}
\emlineto{34.848}{56.572}
\emmoveto{34.848}{56.562}
\emlineto{35.229}{56.454}
\emmoveto{35.229}{56.444}
\emlineto{35.608}{56.336}
\emmoveto{35.608}{56.326}
\emlineto{35.984}{56.217}
\emmoveto{35.984}{56.207}
\emlineto{36.357}{56.096}
\emmoveto{36.357}{56.086}
\emlineto{36.727}{55.975}
\emmoveto{36.727}{55.965}
\emlineto{37.094}{55.852}
\emmoveto{37.094}{55.842}
\emlineto{37.459}{55.728}
\emmoveto{37.459}{55.718}
\emlineto{37.820}{55.603}
\emmoveto{37.820}{55.593}
\emlineto{38.178}{55.477}
\emmoveto{38.178}{55.467}
\emlineto{38.533}{55.350}
\emmoveto{38.533}{55.340}
\emlineto{38.886}{55.221}
\emmoveto{38.886}{55.211}
\emlineto{39.235}{55.091}
\emmoveto{39.235}{55.081}
\emlineto{39.580}{54.960}
\emmoveto{39.580}{54.950}
\emlineto{39.923}{54.827}
\emmoveto{39.923}{54.817}
\emlineto{40.263}{54.694}
\emmoveto{40.263}{54.684}
\emlineto{40.599}{54.559}
\emmoveto{40.599}{54.549}
\emlineto{40.932}{54.422}
\emmoveto{40.932}{54.412}
\emlineto{41.262}{54.285}
\emmoveto{41.262}{54.275}
\emlineto{41.589}{54.146}
\emmoveto{41.589}{54.136}
\emlineto{41.912}{54.006}
\emmoveto{41.912}{53.996}
\emlineto{42.232}{53.865}
\emmoveto{42.232}{53.855}
\emlineto{42.548}{53.722}
\emmoveto{42.548}{53.712}
\emlineto{42.861}{53.579}
\emmoveto{42.861}{53.569}
\emlineto{43.171}{53.434}
\emmoveto{43.171}{53.424}
\emlineto{43.477}{53.287}
\emmoveto{43.477}{53.277}
\emlineto{43.780}{53.140}
\emmoveto{43.780}{53.130}
\emlineto{44.079}{52.991}
\emmoveto{44.079}{52.981}
\emlineto{44.374}{52.841}
\emmoveto{44.374}{52.831}
\emlineto{44.666}{52.690}
\emmoveto{44.666}{52.680}
\emlineto{44.954}{52.538}
\emmoveto{44.954}{52.528}
\emlineto{45.239}{52.385}
\emmoveto{45.239}{52.375}
\emlineto{45.520}{52.231}
\emmoveto{45.520}{52.221}
\emlineto{45.797}{52.075}
\emmoveto{45.797}{52.065}
\emlineto{46.070}{51.918}
\emmoveto{46.070}{51.908}
\emlineto{46.340}{51.761}
\emmoveto{46.340}{51.751}
\emlineto{46.606}{51.602}
\emmoveto{46.606}{51.592}
\emlineto{46.868}{51.442}
\emmoveto{46.868}{51.432}
\emlineto{47.126}{51.282}
\emmoveto{47.126}{51.272}
\emlineto{47.381}{51.120}
\emmoveto{47.381}{51.110}
\emlineto{47.631}{50.957}
\emmoveto{47.631}{50.947}
\emlineto{47.878}{50.794}
\emmoveto{47.878}{50.784}
\emlineto{48.120}{50.629}
\emmoveto{48.120}{50.619}
\emlineto{48.359}{50.464}
\emmoveto{48.359}{50.454}
\emlineto{48.594}{50.298}
\emmoveto{48.594}{50.288}
\emlineto{48.825}{50.131}
\emmoveto{48.825}{50.121}
\emlineto{49.051}{49.963}
\emmoveto{49.051}{49.953}
\emlineto{49.274}{49.794}
\emmoveto{49.274}{49.784}
\emlineto{49.493}{49.625}
\emmoveto{49.493}{49.615}
\emlineto{49.707}{49.455}
\emmoveto{49.707}{49.445}
\emlineto{49.918}{49.284}
\emmoveto{49.918}{49.274}
\emlineto{50.125}{49.113}
\emmoveto{50.125}{49.103}
\emlineto{50.327}{48.941}
\emmoveto{50.327}{48.931}
\emlineto{50.525}{48.768}
\emmoveto{50.525}{48.758}
\emlineto{50.719}{48.595}
\emmoveto{50.719}{48.585}
\emlineto{50.909}{48.421}
\emmoveto{50.909}{48.411}
\emlineto{51.095}{48.247}
\emmoveto{51.095}{48.237}
\emlineto{51.276}{48.072}
\emmoveto{51.276}{48.062}
\emlineto{51.454}{47.897}
\emmoveto{51.454}{47.887}
\emlineto{51.627}{47.721}
\emmoveto{51.627}{47.711}
\emlineto{51.796}{47.545}
\emmoveto{51.796}{47.535}
\emlineto{51.961}{47.368}
\emmoveto{51.961}{47.358}
\emlineto{52.121}{47.191}
\emmoveto{52.121}{47.181}
\emlineto{52.278}{47.014}
\emmoveto{52.278}{47.004}
\emlineto{52.430}{46.836}
\emmoveto{52.430}{46.826}
\emlineto{52.577}{46.658}
\emmoveto{52.577}{46.648}
\emlineto{52.721}{46.479}
\emmoveto{52.721}{46.469}
\emlineto{52.860}{46.300}
\emmoveto{52.860}{46.290}
\emlineto{52.995}{46.121}
\emmoveto{52.995}{46.111}
\emlineto{53.126}{45.942}
\emmoveto{53.126}{45.932}
\emlineto{53.252}{45.762}
\emmoveto{53.252}{45.752}
\emlineto{53.374}{45.582}
\emmoveto{53.374}{45.572}
\emlineto{53.492}{45.402}
\emmoveto{53.492}{45.392}
\emlineto{53.605}{45.222}
\emmoveto{53.605}{45.212}
\emlineto{53.714}{45.041}
\emmoveto{53.714}{45.031}
\emlineto{53.819}{44.860}
\emmoveto{53.819}{44.850}
\emlineto{53.920}{44.680}
\emmoveto{53.920}{44.670}
\emlineto{54.016}{44.498}
\emmoveto{54.016}{44.488}
\emlineto{54.108}{44.317}
\emmoveto{54.108}{44.307}
\emlineto{54.195}{44.136}
\emmoveto{54.195}{44.126}
\emlineto{54.278}{43.954}
\emmoveto{54.278}{43.944}
\emlineto{54.357}{43.773}
\emmoveto{54.357}{43.763}
\emlineto{54.431}{43.591}
\emmoveto{54.431}{43.581}
\emlineto{54.501}{43.409}
\emmoveto{54.501}{43.399}
\emlineto{54.567}{43.227}
\emmoveto{54.567}{43.217}
\emlineto{54.628}{43.046}
\emmoveto{54.628}{43.036}
\emlineto{54.685}{42.864}
\emmoveto{54.685}{42.854}
\emlineto{54.738}{42.682}
\emmoveto{54.738}{42.672}
\emlineto{54.786}{42.500}
\emmoveto{54.786}{42.490}
\emlineto{54.830}{42.318}
\emmoveto{54.830}{42.308}
\emlineto{54.870}{42.136}
\emmoveto{54.870}{42.126}
\emlineto{54.905}{41.955}
\emmoveto{54.905}{41.945}
\emlineto{54.936}{41.773}
\emmoveto{54.936}{41.763}
\emlineto{54.963}{41.591}
\emmoveto{54.963}{41.581}
\emlineto{54.985}{41.410}
\emmoveto{54.985}{41.400}
\emlineto{55.003}{41.229}
\emmoveto{55.003}{41.219}
\emlineto{55.016}{41.047}
\emmoveto{55.016}{41.037}
\emlineto{55.025}{40.866}
\emmoveto{55.025}{40.856}
\emlineto{55.030}{40.685}
\emmoveto{55.030}{40.675}
\emlineto{55.031}{40.505}
\emmoveto{55.031}{40.495}
\emlineto{55.027}{40.324}
\emmoveto{55.027}{40.314}
\emlineto{55.019}{40.144}
\emmoveto{55.019}{40.134}
\emlineto{55.006}{39.964}
\emmoveto{55.006}{39.954}
\emlineto{54.990}{39.785}
\emmoveto{54.990}{39.775}
\emlineto{54.969}{39.605}
\emmoveto{54.969}{39.595}
\emlineto{54.943}{39.426}
\emmoveto{54.943}{39.416}
\emlineto{54.914}{39.247}
\emmoveto{54.914}{39.237}
\emlineto{54.880}{39.069}
\emmoveto{54.880}{39.059}
\emlineto{54.842}{38.891}
\emmoveto{54.842}{38.881}
\emlineto{54.799}{38.713}
\emmoveto{54.799}{38.703}
\emlineto{54.753}{38.536}
\emmoveto{54.753}{38.526}
\emlineto{54.702}{38.359}
\emmoveto{54.702}{38.349}
\emlineto{54.647}{38.183}
\emmoveto{54.647}{38.173}
\emlineto{54.587}{38.007}
\emmoveto{54.587}{37.997}
\emlineto{54.524}{37.832}
\emmoveto{54.524}{37.822}
\emlineto{54.456}{37.657}
\emmoveto{54.456}{37.647}
\emlineto{54.384}{37.482}
\emmoveto{54.384}{37.472}
\emlineto{54.308}{37.308}
\emmoveto{54.308}{37.298}
\emlineto{54.228}{37.135}
\emmoveto{54.228}{37.125}
\emlineto{54.143}{36.962}
\emmoveto{54.143}{36.952}
\emlineto{54.055}{36.790}
\emmoveto{54.055}{36.780}
\emlineto{53.962}{36.618}
\emmoveto{53.962}{36.608}
\emlineto{53.866}{36.447}
\emmoveto{53.866}{36.437}
\emlineto{53.765}{36.277}
\emmoveto{53.765}{36.267}
\emlineto{53.660}{36.107}
\emmoveto{53.660}{36.097}
\emlineto{53.551}{35.938}
\emmoveto{53.551}{35.928}
\emlineto{53.438}{35.770}
\emmoveto{53.438}{35.760}
\emlineto{53.321}{35.602}
\emmoveto{53.321}{35.592}
\emlineto{53.200}{35.435}
\emmoveto{53.200}{35.425}
\emlineto{53.075}{35.269}
\emmoveto{53.075}{35.259}
\emlineto{52.946}{35.103}
\emmoveto{52.946}{35.093}
\emlineto{52.813}{34.939}
\emmoveto{52.813}{34.929}
\emlineto{52.676}{34.775}
\emmoveto{52.676}{34.765}
\emlineto{52.535}{34.612}
\emmoveto{52.535}{34.602}
\emlineto{52.391}{34.449}
\emmoveto{52.391}{34.439}
\emlineto{52.242}{34.288}
\emmoveto{52.242}{34.278}
\emlineto{52.090}{34.127}
\emmoveto{52.090}{34.117}
\emlineto{51.934}{33.967}
\emmoveto{51.934}{33.957}
\emlineto{51.774}{33.808}
\emmoveto{51.774}{33.798}
\emlineto{51.610}{33.650}
\emmoveto{51.610}{33.640}
\emlineto{51.442}{33.493}
\emmoveto{51.442}{33.483}
\emlineto{51.271}{33.337}
\emmoveto{51.271}{33.327}
\emlineto{51.096}{33.182}
\emmoveto{51.096}{33.172}
\emlineto{50.917}{33.027}
\emmoveto{50.917}{33.017}
\emlineto{50.735}{32.874}
\emmoveto{50.735}{32.864}
\emlineto{50.548}{32.721}
\emmoveto{50.548}{32.711}
\emlineto{50.359}{32.570}
\emmoveto{50.359}{32.560}
\emlineto{50.165}{32.420}
\emmoveto{50.165}{32.410}
\emlineto{49.968}{32.270}
\emmoveto{49.968}{32.260}
\emlineto{49.768}{32.122}
\emmoveto{49.768}{32.112}
\emlineto{49.564}{31.974}
\emmoveto{49.564}{31.964}
\emlineto{49.356}{31.828}
\emmoveto{49.356}{31.818}
\emlineto{49.145}{31.683}
\emmoveto{49.145}{31.673}
\emlineto{48.931}{31.539}
\emmoveto{48.931}{31.529}
\emlineto{48.713}{31.396}
\emmoveto{48.713}{31.386}
\emlineto{48.491}{31.254}
\emmoveto{48.491}{31.244}
\emlineto{48.266}{31.113}
\emmoveto{48.266}{31.103}
\emlineto{48.038}{30.973}
\emmoveto{48.038}{30.963}
\emlineto{47.807}{30.834}
\emmoveto{47.807}{30.824}
\emlineto{47.572}{30.697}
\emmoveto{47.572}{30.687}
\emlineto{47.334}{30.561}
\emmoveto{47.334}{30.551}
\emlineto{47.093}{30.425}
\emmoveto{47.093}{30.415}
\emlineto{46.848}{30.291}
\emmoveto{46.848}{30.281}
\emlineto{46.600}{30.159}
\emmoveto{46.600}{30.149}
\emlineto{46.350}{30.027}
\emmoveto{46.350}{30.017}
\emlineto{46.096}{29.897}
\emmoveto{46.096}{29.887}
\emlineto{45.838}{29.768}
\emmoveto{45.838}{29.758}
\emlineto{45.578}{29.640}
\emmoveto{45.578}{29.630}
\emlineto{45.315}{29.513}
\emmoveto{45.315}{29.503}
\emlineto{45.049}{29.388}
\emmoveto{45.049}{29.378}
\emlineto{44.780}{29.264}
\emmoveto{44.780}{29.254}
\emlineto{44.507}{29.141}
\emmoveto{44.507}{29.131}
\emlineto{44.232}{29.020}
\emmoveto{44.232}{29.010}
\emlineto{43.954}{28.900}
\emmoveto{43.954}{28.890}
\emlineto{43.673}{28.781}
\emmoveto{43.673}{28.771}
\emlineto{43.390}{28.664}
\emmoveto{43.390}{28.654}
\emlineto{43.103}{28.548}
\emmoveto{43.103}{28.538}
\emlineto{42.814}{28.433}
\emmoveto{42.814}{28.423}
\emlineto{42.522}{28.319}
\emmoveto{42.522}{28.309}
\emlineto{42.227}{28.207}
\emmoveto{42.227}{28.197}
\emlineto{41.930}{28.097}
\emmoveto{41.930}{28.087}
\emlineto{41.630}{27.988}
\emmoveto{41.630}{27.978}
\emlineto{41.327}{27.880}
\emmoveto{41.327}{27.870}
\emlineto{41.022}{27.773}
\emmoveto{41.022}{27.763}
\emlineto{40.715}{27.668}
\emmoveto{40.715}{27.658}
\emlineto{40.405}{27.565}
\emmoveto{40.405}{27.555}
\emlineto{40.092}{27.463}
\emmoveto{40.092}{27.453}
\emlineto{39.777}{27.362}
\emmoveto{39.777}{27.352}
\emlineto{39.459}{27.263}
\emmoveto{39.459}{27.253}
\emlineto{39.140}{27.165}
\emmoveto{39.140}{27.155}
\emlineto{38.818}{27.069}
\emmoveto{38.818}{27.059}
\emlineto{38.493}{26.974}
\emmoveto{38.493}{26.964}
\emlineto{38.166}{26.880}
\emmoveto{38.166}{26.870}
\emlineto{37.838}{26.788}
\emmoveto{37.838}{26.778}
\emlineto{37.507}{26.698}
\emmoveto{37.507}{26.688}
\emlineto{37.173}{26.609}
\emmoveto{37.173}{26.599}
\emlineto{36.838}{26.522}
\emmoveto{36.838}{26.512}
\emlineto{36.501}{26.436}
\emmoveto{36.501}{26.426}
\emlineto{36.161}{26.352}
\emmoveto{36.161}{26.342}
\emlineto{35.820}{26.269}
\emmoveto{35.820}{26.259}
\emlineto{35.476}{26.188}
\emmoveto{35.476}{26.178}
\emlineto{35.131}{26.108}
\emmoveto{35.131}{26.098}
\emlineto{34.784}{26.030}
\emmoveto{34.784}{26.020}
\emlineto{34.435}{25.954}
\emmoveto{34.435}{25.944}
\emlineto{34.084}{25.879}
\emmoveto{34.084}{25.869}
\emlineto{33.731}{25.805}
\emmoveto{33.731}{25.795}
\emlineto{33.377}{25.734}
\emmoveto{33.377}{25.724}
\emlineto{33.021}{25.663}
\emmoveto{33.021}{25.653}
\emlineto{32.663}{25.595}
\emmoveto{32.663}{25.585}
\emlineto{32.304}{25.528}
\emmoveto{32.304}{25.518}
\emlineto{31.943}{25.462}
\emmoveto{31.943}{25.452}
\emlineto{31.580}{25.398}
\emmoveto{31.580}{25.388}
\emlineto{31.216}{25.336}
\emmoveto{31.216}{25.326}
\emlineto{30.851}{25.275}
\emmoveto{30.851}{25.265}
\emlineto{30.484}{25.216}
\emmoveto{30.484}{25.206}
\emlineto{30.116}{25.159}
\emmoveto{30.116}{25.149}
\emlineto{29.746}{25.103}
\emmoveto{29.746}{25.093}
\emlineto{29.375}{25.049}
\emmoveto{29.375}{25.039}
\emlineto{29.003}{24.996}
\emmoveto{29.003}{24.986}
\emlineto{28.630}{24.946}
\emmoveto{28.630}{24.936}
\emlineto{28.255}{24.896}
\emmoveto{28.255}{24.886}
\emlineto{27.879}{24.849}
\emmoveto{27.879}{24.839}
\emlineto{27.502}{24.803}
\emmoveto{27.502}{24.793}
\emlineto{27.124}{24.758}
\emmoveto{27.124}{24.748}
\emlineto{26.745}{24.716}
\emmoveto{26.745}{24.706}
\emlineto{26.365}{24.674}
\emmoveto{26.365}{24.664}
\emlineto{25.984}{24.635}
\emmoveto{25.984}{24.625}
\emlineto{25.602}{24.597}
\emmoveto{25.602}{24.587}
\emlineto{25.219}{24.561}
\emmoveto{25.219}{24.551}
\emlineto{24.835}{24.527}
\emmoveto{24.835}{24.517}
\emlineto{24.451}{24.494}
\emmoveto{24.451}{24.484}
\emlineto{24.065}{24.463}
\emmoveto{24.065}{24.453}
\emlineto{23.679}{24.433}
\emmoveto{23.679}{24.423}
\emlineto{23.293}{24.406}
\emmoveto{23.293}{24.396}
\emlineto{22.905}{24.379}
\emmoveto{22.905}{24.369}
\emlineto{22.517}{24.355}
\emmoveto{22.517}{24.345}
\emlineto{22.129}{24.332}
\emmoveto{22.129}{24.322}
\emlineto{21.740}{24.311}
\emmoveto{21.740}{24.301}
\emlineto{21.350}{24.291}
\emmoveto{21.350}{24.281}
\emlineto{20.960}{24.273}
\emmoveto{20.960}{24.263}
\emlineto{20.570}{24.257}
\emmoveto{20.570}{24.247}
\emlineto{20.179}{24.243}
\emmoveto{20.179}{24.233}
\emlineto{19.788}{24.230}
\emmoveto{19.788}{24.220}
\emlineto{19.397}{24.219}
\emmoveto{19.397}{24.209}
\emlineto{19.005}{24.209}
\emmoveto{19.005}{24.199}
\emlineto{18.613}{24.201}
\emmoveto{18.613}{24.191}
\emlineto{18.222}{24.195}
\emmoveto{18.222}{24.185}
\emlineto{17.829}{24.190}
\emmoveto{17.829}{24.180}
\emlineto{17.437}{24.187}
\emmoveto{17.437}{24.177}
\emlineto{17.045}{24.176}
\emlineto{17.045}{24.186}
\emmoveto{17.045}{24.176}
\emlineto{16.653}{24.186}
\emmoveto{16.653}{24.176}
\emlineto{16.260}{24.188}
\emmoveto{16.260}{24.178}
\emlineto{15.868}{24.192}
\emmoveto{15.868}{24.182}
\emlineto{15.476}{24.197}
\emmoveto{15.476}{24.187}
\emlineto{15.084}{24.204}
\emmoveto{15.084}{24.194}
\emlineto{14.693}{24.213}
\emmoveto{14.693}{24.203}
\emlineto{14.301}{24.223}
\emmoveto{14.301}{24.213}
\emlineto{13.910}{24.235}
\emmoveto{13.910}{24.225}
\emlineto{13.519}{24.248}
\emmoveto{13.519}{24.238}
\emlineto{13.128}{24.263}
\emmoveto{13.128}{24.253}
\emlineto{12.738}{24.280}
\emshow{48.580}{52.700}{v=0.4}
\emmoveto{12.000}{70.141}
\emlineto{12.708}{70.148}
\emmoveto{12.708}{70.138}
\emlineto{13.416}{70.139}
\emmoveto{13.416}{70.129}
\emlineto{14.123}{70.125}
\emmoveto{14.123}{70.115}
\emlineto{14.831}{70.105}
\emmoveto{14.831}{70.095}
\emlineto{15.537}{70.079}
\emmoveto{15.537}{70.069}
\emlineto{16.243}{70.049}
\emmoveto{16.243}{70.039}
\emlineto{16.948}{70.013}
\emmoveto{16.948}{70.003}
\emlineto{17.652}{69.971}
\emmoveto{17.652}{69.961}
\emlineto{18.356}{69.925}
\emmoveto{18.356}{69.915}
\emlineto{19.058}{69.874}
\emmoveto{19.058}{69.864}
\emlineto{19.758}{69.818}
\emmoveto{19.758}{69.808}
\emlineto{20.458}{69.758}
\emmoveto{20.458}{69.748}
\emlineto{21.155}{69.693}
\emmoveto{21.155}{69.683}
\emlineto{21.852}{69.623}
\emmoveto{21.852}{69.613}
\emlineto{22.546}{69.549}
\emmoveto{22.546}{69.539}
\emlineto{23.239}{69.471}
\emmoveto{23.239}{69.461}
\emlineto{23.930}{69.389}
\emmoveto{23.930}{69.379}
\emlineto{24.618}{69.302}
\emmoveto{24.618}{69.292}
\emlineto{25.305}{69.212}
\emmoveto{25.305}{69.202}
\emlineto{25.989}{69.118}
\emmoveto{25.989}{69.108}
\emlineto{26.671}{69.020}
\emmoveto{26.671}{69.010}
\emlineto{27.351}{68.918}
\emmoveto{27.351}{68.908}
\emlineto{28.028}{68.812}
\emmoveto{28.028}{68.802}
\emlineto{28.703}{68.703}
\emmoveto{28.703}{68.693}
\emlineto{29.375}{68.591}
\emmoveto{29.375}{68.581}
\emlineto{30.044}{68.476}
\emmoveto{30.044}{68.466}
\emlineto{30.711}{68.357}
\emmoveto{30.711}{68.347}
\emlineto{31.374}{68.235}
\emmoveto{31.374}{68.225}
\emlineto{32.035}{68.110}
\emmoveto{32.035}{68.100}
\emlineto{32.693}{67.982}
\emmoveto{32.693}{67.972}
\emlineto{33.347}{67.851}
\emmoveto{33.347}{67.841}
\emlineto{33.999}{67.717}
\emmoveto{33.999}{67.707}
\emlineto{34.647}{67.580}
\emmoveto{34.647}{67.570}
\emlineto{35.291}{67.441}
\emmoveto{35.291}{67.431}
\emlineto{35.933}{67.300}
\emmoveto{35.933}{67.290}
\emlineto{36.571}{67.155}
\emmoveto{36.571}{67.145}
\emlineto{37.205}{67.009}
\emmoveto{37.205}{66.999}
\emlineto{37.837}{66.860}
\emmoveto{37.837}{66.850}
\emlineto{38.464}{66.709}
\emmoveto{38.464}{66.699}
\emlineto{39.088}{66.555}
\emmoveto{39.088}{66.545}
\emlineto{39.708}{66.400}
\emmoveto{39.708}{66.390}
\emlineto{40.324}{66.242}
\emmoveto{40.324}{66.232}
\emlineto{40.937}{66.083}
\emmoveto{40.937}{66.073}
\emlineto{41.545}{65.922}
\emmoveto{41.545}{65.912}
\emlineto{42.150}{65.758}
\emmoveto{42.150}{65.748}
\emlineto{42.751}{65.594}
\emmoveto{42.751}{65.584}
\emlineto{43.348}{65.427}
\emmoveto{43.348}{65.417}
\emlineto{43.941}{65.259}
\emmoveto{43.941}{65.249}
\emlineto{44.530}{65.089}
\emmoveto{44.530}{65.079}
\emlineto{45.115}{64.918}
\emmoveto{45.115}{64.908}
\emlineto{45.695}{64.745}
\emmoveto{45.695}{64.735}
\emlineto{46.272}{64.571}
\emmoveto{46.272}{64.561}
\emlineto{46.844}{64.395}
\emmoveto{46.844}{64.385}
\emlineto{47.412}{64.218}
\emmoveto{47.412}{64.208}
\emlineto{47.976}{64.039}
\emmoveto{47.976}{64.029}
\emlineto{48.536}{63.858}
\emmoveto{48.536}{63.848}
\emlineto{49.091}{63.676}
\emmoveto{49.091}{63.666}
\emlineto{49.641}{63.492}
\emmoveto{49.641}{63.482}
\emlineto{50.188}{63.306}
\emmoveto{50.188}{63.296}
\emlineto{50.730}{63.118}
\emmoveto{50.730}{63.108}
\emlineto{51.267}{62.929}
\emmoveto{51.267}{62.919}
\emlineto{51.800}{62.738}
\emmoveto{51.800}{62.728}
\emlineto{52.328}{62.545}
\emmoveto{52.328}{62.535}
\emlineto{52.852}{62.350}
\emmoveto{52.852}{62.340}
\emlineto{53.371}{62.153}
\emmoveto{53.371}{62.143}
\emlineto{53.885}{61.954}
\emmoveto{53.885}{61.944}
\emlineto{54.394}{61.754}
\emmoveto{54.394}{61.744}
\emlineto{54.899}{61.551}
\emmoveto{54.899}{61.541}
\emlineto{55.399}{61.347}
\emmoveto{55.399}{61.337}
\emlineto{55.893}{61.141}
\emmoveto{55.893}{61.131}
\emlineto{56.383}{60.932}
\emmoveto{56.383}{60.922}
\emlineto{56.868}{60.722}
\emmoveto{56.868}{60.712}
\emlineto{57.348}{60.511}
\emmoveto{57.348}{60.501}
\emlineto{57.823}{60.297}
\emmoveto{57.823}{60.287}
\emlineto{58.292}{60.081}
\emmoveto{58.292}{60.071}
\emlineto{58.756}{59.864}
\emmoveto{58.756}{59.854}
\emlineto{59.216}{59.644}
\emmoveto{59.216}{59.634}
\emlineto{59.669}{59.423}
\emmoveto{59.669}{59.413}
\emlineto{60.118}{59.200}
\emmoveto{60.118}{59.190}
\emlineto{60.561}{58.975}
\emmoveto{60.561}{58.965}
\emlineto{60.999}{58.749}
\emmoveto{60.999}{58.739}
\emlineto{61.431}{58.521}
\emmoveto{61.431}{58.511}
\emlineto{61.858}{58.291}
\emmoveto{61.858}{58.281}
\emlineto{62.280}{58.059}
\emmoveto{62.280}{58.049}
\emlineto{62.695}{57.826}
\emmoveto{62.695}{57.816}
\emlineto{63.105}{57.591}
\emmoveto{63.105}{57.581}
\emlineto{63.510}{57.354}
\emmoveto{63.510}{57.344}
\emlineto{63.909}{57.116}
\emmoveto{63.909}{57.106}
\emlineto{64.302}{56.877}
\emmoveto{64.302}{56.867}
\emlineto{64.689}{56.636}
\emmoveto{64.689}{56.626}
\emlineto{65.071}{56.393}
\emmoveto{65.071}{56.383}
\emlineto{65.447}{56.149}
\emmoveto{65.447}{56.139}
\emlineto{65.817}{55.904}
\emmoveto{65.817}{55.894}
\emlineto{66.181}{55.657}
\emmoveto{66.181}{55.647}
\emlineto{66.539}{55.409}
\emmoveto{66.539}{55.399}
\emlineto{66.891}{55.160}
\emmoveto{66.891}{55.150}
\emlineto{67.237}{54.909}
\emmoveto{67.237}{54.899}
\emlineto{67.577}{54.657}
\emmoveto{67.577}{54.647}
\emlineto{67.911}{54.404}
\emmoveto{67.911}{54.394}
\emlineto{68.239}{54.150}
\emmoveto{68.239}{54.140}
\emlineto{68.561}{53.895}
\emmoveto{68.561}{53.885}
\emlineto{68.877}{53.639}
\emmoveto{68.877}{53.629}
\emlineto{69.187}{53.382}
\emmoveto{69.187}{53.372}
\emlineto{69.490}{53.124}
\emmoveto{69.490}{53.114}
\emlineto{69.788}{52.865}
\emmoveto{69.788}{52.855}
\emlineto{70.079}{52.605}
\emmoveto{70.079}{52.595}
\emlineto{70.364}{52.345}
\emmoveto{70.364}{52.335}
\emlineto{70.642}{52.083}
\emmoveto{70.642}{52.073}
\emlineto{70.915}{51.821}
\emmoveto{70.915}{51.811}
\emlineto{71.181}{51.559}
\emmoveto{71.181}{51.549}
\emlineto{71.441}{51.301}
\emmoveto{71.441}{51.291}
\emlineto{71.694}{51.046}
\emmoveto{71.694}{51.036}
\emlineto{71.942}{50.794}
\emmoveto{71.942}{50.784}
\emlineto{72.184}{50.546}
\emmoveto{72.184}{50.536}
\emlineto{72.419}{50.300}
\emmoveto{72.419}{50.290}
\emlineto{72.649}{50.058}
\emmoveto{72.649}{50.048}
\emlineto{72.873}{49.818}
\emmoveto{72.873}{49.808}
\emlineto{73.092}{49.580}
\emmoveto{73.092}{49.570}
\emlineto{73.305}{49.345}
\emmoveto{73.305}{49.335}
\emlineto{73.512}{49.113}
\emmoveto{73.512}{49.103}
\emlineto{73.714}{48.883}
\emmoveto{73.714}{48.873}
\emlineto{73.910}{48.654}
\emmoveto{73.910}{48.644}
\emlineto{74.100}{48.428}
\emmoveto{74.100}{48.418}
\emlineto{74.286}{48.203}
\emmoveto{74.286}{48.193}
\emlineto{74.466}{47.981}
\emmoveto{74.466}{47.971}
\emlineto{74.640}{47.759}
\emmoveto{74.640}{47.749}
\emlineto{74.810}{47.540}
\emmoveto{74.810}{47.530}
\emlineto{74.974}{47.321}
\emmoveto{74.974}{47.311}
\emlineto{75.133}{47.104}
\emmoveto{75.133}{47.094}
\emlineto{75.287}{46.888}
\emmoveto{75.287}{46.878}
\emlineto{75.435}{46.673}
\emmoveto{75.435}{46.663}
\emlineto{75.579}{46.459}
\emmoveto{75.579}{46.449}
\emlineto{75.717}{46.246}
\emmoveto{75.717}{46.236}
\emlineto{75.850}{46.034}
\emmoveto{75.850}{46.024}
\emlineto{75.978}{45.823}
\emmoveto{75.978}{45.813}
\emlineto{76.102}{45.612}
\emmoveto{76.102}{45.602}
\emlineto{76.220}{45.401}
\emmoveto{76.220}{45.391}
\emlineto{76.333}{45.191}
\emmoveto{76.333}{45.181}
\emlineto{76.441}{44.981}
\emmoveto{76.441}{44.971}
\emlineto{76.544}{44.772}
\emmoveto{76.544}{44.762}
\emlineto{76.642}{44.562}
\emmoveto{76.642}{44.552}
\emlineto{76.735}{44.353}
\emmoveto{76.735}{44.343}
\emlineto{76.823}{44.144}
\emmoveto{76.823}{44.134}
\emlineto{76.906}{43.935}
\emmoveto{76.906}{43.925}
\emlineto{76.984}{43.726}
\emmoveto{76.984}{43.716}
\emlineto{77.057}{43.516}
\emmoveto{77.057}{43.506}
\emlineto{77.124}{43.307}
\emmoveto{77.124}{43.297}
\emlineto{77.187}{43.097}
\emmoveto{77.187}{43.087}
\emlineto{77.245}{42.887}
\emmoveto{77.245}{42.877}
\emlineto{77.298}{42.676}
\emmoveto{77.298}{42.666}
\emlineto{77.346}{42.465}
\emmoveto{77.346}{42.455}
\emlineto{77.389}{42.253}
\emmoveto{77.389}{42.243}
\emlineto{77.426}{42.041}
\emmoveto{77.426}{42.031}
\emlineto{77.459}{41.828}
\emmoveto{77.459}{41.818}
\emlineto{77.486}{41.615}
\emmoveto{77.486}{41.605}
\emlineto{77.509}{41.401}
\emmoveto{77.509}{41.391}
\emlineto{77.526}{41.186}
\emmoveto{77.526}{41.176}
\emlineto{77.538}{40.970}
\emmoveto{77.538}{40.960}
\emlineto{77.545}{40.754}
\emmoveto{77.545}{40.744}
\emlineto{77.547}{40.537}
\emmoveto{77.547}{40.527}
\emlineto{77.543}{40.318}
\emmoveto{77.543}{40.308}
\emlineto{77.535}{40.099}
\emmoveto{77.535}{40.089}
\emlineto{77.521}{39.880}
\emmoveto{77.521}{39.870}
\emlineto{77.501}{39.659}
\emmoveto{77.501}{39.649}
\emlineto{77.477}{39.438}
\emmoveto{77.477}{39.428}
\emlineto{77.447}{39.216}
\emmoveto{77.447}{39.206}
\emlineto{77.412}{38.993}
\emmoveto{77.412}{38.983}
\emlineto{77.371}{38.770}
\emmoveto{77.371}{38.760}
\emlineto{77.326}{38.546}
\emmoveto{77.326}{38.536}
\emlineto{77.274}{38.322}
\emmoveto{77.274}{38.312}
\emlineto{77.218}{38.097}
\emmoveto{77.218}{38.087}
\emlineto{77.156}{37.871}
\emmoveto{77.156}{37.861}
\emlineto{77.088}{37.645}
\emmoveto{77.088}{37.635}
\emlineto{77.016}{37.418}
\emmoveto{77.016}{37.408}
\emlineto{76.937}{37.191}
\emmoveto{76.937}{37.181}
\emlineto{76.854}{36.963}
\emmoveto{76.854}{36.953}
\emlineto{76.764}{36.735}
\emmoveto{76.764}{36.725}
\emlineto{76.670}{36.506}
\emmoveto{76.670}{36.496}
\emlineto{76.570}{36.276}
\emmoveto{76.570}{36.266}
\emlineto{76.464}{36.046}
\emmoveto{76.464}{36.036}
\emlineto{76.353}{35.815}
\emmoveto{76.353}{35.805}
\emlineto{76.236}{35.584}
\emmoveto{76.236}{35.574}
\emlineto{76.114}{35.352}
\emmoveto{76.114}{35.342}
\emlineto{75.986}{35.119}
\emmoveto{75.986}{35.109}
\emlineto{75.853}{34.886}
\emmoveto{75.853}{34.876}
\emlineto{75.714}{34.652}
\emmoveto{75.714}{34.642}
\emlineto{75.569}{34.417}
\emmoveto{75.569}{34.407}
\emlineto{75.419}{34.182}
\emmoveto{75.419}{34.172}
\emlineto{75.263}{33.945}
\emmoveto{75.263}{33.935}
\emlineto{75.102}{33.708}
\emmoveto{75.102}{33.698}
\emlineto{74.935}{33.471}
\emmoveto{74.935}{33.461}
\emlineto{74.762}{33.232}
\emmoveto{74.762}{33.222}
\emlineto{74.583}{32.992}
\emmoveto{74.583}{32.982}
\emlineto{74.399}{32.752}
\emmoveto{74.399}{32.742}
\emlineto{74.209}{32.511}
\emmoveto{74.209}{32.501}
\emlineto{74.013}{32.268}
\emmoveto{74.013}{32.258}
\emlineto{73.811}{32.025}
\emmoveto{73.811}{32.015}
\emlineto{73.604}{31.781}
\emmoveto{73.604}{31.771}
\emlineto{73.390}{31.535}
\emmoveto{73.390}{31.525}
\emlineto{73.171}{31.289}
\emmoveto{73.171}{31.279}
\emlineto{72.946}{31.041}
\emmoveto{72.946}{31.031}
\emlineto{72.715}{30.792}
\emmoveto{72.715}{30.782}
\emlineto{72.478}{30.542}
\emmoveto{72.478}{30.532}
\emlineto{72.234}{30.291}
\emmoveto{72.234}{30.281}
\emlineto{71.985}{30.038}
\emmoveto{71.985}{30.028}
\emlineto{71.730}{29.785}
\emmoveto{71.730}{29.775}
\emlineto{71.469}{29.529}
\emmoveto{71.469}{29.519}
\emlineto{71.201}{29.273}
\emmoveto{71.201}{29.263}
\emlineto{70.928}{29.015}
\emmoveto{70.928}{29.005}
\emlineto{70.648}{28.759}
\emmoveto{70.648}{28.749}
\emlineto{70.362}{28.506}
\emmoveto{70.362}{28.496}
\emlineto{70.070}{28.256}
\emmoveto{70.070}{28.246}
\emlineto{69.772}{28.009}
\emmoveto{69.772}{27.999}
\emlineto{69.469}{27.765}
\emmoveto{69.469}{27.755}
\emlineto{69.159}{27.525}
\emmoveto{69.159}{27.515}
\emlineto{68.844}{27.287}
\emmoveto{68.844}{27.277}
\emlineto{68.523}{27.052}
\emmoveto{68.523}{27.042}
\emlineto{68.197}{26.820}
\emmoveto{68.197}{26.810}
\emlineto{67.865}{26.591}
\emmoveto{67.865}{26.581}
\emlineto{67.527}{26.365}
\emmoveto{67.527}{26.355}
\emlineto{67.185}{26.141}
\emmoveto{67.185}{26.131}
\emlineto{66.836}{25.921}
\emmoveto{66.836}{25.911}
\emlineto{66.483}{25.703}
\emmoveto{66.483}{25.693}
\emlineto{66.125}{25.488}
\emmoveto{66.125}{25.478}
\emlineto{65.761}{25.276}
\emmoveto{65.761}{25.266}
\emlineto{65.392}{25.066}
\emmoveto{65.392}{25.056}
\emlineto{65.019}{24.859}
\emmoveto{65.019}{24.849}
\emlineto{64.640}{24.655}
\emmoveto{64.640}{24.645}
\emlineto{64.257}{24.454}
\emmoveto{64.257}{24.444}
\emlineto{63.868}{24.255}
\emmoveto{63.868}{24.245}
\emlineto{63.475}{24.059}
\emmoveto{63.475}{24.049}
\emlineto{63.078}{23.865}
\emmoveto{63.078}{23.855}
\emlineto{62.675}{23.674}
\emmoveto{62.675}{23.664}
\emlineto{62.269}{23.485}
\emmoveto{62.269}{23.475}
\emlineto{61.857}{23.299}
\emmoveto{61.857}{23.289}
\emlineto{61.442}{23.116}
\emmoveto{61.442}{23.106}
\emlineto{61.022}{22.934}
\emmoveto{61.022}{22.924}
\emlineto{60.597}{22.756}
\emmoveto{60.597}{22.746}
\emlineto{60.169}{22.579}
\emmoveto{60.169}{22.569}
\emlineto{59.736}{22.405}
\emmoveto{59.736}{22.395}
\emlineto{59.299}{22.234}
\emmoveto{59.299}{22.224}
\emlineto{58.858}{22.064}
\emmoveto{58.858}{22.054}
\emlineto{58.413}{21.897}
\emmoveto{58.413}{21.887}
\emlineto{57.964}{21.733}
\emmoveto{57.964}{21.723}
\emlineto{57.511}{21.570}
\emmoveto{57.511}{21.560}
\emlineto{57.054}{21.410}
\emmoveto{57.054}{21.400}
\emlineto{56.593}{21.252}
\emmoveto{56.593}{21.242}
\emlineto{56.129}{21.096}
\emmoveto{56.129}{21.086}
\emlineto{55.661}{20.942}
\emmoveto{55.661}{20.932}
\emlineto{55.189}{20.790}
\emmoveto{55.189}{20.780}
\emlineto{54.714}{20.640}
\emmoveto{54.714}{20.630}
\emlineto{54.235}{20.492}
\emmoveto{54.235}{20.482}
\emlineto{53.752}{20.346}
\emmoveto{53.752}{20.336}
\emlineto{53.267}{20.202}
\emmoveto{53.267}{20.192}
\emlineto{52.777}{20.060}
\emmoveto{52.777}{20.050}
\emlineto{52.284}{19.920}
\emmoveto{52.284}{19.910}
\emlineto{51.788}{19.781}
\emmoveto{51.788}{19.771}
\emlineto{51.289}{19.645}
\emmoveto{51.289}{19.635}
\emlineto{50.786}{19.510}
\emmoveto{50.786}{19.500}
\emlineto{50.281}{19.377}
\emmoveto{50.281}{19.367}
\emlineto{49.772}{19.246}
\emmoveto{49.772}{19.236}
\emlineto{49.260}{19.116}
\emmoveto{49.260}{19.106}
\emlineto{48.744}{18.989}
\emmoveto{48.744}{18.979}
\emlineto{48.226}{18.863}
\emmoveto{48.226}{18.853}
\emlineto{47.705}{18.739}
\emmoveto{47.705}{18.729}
\emlineto{47.181}{18.618}
\emmoveto{47.181}{18.608}
\emlineto{46.654}{18.498}
\emmoveto{46.654}{18.488}
\emlineto{46.124}{18.381}
\emmoveto{46.124}{18.371}
\emlineto{45.592}{18.266}
\emmoveto{45.592}{18.256}
\emlineto{45.056}{18.153}
\emmoveto{45.056}{18.143}
\emlineto{44.518}{18.042}
\emmoveto{44.518}{18.032}
\emlineto{43.978}{17.933}
\emmoveto{43.978}{17.923}
\emlineto{43.434}{17.827}
\emmoveto{43.434}{17.817}
\emlineto{42.889}{17.723}
\emmoveto{42.889}{17.713}
\emlineto{42.340}{17.621}
\emmoveto{42.340}{17.611}
\emlineto{41.790}{17.522}
\emmoveto{41.790}{17.512}
\emlineto{41.237}{17.425}
\emmoveto{41.237}{17.415}
\emlineto{40.682}{17.331}
\emmoveto{40.682}{17.321}
\emlineto{40.124}{17.239}
\emmoveto{40.124}{17.229}
\emlineto{39.565}{17.149}
\emmoveto{39.565}{17.139}
\emlineto{39.003}{17.063}
\emmoveto{39.003}{17.053}
\emlineto{38.439}{16.978}
\emmoveto{38.439}{16.968}
\emlineto{37.873}{16.896}
\emmoveto{37.873}{16.886}
\emlineto{37.306}{16.817}
\emmoveto{37.306}{16.807}
\emlineto{36.736}{16.740}
\emmoveto{36.736}{16.730}
\emlineto{36.165}{16.666}
\emmoveto{36.165}{16.656}
\emlineto{35.591}{16.594}
\emmoveto{35.591}{16.584}
\emlineto{35.017}{16.525}
\emmoveto{35.017}{16.515}
\emlineto{34.440}{16.459}
\emmoveto{34.440}{16.449}
\emlineto{33.862}{16.395}
\emmoveto{33.862}{16.385}
\emlineto{33.283}{16.334}
\emmoveto{33.283}{16.324}
\emlineto{32.702}{16.275}
\emmoveto{32.702}{16.265}
\emlineto{32.119}{16.220}
\emmoveto{32.119}{16.210}
\emlineto{31.536}{16.166}
\emmoveto{31.536}{16.156}
\emlineto{30.951}{16.116}
\emmoveto{30.951}{16.106}
\emlineto{30.365}{16.068}
\emmoveto{30.365}{16.058}
\emlineto{29.778}{16.023}
\emmoveto{29.778}{16.013}
\emlineto{29.190}{15.980}
\emmoveto{29.190}{15.970}
\emlineto{28.600}{15.940}
\emmoveto{28.600}{15.930}
\emlineto{28.010}{15.903}
\emmoveto{28.010}{15.893}
\emlineto{27.419}{15.868}
\emmoveto{27.419}{15.858}
\emlineto{26.828}{15.836}
\emmoveto{26.828}{15.826}
\emlineto{26.235}{15.806}
\emmoveto{26.235}{15.796}
\emlineto{25.642}{15.780}
\emmoveto{25.642}{15.770}
\emlineto{25.048}{15.755}
\emmoveto{25.048}{15.745}
\emlineto{24.454}{15.734}
\emmoveto{24.454}{15.724}
\emlineto{23.859}{15.715}
\emmoveto{23.859}{15.705}
\emlineto{23.264}{15.698}
\emmoveto{23.264}{15.688}
\emlineto{22.668}{15.684}
\emmoveto{22.668}{15.674}
\emlineto{22.072}{15.673}
\emmoveto{22.072}{15.663}
\emlineto{21.476}{15.664}
\emmoveto{21.476}{15.654}
\emlineto{20.880}{15.658}
\emmoveto{20.880}{15.648}
\emlineto{20.283}{15.654}
\emmoveto{20.283}{15.644}
\emlineto{19.687}{15.643}
\emlineto{19.687}{15.653}
\emmoveto{19.687}{15.643}
\emlineto{19.090}{15.654}
\emmoveto{19.090}{15.644}
\emlineto{18.494}{15.658}
\emmoveto{18.494}{15.648}
\emlineto{17.897}{15.664}
\emmoveto{17.897}{15.654}
\emlineto{17.301}{15.673}
\emmoveto{17.301}{15.663}
\emlineto{16.705}{15.684}
\emmoveto{16.705}{15.674}
\emlineto{16.110}{15.697}
\emmoveto{16.110}{15.687}
\emlineto{15.514}{15.713}
\emmoveto{15.514}{15.703}
\emlineto{14.920}{15.732}
\emmoveto{14.920}{15.722}
\emlineto{14.325}{15.752}
\emmoveto{14.325}{15.742}
\emlineto{13.731}{15.776}
\emmoveto{13.731}{15.766}
\emlineto{13.138}{15.801}
\emshow{48.580}{62.500}{v=0.6}
\emmoveto{12.000}{75.070}
\emlineto{12.826}{75.077}
\emmoveto{12.826}{75.067}
\emlineto{13.652}{75.067}
\emmoveto{13.652}{75.057}
\emlineto{14.477}{75.050}
\emmoveto{14.477}{75.040}
\emlineto{15.302}{75.027}
\emmoveto{15.302}{75.017}
\emlineto{16.127}{74.997}
\emmoveto{16.127}{74.987}
\emlineto{16.950}{74.961}
\emmoveto{16.950}{74.951}
\emlineto{17.773}{74.919}
\emmoveto{17.773}{74.909}
\emlineto{18.594}{74.871}
\emmoveto{18.594}{74.861}
\emlineto{19.415}{74.817}
\emmoveto{19.415}{74.807}
\emlineto{20.234}{74.758}
\emmoveto{20.234}{74.748}
\emlineto{21.051}{74.693}
\emmoveto{21.051}{74.683}
\emlineto{21.867}{74.622}
\emmoveto{21.867}{74.612}
\emlineto{22.681}{74.546}
\emmoveto{22.681}{74.536}
\emlineto{23.494}{74.465}
\emmoveto{23.494}{74.455}
\emlineto{24.304}{74.379}
\emmoveto{24.304}{74.369}
\emlineto{25.112}{74.288}
\emmoveto{25.112}{74.278}
\emlineto{25.918}{74.191}
\emmoveto{25.918}{74.181}
\emlineto{26.721}{74.091}
\emmoveto{26.721}{74.081}
\emlineto{27.522}{73.985}
\emmoveto{27.522}{73.975}
\emlineto{28.321}{73.875}
\emmoveto{28.321}{73.865}
\emlineto{29.117}{73.761}
\emmoveto{29.117}{73.751}
\emlineto{29.910}{73.642}
\emmoveto{29.910}{73.632}
\emlineto{30.700}{73.519}
\emmoveto{30.700}{73.509}
\emlineto{31.487}{73.392}
\emmoveto{31.487}{73.382}
\emlineto{32.271}{73.261}
\emmoveto{32.271}{73.251}
\emlineto{33.052}{73.126}
\emmoveto{33.052}{73.116}
\emlineto{33.829}{72.987}
\emmoveto{33.829}{72.977}
\emlineto{34.603}{72.845}
\emmoveto{34.603}{72.835}
\emlineto{35.374}{72.699}
\emmoveto{35.374}{72.689}
\emlineto{36.141}{72.550}
\emmoveto{36.141}{72.540}
\emlineto{36.905}{72.397}
\emmoveto{36.905}{72.387}
\emlineto{37.665}{72.241}
\emmoveto{37.665}{72.231}
\emlineto{38.421}{72.082}
\emmoveto{38.421}{72.072}
\emlineto{39.173}{71.919}
\emmoveto{39.173}{71.909}
\emlineto{39.922}{71.754}
\emmoveto{39.922}{71.744}
\emlineto{40.666}{71.586}
\emmoveto{40.666}{71.576}
\emlineto{41.406}{71.415}
\emmoveto{41.406}{71.405}
\emlineto{42.143}{71.241}
\emmoveto{42.143}{71.231}
\emlineto{42.875}{71.065}
\emmoveto{42.875}{71.055}
\emlineto{43.602}{70.886}
\emmoveto{43.602}{70.876}
\emlineto{44.326}{70.704}
\emmoveto{44.326}{70.694}
\emlineto{45.045}{70.521}
\emmoveto{45.045}{70.511}
\emlineto{45.759}{70.335}
\emmoveto{45.759}{70.325}
\emlineto{46.470}{70.146}
\emmoveto{46.470}{70.136}
\emlineto{47.175}{69.956}
\emmoveto{47.175}{69.946}
\emlineto{47.876}{69.764}
\emmoveto{47.876}{69.754}
\emlineto{48.573}{69.569}
\emmoveto{48.573}{69.559}
\emlineto{49.264}{69.373}
\emmoveto{49.264}{69.363}
\emlineto{49.951}{69.175}
\emmoveto{49.951}{69.165}
\emlineto{50.634}{68.975}
\emmoveto{50.634}{68.965}
\emlineto{51.311}{68.773}
\emmoveto{51.311}{68.763}
\emlineto{51.984}{68.570}
\emmoveto{51.984}{68.560}
\emlineto{52.651}{68.365}
\emmoveto{52.651}{68.355}
\emlineto{53.314}{68.158}
\emmoveto{53.314}{68.148}
\emlineto{53.972}{67.950}
\emmoveto{53.972}{67.940}
\emlineto{54.625}{67.739}
\emmoveto{54.625}{67.729}
\emlineto{55.272}{67.526}
\emmoveto{55.272}{67.516}
\emlineto{55.915}{67.311}
\emmoveto{55.915}{67.301}
\emlineto{56.552}{67.095}
\emmoveto{56.552}{67.085}
\emlineto{57.185}{66.876}
\emmoveto{57.185}{66.866}
\emlineto{57.812}{66.655}
\emmoveto{57.812}{66.645}
\emlineto{58.433}{66.432}
\emmoveto{58.433}{66.422}
\emlineto{59.050}{66.207}
\emmoveto{59.050}{66.197}
\emlineto{59.660}{65.979}
\emmoveto{59.660}{65.969}
\emlineto{60.266}{65.750}
\emmoveto{60.266}{65.740}
\emlineto{60.866}{65.518}
\emmoveto{60.866}{65.508}
\emlineto{61.460}{65.284}
\emmoveto{61.460}{65.274}
\emlineto{62.049}{65.048}
\emmoveto{62.049}{65.038}
\emlineto{62.632}{64.809}
\emmoveto{62.632}{64.799}
\emlineto{63.209}{64.569}
\emmoveto{63.209}{64.559}
\emlineto{63.780}{64.326}
\emmoveto{63.780}{64.316}
\emlineto{64.346}{64.081}
\emmoveto{64.346}{64.071}
\emlineto{64.906}{63.833}
\emmoveto{64.906}{63.823}
\emlineto{65.460}{63.584}
\emmoveto{65.460}{63.574}
\emlineto{66.007}{63.332}
\emmoveto{66.007}{63.322}
\emlineto{66.549}{63.079}
\emmoveto{66.549}{63.069}
\emlineto{67.085}{62.823}
\emmoveto{67.085}{62.813}
\emlineto{67.614}{62.565}
\emmoveto{67.614}{62.555}
\emlineto{68.138}{62.305}
\emmoveto{68.138}{62.295}
\emlineto{68.655}{62.042}
\emmoveto{68.655}{62.032}
\emlineto{69.165}{61.778}
\emmoveto{69.165}{61.768}
\emlineto{69.670}{61.512}
\emmoveto{69.670}{61.502}
\emlineto{70.168}{61.244}
\emmoveto{70.168}{61.234}
\emlineto{70.659}{60.973}
\emmoveto{70.659}{60.963}
\emlineto{71.144}{60.702}
\emmoveto{71.144}{60.692}
\emlineto{71.623}{60.434}
\emmoveto{71.623}{60.424}
\emlineto{72.095}{60.172}
\emmoveto{72.095}{60.162}
\emlineto{72.561}{59.916}
\emmoveto{72.561}{59.906}
\emlineto{73.021}{59.665}
\emmoveto{73.021}{59.655}
\emlineto{73.475}{59.420}
\emmoveto{73.475}{59.410}
\emlineto{73.924}{59.179}
\emmoveto{73.924}{59.169}
\emlineto{74.366}{58.944}
\emmoveto{74.366}{58.934}
\emlineto{74.803}{58.713}
\emmoveto{74.803}{58.703}
\emlineto{75.235}{58.486}
\emmoveto{75.235}{58.476}
\emlineto{75.661}{58.264}
\emmoveto{75.661}{58.254}
\emlineto{76.082}{58.046}
\emmoveto{76.082}{58.036}
\emlineto{76.497}{57.831}
\emmoveto{76.497}{57.821}
\emlineto{76.908}{57.621}
\emmoveto{76.908}{57.611}
\emlineto{77.313}{57.414}
\emmoveto{77.313}{57.404}
\emlineto{77.714}{57.211}
\emmoveto{77.714}{57.201}
\emlineto{78.110}{57.011}
\emmoveto{78.110}{57.001}
\emlineto{78.501}{56.814}
\emmoveto{78.501}{56.804}
\emlineto{78.888}{56.621}
\emmoveto{78.888}{56.611}
\emlineto{79.269}{56.430}
\emmoveto{79.269}{56.420}
\emlineto{79.647}{56.242}
\emmoveto{79.647}{56.232}
\emlineto{80.020}{56.057}
\emmoveto{80.020}{56.047}
\emlineto{80.388}{55.874}
\emmoveto{80.388}{55.864}
\emlineto{80.752}{55.694}
\emmoveto{80.752}{55.684}
\emlineto{81.112}{55.516}
\emmoveto{81.112}{55.506}
\emlineto{81.467}{55.340}
\emmoveto{81.467}{55.330}
\emlineto{81.819}{55.166}
\emmoveto{81.819}{55.156}
\emlineto{82.166}{54.994}
\emmoveto{82.166}{54.984}
\emlineto{82.509}{54.824}
\emmoveto{82.509}{54.814}
\emlineto{82.848}{54.656}
\emmoveto{82.848}{54.646}
\emlineto{83.184}{54.489}
\emmoveto{83.184}{54.479}
\emlineto{83.515}{54.324}
\emmoveto{83.515}{54.314}
\emlineto{83.842}{54.160}
\emmoveto{83.842}{54.150}
\emlineto{84.165}{53.998}
\emmoveto{84.165}{53.988}
\emlineto{84.485}{53.836}
\emmoveto{84.485}{53.826}
\emlineto{84.800}{53.676}
\emmoveto{84.800}{53.666}
\emlineto{85.112}{53.517}
\emmoveto{85.112}{53.507}
\emlineto{85.420}{53.358}
\emmoveto{85.420}{53.348}
\emlineto{85.724}{53.201}
\emmoveto{85.724}{53.191}
\emlineto{86.024}{53.044}
\emmoveto{86.024}{53.034}
\emlineto{86.321}{52.887}
\emmoveto{86.321}{52.877}
\emlineto{86.614}{52.731}
\emmoveto{86.614}{52.721}
\emlineto{86.903}{52.576}
\emmoveto{86.903}{52.566}
\emlineto{87.188}{52.421}
\emmoveto{87.188}{52.411}
\emlineto{87.470}{52.266}
\emmoveto{87.470}{52.256}
\emlineto{87.748}{52.112}
\emmoveto{87.748}{52.102}
\emlineto{88.023}{51.958}
\emmoveto{88.023}{51.948}
\emlineto{88.293}{51.803}
\emmoveto{88.293}{51.793}
\emlineto{88.560}{51.649}
\emmoveto{88.560}{51.639}
\emlineto{88.823}{51.495}
\emmoveto{88.823}{51.485}
\emlineto{89.083}{51.340}
\emmoveto{89.083}{51.330}
\emlineto{89.339}{51.185}
\emmoveto{89.339}{51.175}
\emlineto{89.591}{51.031}
\emmoveto{89.591}{51.021}
\emlineto{89.839}{50.876}
\emmoveto{89.839}{50.866}
\emlineto{90.084}{50.721}
\emmoveto{90.084}{50.711}
\emlineto{90.325}{50.567}
\emmoveto{90.325}{50.557}
\emlineto{90.563}{50.412}
\emmoveto{90.563}{50.402}
\emlineto{90.796}{50.258}
\emmoveto{90.796}{50.248}
\emlineto{91.026}{50.105}
\emmoveto{91.026}{50.095}
\emlineto{91.253}{49.951}
\emmoveto{91.253}{49.941}
\emlineto{91.475}{49.799}
\emmoveto{91.475}{49.789}
\emlineto{91.694}{49.646}
\emmoveto{91.694}{49.636}
\emlineto{91.910}{49.494}
\emmoveto{91.910}{49.484}
\emlineto{92.121}{49.343}
\emmoveto{92.121}{49.333}
\emlineto{92.330}{49.193}
\emmoveto{92.330}{49.183}
\emlineto{92.534}{49.043}
\emmoveto{92.534}{49.033}
\emlineto{92.735}{48.893}
\emmoveto{92.735}{48.883}
\emlineto{92.932}{48.745}
\emmoveto{92.932}{48.735}
\emlineto{93.126}{48.597}
\emmoveto{93.126}{48.587}
\emlineto{93.316}{48.449}
\emmoveto{93.316}{48.439}
\emlineto{93.503}{48.303}
\emmoveto{93.503}{48.293}
\emlineto{93.687}{48.157}
\emmoveto{93.687}{48.147}
\emlineto{93.866}{48.012}
\emmoveto{93.866}{48.002}
\emlineto{94.043}{47.868}
\emmoveto{94.043}{47.858}
\emlineto{94.215}{47.724}
\emmoveto{94.215}{47.714}
\emlineto{94.385}{47.581}
\emmoveto{94.385}{47.571}
\emlineto{94.551}{47.439}
\emmoveto{94.551}{47.429}
\emlineto{94.714}{47.298}
\emmoveto{94.714}{47.288}
\emlineto{94.873}{47.158}
\emmoveto{94.873}{47.148}
\emlineto{95.029}{47.018}
\emmoveto{95.029}{47.008}
\emlineto{95.182}{46.879}
\emmoveto{95.182}{46.869}
\emlineto{95.331}{46.740}
\emmoveto{95.331}{46.730}
\emlineto{95.477}{46.603}
\emmoveto{95.477}{46.593}
\emlineto{95.619}{46.466}
\emmoveto{95.619}{46.456}
\emlineto{95.759}{46.330}
\emmoveto{95.759}{46.320}
\emlineto{95.895}{46.194}
\emmoveto{95.895}{46.184}
\emlineto{96.028}{46.059}
\emmoveto{96.028}{46.049}
\emlineto{96.158}{45.924}
\emmoveto{96.158}{45.914}
\emlineto{96.284}{45.791}
\emmoveto{96.284}{45.781}
\emlineto{96.407}{45.657}
\emmoveto{96.407}{45.647}
\emlineto{96.528}{45.525}
\emmoveto{96.528}{45.515}
\emlineto{96.644}{45.393}
\emmoveto{96.644}{45.383}
\emlineto{96.758}{45.261}
\emmoveto{96.758}{45.251}
\emlineto{96.869}{45.130}
\emmoveto{96.869}{45.120}
\emlineto{96.976}{44.999}
\emmoveto{96.976}{44.989}
\emlineto{97.081}{44.869}
\emmoveto{97.081}{44.859}
\emlineto{97.182}{44.739}
\emmoveto{97.182}{44.729}
\emlineto{97.280}{44.609}
\emmoveto{97.280}{44.599}
\emlineto{97.375}{44.480}
\emmoveto{97.375}{44.470}
\emlineto{97.467}{44.351}
\emmoveto{97.467}{44.341}
\emlineto{97.556}{44.222}
\emmoveto{97.556}{44.212}
\emlineto{97.642}{44.093}
\emmoveto{97.642}{44.083}
\emlineto{97.725}{43.965}
\emmoveto{97.725}{43.955}
\emlineto{97.804}{43.836}
\emmoveto{97.804}{43.826}
\emlineto{97.881}{43.708}
\emmoveto{97.881}{43.698}
\emlineto{97.954}{43.580}
\emmoveto{97.954}{43.570}
\emlineto{98.025}{43.452}
\emmoveto{98.025}{43.442}
\emlineto{98.092}{43.324}
\emmoveto{98.092}{43.314}
\emlineto{98.156}{43.196}
\emmoveto{98.156}{43.186}
\emlineto{98.218}{43.068}
\emmoveto{98.218}{43.058}
\emlineto{98.276}{42.941}
\emmoveto{98.276}{42.931}
\emlineto{98.331}{42.813}
\emmoveto{98.331}{42.803}
\emlineto{98.383}{42.685}
\emmoveto{98.432}{42.547}
\emlineto{98.521}{42.300}
\emmoveto{98.561}{42.162}
\emlineto{98.631}{41.916}
\emmoveto{98.662}{41.777}
\emlineto{98.714}{41.530}
\emmoveto{98.735}{41.391}
\emlineto{98.768}{41.142}
\emmoveto{98.780}{41.003}
\emlineto{98.795}{40.754}
\emmoveto{98.798}{40.614}
\emlineto{98.794}{40.363}
\emmoveto{98.788}{40.223}
\emlineto{98.765}{39.971}
\emmoveto{98.749}{39.830}
\emlineto{98.708}{39.577}
\emmoveto{98.682}{39.435}
\emlineto{98.622}{39.180}
\emmoveto{98.622}{39.170}
\emlineto{98.587}{39.047}
\emmoveto{98.587}{39.037}
\emlineto{98.549}{38.914}
\emmoveto{98.549}{38.904}
\emlineto{98.508}{38.781}
\emmoveto{98.508}{38.771}
\emlineto{98.463}{38.647}
\emmoveto{98.463}{38.637}
\emlineto{98.415}{38.513}
\emmoveto{98.415}{38.503}
\emlineto{98.364}{38.379}
\emmoveto{98.364}{38.369}
\emlineto{98.310}{38.244}
\emmoveto{98.310}{38.234}
\emlineto{98.253}{38.109}
\emmoveto{98.253}{38.099}
\emlineto{98.192}{37.974}
\emmoveto{98.192}{37.964}
\emlineto{98.128}{37.838}
\emmoveto{98.128}{37.828}
\emlineto{98.061}{37.702}
\emmoveto{98.061}{37.692}
\emlineto{97.991}{37.565}
\emmoveto{97.991}{37.555}
\emlineto{97.917}{37.428}
\emmoveto{97.917}{37.418}
\emlineto{97.840}{37.290}
\emmoveto{97.840}{37.280}
\emlineto{97.760}{37.152}
\emmoveto{97.760}{37.142}
\emlineto{97.676}{37.014}
\emmoveto{97.676}{37.004}
\emlineto{97.590}{36.875}
\emmoveto{97.590}{36.865}
\emlineto{97.499}{36.735}
\emmoveto{97.499}{36.725}
\emlineto{97.406}{36.595}
\emmoveto{97.406}{36.585}
\emlineto{97.309}{36.455}
\emmoveto{97.309}{36.445}
\emlineto{97.209}{36.313}
\emmoveto{97.209}{36.303}
\emlineto{97.105}{36.172}
\emmoveto{97.105}{36.162}
\emlineto{96.998}{36.029}
\emmoveto{96.998}{36.019}
\emlineto{96.887}{35.886}
\emmoveto{96.887}{35.876}
\emlineto{96.774}{35.743}
\emmoveto{96.774}{35.733}
\emlineto{96.656}{35.599}
\emmoveto{96.656}{35.589}
\emlineto{96.535}{35.454}
\emmoveto{96.535}{35.444}
\emlineto{96.411}{35.308}
\emmoveto{96.411}{35.298}
\emlineto{96.283}{35.162}
\emmoveto{96.283}{35.152}
\emlineto{96.152}{35.015}
\emmoveto{96.152}{35.005}
\emlineto{96.017}{34.867}
\emmoveto{96.017}{34.857}
\emlineto{95.879}{34.719}
\emmoveto{95.879}{34.709}
\emlineto{95.737}{34.569}
\emmoveto{95.737}{34.559}
\emlineto{95.591}{34.419}
\emmoveto{95.591}{34.409}
\emlineto{95.442}{34.269}
\emmoveto{95.442}{34.259}
\emlineto{95.290}{34.117}
\emmoveto{95.290}{34.107}
\emlineto{95.133}{33.964}
\emmoveto{95.133}{33.954}
\emlineto{94.973}{33.811}
\emmoveto{94.973}{33.801}
\emlineto{94.809}{33.657}
\emmoveto{94.809}{33.647}
\emlineto{94.642}{33.502}
\emmoveto{94.642}{33.492}
\emlineto{94.471}{33.346}
\emmoveto{94.471}{33.336}
\emlineto{94.296}{33.189}
\emmoveto{94.296}{33.179}
\emlineto{94.117}{33.031}
\emmoveto{94.117}{33.021}
\emlineto{93.935}{32.872}
\emmoveto{93.935}{32.862}
\emlineto{93.749}{32.712}
\emmoveto{93.749}{32.702}
\emlineto{93.558}{32.551}
\emmoveto{93.558}{32.541}
\emlineto{93.365}{32.389}
\emmoveto{93.365}{32.379}
\emlineto{93.167}{32.226}
\emmoveto{93.167}{32.216}
\emlineto{92.965}{32.063}
\emmoveto{92.965}{32.053}
\emlineto{92.759}{31.898}
\emmoveto{92.759}{31.888}
\emlineto{92.550}{31.732}
\emmoveto{92.550}{31.722}
\emlineto{92.336}{31.564}
\emmoveto{92.336}{31.554}
\emlineto{92.118}{31.396}
\emmoveto{92.118}{31.386}
\emlineto{91.897}{31.227}
\emmoveto{91.897}{31.217}
\emlineto{91.671}{31.056}
\emmoveto{91.671}{31.046}
\emlineto{91.441}{30.885}
\emmoveto{91.441}{30.875}
\emlineto{91.207}{30.712}
\emmoveto{91.207}{30.702}
\emlineto{90.969}{30.538}
\emmoveto{90.969}{30.528}
\emlineto{90.726}{30.362}
\emmoveto{90.726}{30.352}
\emlineto{90.480}{30.186}
\emmoveto{90.480}{30.176}
\emlineto{90.229}{30.008}
\emmoveto{90.229}{29.998}
\emlineto{89.974}{29.829}
\emmoveto{89.974}{29.819}
\emlineto{89.715}{29.649}
\emmoveto{89.715}{29.639}
\emlineto{89.451}{29.467}
\emmoveto{89.451}{29.457}
\emlineto{89.183}{29.284}
\emmoveto{89.183}{29.274}
\emlineto{88.911}{29.100}
\emmoveto{88.911}{29.090}
\emlineto{88.634}{28.914}
\emmoveto{88.634}{28.904}
\emlineto{88.352}{28.727}
\emmoveto{88.352}{28.717}
\emlineto{88.067}{28.538}
\emmoveto{88.067}{28.528}
\emlineto{87.776}{28.348}
\emmoveto{87.776}{28.338}
\emlineto{87.481}{28.157}
\emmoveto{87.481}{28.147}
\emlineto{87.182}{27.964}
\emmoveto{87.182}{27.954}
\emlineto{86.878}{27.770}
\emmoveto{86.878}{27.760}
\emlineto{86.569}{27.574}
\emmoveto{86.569}{27.564}
\emlineto{86.255}{27.376}
\emmoveto{86.255}{27.366}
\emlineto{85.937}{27.177}
\emmoveto{85.937}{27.167}
\emlineto{85.614}{26.977}
\emmoveto{85.614}{26.967}
\emlineto{85.286}{26.775}
\emmoveto{85.286}{26.765}
\emlineto{84.953}{26.571}
\emmoveto{84.953}{26.561}
\emlineto{84.616}{26.366}
\emmoveto{84.616}{26.356}
\emlineto{84.273}{26.159}
\emmoveto{84.273}{26.149}
\emlineto{83.926}{25.950}
\emmoveto{83.926}{25.940}
\emlineto{83.573}{25.739}
\emmoveto{83.573}{25.729}
\emlineto{83.215}{25.527}
\emmoveto{83.215}{25.517}
\emlineto{82.853}{25.313}
\emmoveto{82.853}{25.303}
\emlineto{82.485}{25.098}
\emmoveto{82.485}{25.088}
\emlineto{82.112}{24.880}
\emmoveto{82.112}{24.870}
\emlineto{81.733}{24.661}
\emmoveto{81.733}{24.651}
\emlineto{81.350}{24.440}
\emmoveto{81.350}{24.430}
\emlineto{80.961}{24.216}
\emmoveto{80.961}{24.206}
\emlineto{80.567}{23.991}
\emmoveto{80.567}{23.981}
\emlineto{80.167}{23.765}
\emmoveto{80.167}{23.755}
\emlineto{79.762}{23.536}
\emmoveto{79.762}{23.526}
\emlineto{79.352}{23.305}
\emmoveto{79.352}{23.295}
\emlineto{78.935}{23.072}
\emmoveto{78.935}{23.062}
\emlineto{78.514}{22.837}
\emmoveto{78.514}{22.827}
\emlineto{78.086}{22.600}
\emmoveto{78.086}{22.590}
\emlineto{77.653}{22.361}
\emmoveto{77.653}{22.351}
\emlineto{77.214}{22.120}
\emmoveto{77.214}{22.110}
\emlineto{76.770}{21.877}
\emmoveto{76.770}{21.867}
\emlineto{76.319}{21.631}
\emmoveto{76.319}{21.621}
\emlineto{75.863}{21.384}
\emmoveto{75.863}{21.374}
\emlineto{75.401}{21.134}
\emmoveto{75.401}{21.124}
\emlineto{74.932}{20.882}
\emmoveto{74.932}{20.872}
\emlineto{74.458}{20.628}
\emmoveto{74.458}{20.618}
\emlineto{73.977}{20.371}
\emmoveto{73.977}{20.361}
\emlineto{73.491}{20.112}
\emmoveto{73.491}{20.102}
\emlineto{72.998}{19.851}
\emmoveto{72.998}{19.841}
\emlineto{72.498}{19.587}
\emmoveto{72.498}{19.577}
\emlineto{71.993}{19.321}
\emmoveto{71.993}{19.311}
\emlineto{71.481}{19.052}
\emmoveto{71.481}{19.042}
\emlineto{70.963}{18.781}
\emmoveto{70.963}{18.771}
\emlineto{70.438}{18.512}
\emmoveto{70.438}{18.502}
\emlineto{69.906}{18.250}
\emmoveto{69.906}{18.240}
\emlineto{69.369}{17.993}
\emmoveto{69.369}{17.983}
\emlineto{68.826}{17.743}
\emmoveto{68.826}{17.733}
\emlineto{68.276}{17.498}
\emmoveto{68.276}{17.488}
\emlineto{67.721}{17.259}
\emmoveto{67.721}{17.249}
\emlineto{67.160}{17.026}
\emmoveto{67.160}{17.016}
\emlineto{66.594}{16.798}
\emmoveto{66.594}{16.788}
\emlineto{66.022}{16.575}
\emmoveto{66.022}{16.565}
\emlineto{65.445}{16.358}
\emmoveto{65.445}{16.348}
\emlineto{64.863}{16.146}
\emmoveto{64.863}{16.136}
\emlineto{64.275}{15.940}
\emmoveto{64.275}{15.930}
\emlineto{63.683}{15.738}
\emmoveto{63.683}{15.728}
\emlineto{63.086}{15.542}
\emmoveto{63.086}{15.532}
\emlineto{62.485}{15.350}
\emmoveto{62.485}{15.340}
\emlineto{61.879}{15.163}
\emmoveto{61.879}{15.153}
\emlineto{61.268}{14.981}
\emmoveto{61.268}{14.971}
\emlineto{60.654}{14.803}
\emmoveto{60.654}{14.793}
\emlineto{60.035}{14.630}
\emmoveto{60.035}{14.620}
\emlineto{59.412}{14.462}
\emmoveto{59.412}{14.452}
\emlineto{58.785}{14.298}
\emmoveto{58.785}{14.288}
\emlineto{58.154}{14.138}
\emmoveto{58.154}{14.128}
\emlineto{57.519}{13.982}
\emmoveto{57.519}{13.972}
\emlineto{56.881}{13.830}
\emmoveto{56.881}{13.820}
\emlineto{56.239}{13.683}
\emmoveto{56.239}{13.673}
\emlineto{55.594}{13.539}
\emmoveto{55.594}{13.529}
\emlineto{54.945}{13.399}
\emmoveto{54.945}{13.389}
\emlineto{54.293}{13.263}
\emmoveto{54.293}{13.253}
\emlineto{53.637}{13.130}
\emmoveto{53.637}{13.120}
\emlineto{52.979}{13.001}
\emmoveto{52.979}{12.991}
\emlineto{52.317}{12.876}
\emmoveto{52.317}{12.866}
\emlineto{51.653}{12.754}
\emmoveto{51.653}{12.744}
\emlineto{50.985}{12.635}
\emmoveto{50.985}{12.625}
\emlineto{50.315}{12.520}
\emmoveto{50.315}{12.510}
\emlineto{49.642}{12.408}
\emmoveto{49.642}{12.398}
\emlineto{48.967}{12.299}
\emmoveto{48.967}{12.289}
\emlineto{48.289}{12.193}
\emmoveto{48.289}{12.183}
\emlineto{47.608}{12.089}
\emmoveto{47.608}{12.079}
\emlineto{46.925}{11.989}
\emmoveto{46.925}{11.979}
\emlineto{46.240}{11.891}
\emmoveto{46.240}{11.881}
\emlineto{45.552}{11.797}
\emmoveto{45.552}{11.787}
\emlineto{44.862}{11.704}
\emmoveto{44.862}{11.694}
\emlineto{44.170}{11.614}
\emmoveto{44.170}{11.604}
\emlineto{43.476}{11.527}
\emmoveto{43.476}{11.517}
\emlineto{42.780}{11.442}
\emmoveto{42.780}{11.432}
\emlineto{42.081}{11.360}
\emmoveto{42.081}{11.350}
\emlineto{41.381}{11.279}
\emmoveto{41.381}{11.269}
\emlineto{40.679}{11.201}
\emmoveto{40.679}{11.191}
\emlineto{39.975}{11.125}
\emmoveto{39.975}{11.115}
\emlineto{39.269}{11.050}
\emmoveto{39.269}{11.040}
\emlineto{38.561}{10.978}
\emmoveto{38.561}{10.968}
\emlineto{37.852}{10.908}
\emmoveto{37.852}{10.898}
\emlineto{37.141}{10.840}
\emmoveto{37.141}{10.830}
\emlineto{36.429}{10.774}
\emmoveto{36.429}{10.764}
\emlineto{35.715}{10.711}
\emmoveto{35.715}{10.701}
\emlineto{34.999}{10.649}
\emmoveto{34.999}{10.639}
\emlineto{34.282}{10.591}
\emmoveto{34.282}{10.581}
\emlineto{33.564}{10.534}
\emmoveto{33.564}{10.524}
\emlineto{32.844}{10.481}
\emmoveto{32.844}{10.471}
\emlineto{32.123}{10.430}
\emmoveto{32.123}{10.420}
\emlineto{31.401}{10.382}
\emmoveto{31.401}{10.372}
\emlineto{30.678}{10.336}
\emmoveto{30.678}{10.326}
\emlineto{29.953}{10.294}
\emmoveto{29.953}{10.284}
\emlineto{29.228}{10.254}
\emmoveto{29.228}{10.244}
\emlineto{28.502}{10.217}
\emmoveto{28.502}{10.207}
\emlineto{27.775}{10.184}
\emmoveto{27.775}{10.174}
\emlineto{27.047}{10.153}
\emmoveto{27.047}{10.143}
\emlineto{26.318}{10.125}
\emmoveto{26.318}{10.115}
\emlineto{25.589}{10.101}
\emmoveto{25.589}{10.091}
\emlineto{24.860}{10.079}
\emmoveto{24.860}{10.069}
\emlineto{24.130}{10.061}
\emmoveto{24.130}{10.051}
\emlineto{23.399}{10.046}
\emmoveto{23.399}{10.036}
\emlineto{22.668}{10.034}
\emmoveto{22.668}{10.024}
\emlineto{21.937}{10.026}
\emmoveto{21.937}{10.016}
\emlineto{21.206}{10.021}
\emmoveto{21.206}{10.011}
\emlineto{20.474}{10.009}
\emlineto{20.474}{10.019}
\emmoveto{20.474}{10.009}
\emlineto{19.743}{10.020}
\emmoveto{19.743}{10.010}
\emlineto{19.012}{10.024}
\emmoveto{19.012}{10.014}
\emlineto{18.280}{10.032}
\emmoveto{18.280}{10.022}
\emlineto{17.549}{10.043}
\emmoveto{17.549}{10.033}
\emlineto{16.819}{10.058}
\emmoveto{16.819}{10.048}
\emlineto{16.089}{10.076}
\emmoveto{16.089}{10.066}
\emlineto{15.359}{10.097}
\emmoveto{15.359}{10.087}
\emlineto{14.630}{10.121}
\emmoveto{14.630}{10.111}
\emlineto{13.901}{10.149}
\emmoveto{13.901}{10.139}
\emlineto{13.173}{10.180}
\emshow{48.580}{68.800}{v=0.7}
\emmoveto{12.000}{80.000}
\emlineto{12.944}{80.006}
\emmoveto{12.944}{79.996}
\emlineto{13.888}{79.994}
\emmoveto{13.888}{79.984}
\emlineto{14.831}{79.975}
\emmoveto{14.831}{79.965}
\emlineto{15.774}{79.949}
\emmoveto{15.774}{79.939}
\emlineto{16.716}{79.915}
\emmoveto{16.716}{79.905}
\emlineto{17.657}{79.874}
\emmoveto{17.657}{79.864}
\emlineto{18.598}{79.826}
\emmoveto{18.598}{79.816}
\emlineto{19.537}{79.771}
\emmoveto{19.537}{79.761}
\emlineto{20.474}{79.709}
\emmoveto{20.474}{79.699}
\emlineto{21.410}{79.641}
\emmoveto{21.410}{79.631}
\emlineto{22.344}{79.567}
\emmoveto{22.344}{79.557}
\emlineto{23.277}{79.486}
\emmoveto{23.277}{79.476}
\emlineto{24.207}{79.399}
\emmoveto{24.207}{79.389}
\emlineto{25.136}{79.307}
\emmoveto{25.136}{79.297}
\emlineto{26.062}{79.208}
\emmoveto{26.062}{79.198}
\emlineto{26.985}{79.104}
\emmoveto{26.985}{79.094}
\emlineto{27.906}{78.994}
\emmoveto{27.906}{78.984}
\emlineto{28.824}{78.879}
\emmoveto{28.824}{78.869}
\emlineto{29.740}{78.758}
\emmoveto{29.740}{78.748}
\emlineto{30.652}{78.632}
\emmoveto{30.652}{78.622}
\emlineto{31.562}{78.502}
\emmoveto{31.562}{78.492}
\emlineto{32.468}{78.366}
\emmoveto{32.468}{78.356}
\emlineto{33.371}{78.225}
\emmoveto{33.371}{78.215}
\emlineto{34.271}{78.080}
\emmoveto{34.271}{78.070}
\emlineto{35.167}{77.930}
\emmoveto{35.167}{77.920}
\emlineto{36.059}{77.776}
\emmoveto{36.059}{77.766}
\emlineto{36.948}{77.618}
\emmoveto{36.948}{77.608}
\emlineto{37.833}{77.455}
\emmoveto{37.833}{77.445}
\emlineto{38.713}{77.288}
\emmoveto{38.713}{77.278}
\emlineto{39.590}{77.118}
\emmoveto{39.590}{77.108}
\emlineto{40.463}{76.943}
\emmoveto{40.463}{76.933}
\emlineto{41.331}{76.765}
\emmoveto{41.331}{76.755}
\emlineto{42.195}{76.583}
\emmoveto{42.195}{76.573}
\emlineto{43.055}{76.397}
\emmoveto{43.055}{76.387}
\emlineto{43.911}{76.208}
\emmoveto{43.911}{76.198}
\emlineto{44.761}{76.016}
\emmoveto{44.761}{76.006}
\emlineto{45.607}{75.821}
\emmoveto{45.607}{75.811}
\emlineto{46.449}{75.622}
\emmoveto{46.449}{75.612}
\emlineto{47.285}{75.421}
\emmoveto{47.285}{75.411}
\emlineto{48.117}{75.216}
\emmoveto{48.117}{75.206}
\emlineto{48.944}{75.009}
\emmoveto{48.944}{74.999}
\emlineto{49.765}{74.799}
\emmoveto{49.765}{74.789}
\emlineto{50.582}{74.586}
\emmoveto{50.582}{74.576}
\emlineto{51.394}{74.371}
\emmoveto{51.394}{74.361}
\emlineto{52.200}{74.154}
\emmoveto{52.200}{74.144}
\emlineto{53.001}{73.934}
\emmoveto{53.001}{73.924}
\emlineto{53.797}{73.711}
\emmoveto{53.797}{73.701}
\emlineto{54.588}{73.487}
\emmoveto{54.588}{73.477}
\emlineto{55.373}{73.261}
\emmoveto{55.373}{73.251}
\emlineto{56.153}{73.032}
\emmoveto{56.153}{73.022}
\emlineto{56.927}{72.802}
\emmoveto{56.927}{72.792}
\emlineto{57.696}{72.570}
\emmoveto{57.696}{72.560}
\emlineto{58.459}{72.335}
\emmoveto{58.459}{72.325}
\emlineto{59.216}{72.099}
\emmoveto{59.216}{72.089}
\emlineto{59.968}{71.860}
\emmoveto{59.968}{71.850}
\emlineto{60.714}{71.620}
\emmoveto{60.714}{71.610}
\emlineto{61.454}{71.376}
\emmoveto{61.454}{71.366}
\emlineto{62.189}{71.131}
\emmoveto{62.189}{71.121}
\emlineto{62.917}{70.883}
\emmoveto{62.917}{70.873}
\emlineto{63.640}{70.633}
\emmoveto{63.640}{70.623}
\emlineto{64.356}{70.381}
\emmoveto{64.356}{70.371}
\emlineto{65.067}{70.126}
\emmoveto{65.067}{70.116}
\emlineto{65.771}{69.868}
\emmoveto{65.771}{69.858}
\emlineto{66.469}{69.609}
\emmoveto{66.469}{69.599}
\emlineto{67.161}{69.346}
\emmoveto{67.161}{69.336}
\emlineto{67.847}{69.081}
\emmoveto{67.847}{69.071}
\emlineto{68.526}{68.814}
\emmoveto{68.526}{68.804}
\emlineto{69.199}{68.544}
\emmoveto{69.199}{68.534}
\emlineto{69.865}{68.271}
\emmoveto{69.865}{68.261}
\emlineto{70.525}{67.996}
\emmoveto{70.525}{67.986}
\emlineto{71.178}{67.719}
\emmoveto{71.178}{67.709}
\emlineto{71.824}{67.448}
\emmoveto{71.824}{67.438}
\emlineto{72.464}{67.184}
\emmoveto{72.464}{67.174}
\emlineto{73.098}{66.928}
\emmoveto{73.098}{66.918}
\emlineto{73.726}{66.679}
\emmoveto{73.726}{66.669}
\emlineto{74.348}{66.437}
\emmoveto{74.348}{66.427}
\emlineto{74.964}{66.202}
\emmoveto{74.964}{66.192}
\emlineto{75.575}{65.974}
\emmoveto{75.575}{65.964}
\emlineto{76.180}{65.752}
\emmoveto{76.180}{65.742}
\emlineto{76.781}{65.536}
\emmoveto{76.781}{65.526}
\emlineto{77.376}{65.326}
\emmoveto{77.376}{65.316}
\emlineto{77.966}{65.121}
\emmoveto{77.966}{65.111}
\emlineto{78.551}{64.922}
\emmoveto{78.551}{64.912}
\emlineto{79.131}{64.728}
\emmoveto{79.131}{64.718}
\emlineto{79.707}{64.539}
\emmoveto{79.707}{64.529}
\emlineto{80.279}{64.354}
\emmoveto{80.279}{64.344}
\emlineto{80.846}{64.174}
\emmoveto{80.846}{64.164}
\emlineto{81.409}{63.999}
\emmoveto{81.409}{63.989}
\emlineto{81.967}{63.827}
\emmoveto{81.967}{63.817}
\emlineto{82.522}{63.660}
\emmoveto{82.522}{63.650}
\emlineto{83.073}{63.496}
\emmoveto{83.073}{63.486}
\emlineto{83.619}{63.336}
\emmoveto{83.619}{63.326}
\emlineto{84.162}{63.179}
\emmoveto{84.162}{63.169}
\emlineto{84.702}{63.026}
\emmoveto{84.702}{63.016}
\emlineto{85.237}{62.876}
\emmoveto{85.237}{62.866}
\emlineto{85.769}{62.729}
\emmoveto{85.769}{62.719}
\emlineto{86.298}{62.585}
\emmoveto{86.298}{62.575}
\emlineto{86.823}{62.444}
\emmoveto{86.823}{62.434}
\emlineto{87.345}{62.305}
\emmoveto{87.345}{62.295}
\emlineto{87.864}{62.169}
\emmoveto{87.864}{62.159}
\emlineto{88.379}{62.036}
\emmoveto{88.379}{62.026}
\emlineto{88.891}{61.905}
\emmoveto{88.891}{61.895}
\emlineto{89.400}{61.775}
\emmoveto{89.400}{61.765}
\emlineto{89.906}{61.648}
\emmoveto{89.906}{61.638}
\emlineto{90.409}{61.523}
\emmoveto{90.409}{61.513}
\emlineto{90.909}{61.400}
\emmoveto{90.909}{61.390}
\emlineto{91.406}{61.279}
\emmoveto{91.406}{61.269}
\emlineto{91.900}{61.159}
\emmoveto{91.900}{61.149}
\emlineto{92.392}{61.041}
\emmoveto{92.392}{61.031}
\emlineto{92.880}{60.924}
\emmoveto{92.880}{60.914}
\emlineto{93.366}{60.809}
\emmoveto{93.366}{60.799}
\emlineto{93.849}{60.695}
\emmoveto{93.849}{60.685}
\emlineto{94.329}{60.582}
\emmoveto{94.329}{60.572}
\emlineto{94.807}{60.470}
\emmoveto{94.807}{60.460}
\emlineto{95.282}{60.360}
\emmoveto{95.282}{60.350}
\emlineto{95.754}{60.250}
\emmoveto{95.754}{60.240}
\emlineto{96.224}{60.141}
\emmoveto{96.224}{60.131}
\emlineto{96.691}{60.032}
\emmoveto{96.691}{60.022}
\emlineto{97.156}{59.924}
\emmoveto{97.156}{59.914}
\emlineto{97.617}{59.817}
\emmoveto{97.617}{59.807}
\emlineto{98.077}{59.710}
\emmoveto{98.077}{59.700}
\emlineto{98.534}{59.604}
\emmoveto{98.534}{59.594}
\emlineto{98.988}{59.498}
\emmoveto{98.988}{59.488}
\emlineto{99.440}{59.393}
\emmoveto{99.440}{59.383}
\emlineto{99.889}{59.289}
\emmoveto{99.889}{59.279}
\emlineto{100.336}{59.186}
\emmoveto{100.336}{59.176}
\emlineto{100.780}{59.084}
\emmoveto{100.780}{59.074}
\emlineto{101.222}{58.982}
\emmoveto{101.222}{58.972}
\emlineto{101.661}{58.882}
\emmoveto{101.661}{58.872}
\emlineto{102.098}{58.784}
\emmoveto{102.098}{58.774}
\emlineto{102.533}{58.686}
\emmoveto{102.533}{58.676}
\emlineto{102.966}{58.590}
\emmoveto{102.966}{58.580}
\emlineto{103.396}{58.496}
\emmoveto{103.396}{58.486}
\emlineto{103.824}{58.403}
\emmoveto{103.824}{58.393}
\emlineto{104.249}{58.311}
\emmoveto{104.249}{58.301}
\emlineto{104.673}{58.222}
\emmoveto{104.673}{58.212}
\emlineto{105.094}{58.133}
\emmoveto{105.094}{58.123}
\emlineto{105.514}{58.047}
\emmoveto{105.514}{58.037}
\emlineto{105.931}{57.962}
\emmoveto{105.931}{57.952}
\emlineto{106.346}{57.879}
\emmoveto{106.346}{57.869}
\emlineto{106.759}{57.798}
\emmoveto{106.759}{57.788}
\emlineto{107.171}{57.719}
\emmoveto{107.171}{57.709}
\emlineto{107.580}{57.641}
\emmoveto{107.580}{57.631}
\emlineto{107.988}{57.566}
\emmoveto{107.988}{57.556}
\emlineto{108.394}{57.492}
\emmoveto{108.394}{57.482}
\emlineto{108.798}{57.420}
\emmoveto{108.798}{57.410}
\emlineto{109.200}{57.350}
\emmoveto{109.200}{57.340}
\emlineto{109.601}{57.282}
\emmoveto{109.601}{57.272}
\emlineto{110.000}{57.216}
\emmoveto{110.000}{57.206}
\emlineto{110.398}{57.151}
\emmoveto{110.398}{57.141}
\emlineto{110.794}{57.089}
\emmoveto{110.794}{57.079}
\emlineto{111.188}{57.028}
\emmoveto{111.188}{57.018}
\emlineto{111.582}{56.970}
\emmoveto{111.582}{56.960}
\emlineto{111.973}{56.913}
\emmoveto{111.973}{56.903}
\emlineto{112.364}{56.858}
\emmoveto{112.364}{56.848}
\emlineto{112.753}{56.805}
\emmoveto{112.753}{56.795}
\emlineto{113.141}{56.753}
\emmoveto{113.141}{56.743}
\emlineto{113.528}{56.704}
\emmoveto{113.528}{56.694}
\emlineto{113.913}{56.656}
\emmoveto{113.913}{56.646}
\emlineto{114.298}{56.611}
\emmoveto{114.298}{56.601}
\emlineto{114.681}{56.567}
\emmoveto{114.681}{56.557}
\emlineto{115.063}{56.524}
\emmoveto{115.063}{56.514}
\emlineto{115.445}{56.484}
\emmoveto{115.445}{56.474}
\emlineto{115.825}{56.445}
\emmoveto{115.825}{56.435}
\emlineto{116.205}{56.408}
\emmoveto{116.205}{56.398}
\emlineto{116.583}{56.373}
\emmoveto{116.583}{56.363}
\emlineto{116.961}{56.339}
\emmoveto{116.961}{56.329}
\emlineto{117.338}{56.307}
\emmoveto{117.338}{56.297}
\emlineto{117.714}{56.276}
\emmoveto{117.714}{56.266}
\emlineto{118.090}{56.247}
\emmoveto{118.090}{56.237}
\emlineto{118.464}{56.220}
\emmoveto{118.464}{56.210}
\emlineto{118.839}{56.194}
\emmoveto{118.839}{56.184}
\emlineto{119.212}{56.170}
\emmoveto{119.212}{56.160}
\emlineto{119.585}{56.147}
\emmoveto{119.585}{56.137}
\emlineto{119.958}{56.126}
\emmoveto{119.958}{56.116}
\emlineto{120.330}{56.106}
\emmoveto{120.330}{56.096}
\emlineto{120.702}{56.087}
\emmoveto{120.702}{56.077}
\emlineto{121.073}{56.070}
\emmoveto{121.073}{56.060}
\emlineto{121.443}{56.055}
\emmoveto{121.443}{56.045}
\emlineto{121.814}{56.040}
\emmoveto{121.814}{56.030}
\emlineto{122.184}{56.028}
\emmoveto{122.184}{56.018}
\emlineto{122.554}{56.016}
\emmoveto{122.554}{56.006}
\emlineto{122.923}{56.006}
\emmoveto{122.923}{55.996}
\emlineto{123.293}{55.997}
\emmoveto{123.293}{55.987}
\emlineto{123.662}{55.990}
\emmoveto{123.662}{55.980}
\emlineto{124.031}{55.984}
\emmoveto{124.031}{55.974}
\emlineto{124.399}{55.980}
\emmoveto{124.399}{55.970}
\emlineto{124.768}{55.977}
\emmoveto{124.768}{55.967}
\emlineto{125.137}{55.975}
\emmoveto{125.137}{55.965}
\emlineto{125.506}{55.975}
\emmoveto{125.506}{55.965}
\emlineto{125.874}{55.976}
\emmoveto{125.874}{55.966}
\emlineto{126.243}{55.979}
\emmoveto{126.243}{55.969}
\emlineto{126.612}{55.983}
\emmoveto{126.612}{55.973}
\emlineto{126.981}{55.989}
\emmoveto{126.981}{55.979}
\emlineto{127.350}{55.996}
\emmoveto{127.350}{55.986}
\emlineto{127.719}{56.004}
\emmoveto{127.719}{55.994}
\emlineto{128.089}{56.014}
\emmoveto{128.089}{56.004}
\emlineto{128.458}{56.026}
\emmoveto{128.458}{56.016}
\emlineto{128.828}{56.039}
\emmoveto{128.828}{56.029}
\emlineto{129.199}{56.053}
\emmoveto{129.199}{56.043}
\emlineto{129.569}{56.069}
\emshow{48.580}{75.100}{v=0.8}
\emshow{1.000}{10.000}{-6.20e-1}
\emshow{1.000}{17.000}{-4.78e-1}
\emshow{1.000}{24.000}{-3.36e-1}
\emshow{1.000}{31.000}{-1.94e-1}
\emshow{1.000}{38.000}{-5.20e-2}
\emshow{1.000}{45.000}{9.00e-2}
\emshow{1.000}{52.000}{2.32e-1}
\emshow{1.000}{59.000}{3.74e-1}
\emshow{1.000}{66.000}{5.16e-1}
\emshow{1.000}{73.000}{6.58e-1}
\emshow{1.000}{80.000}{8.00e-1}
\emshow{12.000}{5.000}{-5.00e-1}
\emshow{23.800}{5.000}{-3.00e-1}
\emshow{35.600}{5.000}{-1.00e-1}
\emshow{47.400}{5.000}{1.00e-1}
\emshow{59.200}{5.000}{3.00e-1}
\emshow{71.000}{5.000}{5.00e-1}
\emshow{82.800}{5.000}{7.00e-1}
\emshow{94.600}{5.000}{9.00e-1}
\emshow{106.400}{5.000}{1.10e0}
\emshow{118.200}{5.000}{1.30e0}
\emshow{130.000}{5.000}{1.50e0}

\centerline{\bf{Fig.A.1}}
\eject
\newcount\numpoint
\newcount\numpointo
\numpoint=1 \numpointo=1
\def\emmoveto#1#2{\offinterlineskip
\hbox to 0 true cm{\vbox to 0
true cm{\vskip - #2 true mm
\hskip #1 true mm \special{em:point
\the\numpoint}\vss}\hss}
\numpointo=\numpoint
\global\advance \numpoint by 1}
\def\emlineto#1#2{\offinterlineskip
\hbox to 0 true cm{\vbox to 0
true cm{\vskip - #2 true mm
\hskip #1 true mm \special{em:point
\the\numpoint}\vss}\hss}
\special{em:line
\the\numpointo,\the\numpoint}
\numpointo=\numpoint
\global\advance \numpoint by 1}
\def\emshow#1#2#3{\offinterlineskip
\hbox to 0 true cm{\vbox to 0
true cm{\vskip - #2 true mm
\hskip #1 true mm \vbox to 0
true cm{\vss\hbox{#3\hss
}}\vss}\hss}}
\special{em:linewidth 0.8pt}

\vrule width 0 mm height                0 mm depth 90.000 true mm

\special{em:linewidth 0.8pt}
\emmoveto{130.000}{10.000}
\emlineto{12.000}{10.000}
\emlineto{12.000}{80.000}
\emmoveto{71.000}{10.000}
\emlineto{71.000}{80.000}
\emmoveto{12.000}{45.000}
\emlineto{130.000}{45.000}
\emmoveto{130.000}{10.000}
\emlineto{130.000}{80.000}
\emlineto{12.000}{80.000}
\emlineto{12.000}{10.000}
\emlineto{130.000}{10.000}
\special{em:linewidth 0.4pt}
\emmoveto{12.000}{17.000}
\emlineto{130.000}{17.000}
\emmoveto{12.000}{24.000}
\emlineto{130.000}{24.000}
\emmoveto{12.000}{31.000}
\emlineto{130.000}{31.000}
\emmoveto{12.000}{38.000}
\emlineto{130.000}{38.000}
\emmoveto{12.000}{45.000}
\emlineto{130.000}{45.000}
\emmoveto{12.000}{52.000}
\emlineto{130.000}{52.000}
\emmoveto{12.000}{59.000}
\emlineto{130.000}{59.000}
\emmoveto{12.000}{66.000}
\emlineto{130.000}{66.000}
\emmoveto{12.000}{73.000}
\emlineto{130.000}{73.000}
\emmoveto{23.800}{10.000}
\emlineto{23.800}{80.000}
\emmoveto{35.600}{10.000}
\emlineto{35.600}{80.000}
\emmoveto{47.400}{10.000}
\emlineto{47.400}{80.000}
\emmoveto{59.200}{10.000}
\emlineto{59.200}{80.000}
\emmoveto{71.000}{10.000}
\emlineto{71.000}{80.000}
\emmoveto{82.800}{10.000}
\emlineto{82.800}{80.000}
\emmoveto{94.600}{10.000}
\emlineto{94.600}{80.000}
\emmoveto{106.400}{10.000}
\emlineto{106.400}{80.000}
\emmoveto{118.200}{10.000}
\emlineto{118.200}{80.000}
\special{em:linewidth 0.8pt}
\emmoveto{12.000}{52.000}
\emlineto{12.142}{52.010}
\emmoveto{12.142}{52.000}
\emlineto{12.283}{52.009}
\emmoveto{12.283}{51.999}
\emlineto{12.425}{52.008}
\emmoveto{12.425}{51.998}
\emlineto{12.566}{52.006}
\emmoveto{12.566}{51.996}
\emlineto{12.708}{52.004}
\emmoveto{12.708}{51.994}
\emlineto{12.849}{52.002}
\emmoveto{12.849}{51.992}
\emlineto{12.991}{51.999}
\emmoveto{12.991}{51.989}
\emlineto{13.132}{51.995}
\emmoveto{13.132}{51.985}
\emlineto{13.274}{51.992}
\emmoveto{13.274}{51.982}
\emlineto{13.415}{51.988}
\emmoveto{13.415}{51.978}
\emlineto{13.556}{51.983}
\emmoveto{13.556}{51.973}
\emlineto{13.698}{51.978}
\emmoveto{13.698}{51.968}
\emlineto{13.839}{51.973}
\emmoveto{13.839}{51.963}
\emlineto{13.980}{51.967}
\emmoveto{13.980}{51.957}
\emlineto{14.121}{51.961}
\emmoveto{14.121}{51.951}
\emlineto{14.262}{51.955}
\emmoveto{14.262}{51.945}
\emlineto{14.403}{51.948}
\emmoveto{14.403}{51.938}
\emlineto{14.544}{51.941}
\emmoveto{14.544}{51.931}
\emlineto{14.685}{51.933}
\emmoveto{14.685}{51.923}
\emlineto{14.825}{51.926}
\emmoveto{14.825}{51.916}
\emlineto{14.966}{51.918}
\emmoveto{14.966}{51.908}
\emlineto{15.106}{51.909}
\emmoveto{15.106}{51.899}
\emlineto{15.247}{51.901}
\emmoveto{15.247}{51.891}
\emlineto{15.387}{51.892}
\emmoveto{15.387}{51.882}
\emlineto{15.527}{51.882}
\emmoveto{15.527}{51.872}
\emlineto{15.667}{51.873}
\emmoveto{15.667}{51.863}
\emlineto{15.807}{51.863}
\emmoveto{15.807}{51.853}
\emlineto{15.947}{51.853}
\emmoveto{15.947}{51.843}
\emlineto{16.086}{51.843}
\emmoveto{16.086}{51.833}
\emlineto{16.226}{51.832}
\emmoveto{16.226}{51.822}
\emlineto{16.365}{51.821}
\emmoveto{16.365}{51.811}
\emlineto{16.505}{51.810}
\emmoveto{16.505}{51.800}
\emlineto{16.644}{51.799}
\emmoveto{16.644}{51.789}
\emlineto{16.783}{51.787}
\emmoveto{16.783}{51.777}
\emlineto{16.922}{51.775}
\emmoveto{16.922}{51.765}
\emlineto{17.061}{51.763}
\emmoveto{17.061}{51.753}
\emlineto{17.199}{51.751}
\emmoveto{17.199}{51.741}
\emlineto{17.338}{51.738}
\emmoveto{17.338}{51.728}
\emlineto{17.476}{51.725}
\emmoveto{17.476}{51.715}
\emlineto{17.614}{51.712}
\emmoveto{17.614}{51.702}
\emlineto{17.752}{51.699}
\emmoveto{17.752}{51.689}
\emlineto{17.890}{51.686}
\emmoveto{17.890}{51.676}
\emlineto{18.028}{51.672}
\emmoveto{18.028}{51.662}
\emlineto{18.165}{51.658}
\emmoveto{18.165}{51.648}
\emlineto{18.303}{51.644}
\emmoveto{18.303}{51.634}
\emlineto{18.440}{51.630}
\emmoveto{18.440}{51.620}
\emlineto{18.577}{51.616}
\emmoveto{18.577}{51.606}
\emlineto{18.714}{51.601}
\emmoveto{18.714}{51.591}
\emlineto{18.850}{51.587}
\emmoveto{18.850}{51.577}
\emlineto{18.987}{51.572}
\emmoveto{19.123}{51.547}
\emlineto{19.395}{51.526}
\emmoveto{19.531}{51.501}
\emlineto{19.802}{51.479}
\emmoveto{19.937}{51.453}
\emlineto{20.207}{51.430}
\emmoveto{20.342}{51.404}
\emlineto{20.611}{51.380}
\emmoveto{20.745}{51.352}
\emlineto{21.012}{51.327}
\emmoveto{21.146}{51.299}
\emlineto{21.412}{51.272}
\emmoveto{21.545}{51.244}
\emlineto{21.810}{51.216}
\emmoveto{21.942}{51.186}
\emlineto{22.205}{51.157}
\emmoveto{22.337}{51.127}
\emlineto{22.599}{51.096}
\emmoveto{22.730}{51.066}
\emlineto{22.990}{51.034}
\emmoveto{23.120}{51.002}
\emlineto{23.379}{50.969}
\emmoveto{23.508}{50.937}
\emlineto{23.766}{50.902}
\emmoveto{23.894}{50.869}
\emlineto{24.150}{50.833}
\emmoveto{24.278}{50.799}
\emlineto{24.532}{50.762}
\emmoveto{24.659}{50.728}
\emlineto{24.912}{50.689}
\emmoveto{25.037}{50.654}
\emlineto{25.288}{50.614}
\emmoveto{25.413}{50.578}
\emlineto{25.662}{50.537}
\emmoveto{25.786}{50.501}
\emlineto{26.034}{50.458}
\emmoveto{26.157}{50.422}
\emlineto{26.402}{50.378}
\emmoveto{26.524}{50.341}
\emlineto{26.768}{50.296}
\emmoveto{26.889}{50.258}
\emlineto{27.130}{50.212}
\emmoveto{27.250}{50.173}
\emlineto{27.490}{50.126}
\emmoveto{27.609}{50.087}
\emlineto{27.846}{50.039}
\emmoveto{27.965}{49.999}
\emlineto{28.200}{49.950}
\emmoveto{28.317}{49.909}
\emlineto{28.550}{49.859}
\emmoveto{28.666}{49.818}
\emlineto{28.897}{49.766}
\emmoveto{29.012}{49.725}
\emlineto{29.241}{49.672}
\emmoveto{29.355}{49.630}
\emlineto{29.581}{49.576}
\emmoveto{29.694}{49.534}
\emlineto{29.918}{49.479}
\emmoveto{30.030}{49.436}
\emlineto{30.252}{49.380}
\emmoveto{30.362}{49.337}
\emlineto{30.582}{49.279}
\emmoveto{30.691}{49.235}
\emlineto{30.908}{49.177}
\emmoveto{31.016}{49.133}
\emlineto{31.231}{49.073}
\emmoveto{31.337}{49.028}
\emlineto{31.550}{48.968}
\emmoveto{31.655}{48.922}
\emlineto{31.865}{48.861}
\emmoveto{31.969}{48.815}
\emlineto{32.176}{48.752}
\emmoveto{32.279}{48.706}
\emlineto{32.484}{48.642}
\emmoveto{32.585}{48.595}
\emlineto{32.787}{48.531}
\emmoveto{32.888}{48.483}
\emlineto{33.087}{48.418}
\emmoveto{33.186}{48.370}
\emlineto{33.383}{48.303}
\emmoveto{33.480}{48.255}
\emlineto{33.674}{48.188}
\emmoveto{33.770}{48.139}
\emlineto{33.962}{48.070}
\emmoveto{34.056}{48.021}
\emlineto{34.245}{47.952}
\emmoveto{34.338}{47.902}
\emlineto{34.524}{47.832}
\emmoveto{34.616}{47.782}
\emlineto{34.799}{47.711}
\emmoveto{34.889}{47.660}
\emlineto{35.069}{47.588}
\emmoveto{35.158}{47.537}
\emlineto{35.335}{47.464}
\emmoveto{35.423}{47.413}
\emlineto{35.597}{47.340}
\emmoveto{35.683}{47.288}
\emlineto{35.854}{47.213}
\emmoveto{35.939}{47.161}
\emlineto{36.107}{47.086}
\emmoveto{36.190}{47.033}
\emlineto{36.355}{46.957}
\emmoveto{36.437}{46.904}
\emlineto{36.599}{46.828}
\emmoveto{36.679}{46.774}
\emlineto{36.838}{46.697}
\emmoveto{36.916}{46.643}
\emlineto{37.072}{46.565}
\emmoveto{37.149}{46.510}
\emlineto{37.302}{46.432}
\emmoveto{37.377}{46.377}
\emlineto{37.527}{46.297}
\emmoveto{37.601}{46.243}
\emlineto{37.747}{46.162}
\emmoveto{37.819}{46.107}
\emlineto{37.962}{46.026}
\emmoveto{38.033}{45.970}
\emlineto{38.173}{45.889}
\emmoveto{38.242}{45.833}
\emlineto{38.378}{45.751}
\emmoveto{38.446}{45.695}
\emlineto{38.579}{45.612}
\emmoveto{38.645}{45.555}
\emlineto{38.775}{45.472}
\emmoveto{38.839}{45.415}
\emlineto{38.966}{45.331}
\emmoveto{39.028}{45.274}
\emlineto{39.152}{45.189}
\emmoveto{39.213}{45.132}
\emlineto{39.333}{45.047}
\emmoveto{39.392}{44.989}
\emlineto{39.508}{44.904}
\emmoveto{39.566}{44.846}
\emlineto{39.679}{44.759}
\emmoveto{39.735}{44.701}
\emlineto{39.845}{44.615}
\emmoveto{39.899}{44.556}
\emlineto{40.005}{44.469}
\emmoveto{40.057}{44.410}
\emlineto{40.160}{44.323}
\emmoveto{40.211}{44.264}
\emlineto{40.310}{44.176}
\emmoveto{40.359}{44.117}
\emlineto{40.455}{44.028}
\emmoveto{40.549}{43.920}
\emlineto{40.685}{43.781}
\emmoveto{40.773}{43.672}
\emlineto{40.900}{43.532}
\emmoveto{40.982}{43.422}
\emlineto{41.101}{43.282}
\emmoveto{41.177}{43.171}
\emlineto{41.286}{43.030}
\emmoveto{41.356}{42.919}
\emlineto{41.457}{42.778}
\emmoveto{41.521}{42.666}
\emlineto{41.613}{42.524}
\emmoveto{41.671}{42.412}
\emlineto{41.754}{42.269}
\emmoveto{41.806}{42.157}
\emlineto{41.879}{42.013}
\emmoveto{41.925}{41.901}
\emlineto{41.990}{41.757}
\emmoveto{42.030}{41.644}
\emlineto{42.085}{41.500}
\emmoveto{42.119}{41.387}
\emlineto{42.166}{41.243}
\emmoveto{42.194}{41.130}
\emlineto{42.231}{40.985}
\emmoveto{42.253}{40.872}
\emlineto{42.281}{40.727}
\emmoveto{42.297}{40.614}
\emlineto{42.316}{40.470}
\emmoveto{42.325}{40.357}
\emlineto{42.335}{40.212}
\emmoveto{42.339}{40.099}
\emlineto{42.339}{39.955}
\emmoveto{42.337}{39.842}
\emlineto{42.329}{39.697}
\emmoveto{42.320}{39.585}
\emlineto{42.303}{39.441}
\emmoveto{42.288}{39.328}
\emlineto{42.261}{39.185}
\emmoveto{42.241}{39.073}
\emlineto{42.205}{38.929}
\emmoveto{42.178}{38.818}
\emlineto{42.134}{38.675}
\emmoveto{42.101}{38.564}
\emlineto{42.048}{38.422}
\emmoveto{42.009}{38.310}
\emlineto{41.946}{38.169}
\emmoveto{41.902}{38.059}
\emlineto{41.830}{37.918}
\emmoveto{41.780}{37.808}
\emlineto{41.700}{37.668}
\emmoveto{41.643}{37.559}
\emlineto{41.554}{37.420}
\emmoveto{41.492}{37.311}
\emlineto{41.394}{37.173}
\emmoveto{41.326}{37.065}
\emlineto{41.219}{36.929}
\emmoveto{41.146}{36.821}
\emlineto{41.030}{36.685}
\emmoveto{40.951}{36.579}
\emlineto{40.827}{36.444}
\emmoveto{40.742}{36.338}
\emlineto{40.610}{36.205}
\emmoveto{40.519}{36.100}
\emlineto{40.378}{35.969}
\emmoveto{40.282}{35.864}
\emlineto{40.133}{35.734}
\emmoveto{40.082}{35.677}
\emlineto{39.979}{35.594}
\emmoveto{39.927}{35.538}
\emlineto{39.820}{35.456}
\emmoveto{39.766}{35.400}
\emlineto{39.657}{35.318}
\emmoveto{39.601}{35.262}
\emlineto{39.488}{35.181}
\emmoveto{39.431}{35.126}
\emlineto{39.315}{35.046}
\emmoveto{39.256}{34.990}
\emlineto{39.137}{34.911}
\emmoveto{39.076}{34.856}
\emlineto{38.954}{34.777}
\emmoveto{38.892}{34.723}
\emlineto{38.766}{34.644}
\emmoveto{38.703}{34.590}
\emlineto{38.574}{34.513}
\emmoveto{38.509}{34.459}
\emlineto{38.377}{34.382}
\emmoveto{38.310}{34.329}
\emlineto{38.175}{34.252}
\emmoveto{38.107}{34.200}
\emlineto{37.969}{34.124}
\emmoveto{37.900}{34.072}
\emlineto{37.759}{33.997}
\emmoveto{37.688}{33.945}
\emlineto{37.544}{33.871}
\emmoveto{37.471}{33.819}
\emlineto{37.324}{33.746}
\emmoveto{37.250}{33.695}
\emlineto{37.100}{33.623}
\emmoveto{37.025}{33.572}
\emlineto{36.872}{33.500}
\emmoveto{36.795}{33.450}
\emlineto{36.639}{33.379}
\emmoveto{36.561}{33.329}
\emlineto{36.402}{33.259}
\emmoveto{36.323}{33.210}
\emlineto{36.161}{33.141}
\emmoveto{36.080}{33.092}
\emlineto{35.916}{33.024}
\emmoveto{35.833}{32.975}
\emlineto{35.667}{32.908}
\emmoveto{35.583}{32.860}
\emlineto{35.413}{32.794}
\emmoveto{35.328}{32.746}
\emlineto{35.156}{32.680}
\emmoveto{35.069}{32.633}
\emlineto{34.894}{32.569}
\emmoveto{34.806}{32.522}
\emlineto{34.629}{32.459}
\emmoveto{34.540}{32.412}
\emlineto{34.360}{32.350}
\emmoveto{34.269}{32.304}
\emlineto{34.087}{32.242}
\emmoveto{33.995}{32.197}
\emlineto{33.810}{32.137}
\emmoveto{33.717}{32.092}
\emlineto{33.529}{32.032}
\emmoveto{33.435}{31.988}
\emlineto{33.245}{31.929}
\emmoveto{33.149}{31.886}
\emlineto{32.957}{31.828}
\emmoveto{32.860}{31.785}
\emlineto{32.666}{31.728}
\emmoveto{32.568}{31.686}
\emlineto{32.371}{31.630}
\emmoveto{32.272}{31.588}
\emlineto{32.073}{31.534}
\emmoveto{31.972}{31.492}
\emlineto{31.771}{31.439}
\emmoveto{31.669}{31.397}
\emlineto{31.466}{31.345}
\emmoveto{31.363}{31.305}
\emlineto{31.157}{31.254}
\emmoveto{31.054}{31.213}
\emlineto{30.846}{31.163}
\emmoveto{30.741}{31.124}
\emlineto{30.531}{31.075}
\emmoveto{30.425}{31.036}
\emlineto{30.213}{30.988}
\emmoveto{30.107}{30.950}
\emlineto{29.892}{30.903}
\emmoveto{29.785}{30.865}
\emlineto{29.568}{30.820}
\emmoveto{29.460}{30.782}
\emlineto{29.242}{30.738}
\emmoveto{29.132}{30.701}
\emlineto{28.912}{30.658}
\emmoveto{28.801}{30.622}
\emlineto{28.580}{30.580}
\emmoveto{28.468}{30.544}
\emlineto{28.244}{30.504}
\emmoveto{28.132}{30.468}
\emlineto{27.907}{30.429}
\emmoveto{27.793}{30.394}
\emlineto{27.566}{30.356}
\emmoveto{27.452}{30.322}
\emlineto{27.223}{30.285}
\emmoveto{27.108}{30.252}
\emlineto{26.878}{30.216}
\emmoveto{26.762}{30.183}
\emlineto{26.530}{30.148}
\emmoveto{26.413}{30.116}
\emlineto{26.179}{30.083}
\emmoveto{26.062}{30.051}
\emlineto{25.827}{30.019}
\emmoveto{25.709}{29.988}
\emlineto{25.472}{29.957}
\emmoveto{25.353}{29.927}
\emlineto{25.115}{29.897}
\emmoveto{24.996}{29.867}
\emlineto{24.756}{29.839}
\emmoveto{24.636}{29.810}
\emlineto{24.395}{29.782}
\emmoveto{24.274}{29.754}
\emlineto{24.032}{29.728}
\emmoveto{23.911}{29.700}
\emlineto{23.667}{29.675}
\emmoveto{23.545}{29.648}
\emlineto{23.300}{29.625}
\emmoveto{23.178}{29.598}
\emlineto{22.932}{29.576}
\emmoveto{22.809}{29.550}
\emlineto{22.562}{29.529}
\emmoveto{22.438}{29.504}
\emlineto{22.190}{29.484}
\emmoveto{22.065}{29.459}
\emlineto{21.816}{29.441}
\emmoveto{21.692}{29.417}
\emlineto{21.441}{29.400}
\emmoveto{21.316}{29.377}
\emlineto{21.065}{29.361}
\emmoveto{20.939}{29.338}
\emlineto{20.688}{29.324}
\emmoveto{20.561}{29.302}
\emlineto{20.309}{29.288}
\emmoveto{20.182}{29.267}
\emlineto{19.928}{29.255}
\emmoveto{19.801}{29.235}
\emlineto{19.547}{29.224}
\emmoveto{19.420}{29.204}
\emlineto{19.165}{29.195}
\emmoveto{19.037}{29.175}
\emlineto{18.781}{29.167}
\emmoveto{18.653}{29.149}
\emlineto{18.397}{29.142}
\emmoveto{18.269}{29.124}
\emlineto{18.012}{29.118}
\emmoveto{17.883}{29.101}
\emlineto{17.626}{29.097}
\emmoveto{17.497}{29.080}
\emlineto{17.239}{29.078}
\emmoveto{17.110}{29.062}
\emlineto{16.852}{29.060}
\emmoveto{16.723}{29.045}
\emlineto{16.464}{29.045}
\emmoveto{16.335}{29.030}
\emlineto{16.076}{29.031}
\emmoveto{15.946}{29.017}
\emlineto{15.687}{29.020}
\emmoveto{15.557}{29.006}
\emlineto{15.298}{29.010}
\emmoveto{15.168}{28.997}
\emlineto{14.908}{29.003}
\emmoveto{14.778}{28.991}
\emlineto{14.518}{28.997}
\emmoveto{14.388}{28.986}
\emlineto{14.128}{28.994}
\emmoveto{13.998}{28.983}
\emlineto{13.738}{28.992}
\emmoveto{13.608}{28.982}
\emlineto{13.348}{28.992}
\emmoveto{13.218}{28.983}
\emlineto{12.958}{28.995}
\emmoveto{12.828}{28.986}
\emlineto{12.569}{28.999}
\emmoveto{12.439}{28.991}
\emlineto{12.179}{29.006}
\emshow{24.980}{24.700}{k=1,v=0.12}
\emmoveto{12.000}{70.000}
\emlineto{12.354}{70.009}
\emmoveto{12.354}{69.999}
\emlineto{12.708}{70.008}
\emmoveto{12.708}{69.998}
\emlineto{13.062}{70.005}
\emmoveto{13.062}{69.995}
\emlineto{13.416}{70.001}
\emmoveto{13.416}{69.991}
\emlineto{13.770}{69.995}
\emmoveto{13.770}{69.985}
\emlineto{14.124}{69.989}
\emmoveto{14.124}{69.979}
\emlineto{14.477}{69.982}
\emmoveto{14.477}{69.972}
\emlineto{14.831}{69.974}
\emmoveto{14.831}{69.964}
\emlineto{15.184}{69.964}
\emmoveto{15.184}{69.954}
\emlineto{15.538}{69.954}
\emmoveto{15.538}{69.944}
\emlineto{15.891}{69.942}
\emmoveto{15.891}{69.932}
\emlineto{16.244}{69.930}
\emmoveto{16.244}{69.920}
\emlineto{16.597}{69.917}
\emmoveto{16.597}{69.907}
\emlineto{16.950}{69.903}
\emmoveto{16.950}{69.893}
\emlineto{17.303}{69.888}
\emmoveto{17.303}{69.878}
\emlineto{17.655}{69.872}
\emmoveto{17.655}{69.862}
\emlineto{18.007}{69.855}
\emmoveto{18.007}{69.845}
\emlineto{18.359}{69.837}
\emmoveto{18.359}{69.827}
\emlineto{18.711}{69.819}
\emmoveto{18.711}{69.809}
\emlineto{19.063}{69.799}
\emmoveto{19.063}{69.789}
\emlineto{19.414}{69.779}
\emmoveto{19.414}{69.769}
\emlineto{19.765}{69.758}
\emmoveto{19.765}{69.748}
\emlineto{20.116}{69.737}
\emmoveto{20.116}{69.727}
\emlineto{20.467}{69.714}
\emmoveto{20.467}{69.704}
\emlineto{20.817}{69.691}
\emmoveto{20.817}{69.681}
\emlineto{21.167}{69.667}
\emmoveto{21.167}{69.657}
\emlineto{21.517}{69.643}
\emmoveto{21.517}{69.633}
\emlineto{21.867}{69.617}
\emmoveto{21.867}{69.607}
\emlineto{22.216}{69.591}
\emmoveto{22.216}{69.581}
\emlineto{22.565}{69.565}
\emmoveto{22.565}{69.555}
\emlineto{22.913}{69.538}
\emmoveto{22.913}{69.528}
\emlineto{23.262}{69.510}
\emmoveto{23.262}{69.500}
\emlineto{23.610}{69.481}
\emmoveto{23.610}{69.471}
\emlineto{23.957}{69.452}
\emmoveto{23.957}{69.442}
\emlineto{24.305}{69.423}
\emmoveto{24.305}{69.413}
\emlineto{24.651}{69.392}
\emmoveto{24.651}{69.382}
\emlineto{24.998}{69.361}
\emmoveto{24.998}{69.351}
\emlineto{25.344}{69.330}
\emmoveto{25.344}{69.320}
\emlineto{25.690}{69.298}
\emmoveto{25.690}{69.288}
\emlineto{26.035}{69.266}
\emmoveto{26.035}{69.256}
\emlineto{26.380}{69.233}
\emmoveto{26.380}{69.223}
\emlineto{26.725}{69.199}
\emmoveto{26.725}{69.189}
\emlineto{27.069}{69.165}
\emmoveto{27.069}{69.155}
\emlineto{27.413}{69.131}
\emmoveto{27.413}{69.121}
\emlineto{27.756}{69.096}
\emmoveto{27.756}{69.086}
\emlineto{28.099}{69.061}
\emmoveto{28.099}{69.051}
\emlineto{28.442}{69.025}
\emmoveto{28.442}{69.015}
\emlineto{28.784}{68.989}
\emmoveto{28.784}{68.979}
\emlineto{29.126}{68.952}
\emmoveto{29.126}{68.942}
\emlineto{29.467}{68.915}
\emmoveto{29.467}{68.905}
\emlineto{29.808}{68.877}
\emmoveto{29.808}{68.867}
\emlineto{30.148}{68.839}
\emmoveto{30.148}{68.829}
\emlineto{30.488}{68.801}
\emmoveto{30.488}{68.791}
\emlineto{30.828}{68.762}
\emmoveto{30.828}{68.752}
\emlineto{31.167}{68.723}
\emmoveto{31.167}{68.713}
\emlineto{31.506}{68.683}
\emmoveto{31.506}{68.673}
\emlineto{31.844}{68.643}
\emmoveto{31.844}{68.633}
\emlineto{32.181}{68.602}
\emmoveto{32.181}{68.592}
\emlineto{32.518}{68.561}
\emmoveto{32.518}{68.551}
\emlineto{32.855}{68.519}
\emmoveto{32.855}{68.509}
\emlineto{33.191}{68.477}
\emmoveto{33.191}{68.467}
\emlineto{33.527}{68.434}
\emmoveto{33.527}{68.424}
\emlineto{33.862}{68.391}
\emmoveto{33.862}{68.381}
\emlineto{34.197}{68.347}
\emmoveto{34.197}{68.337}
\emlineto{34.531}{68.303}
\emmoveto{34.531}{68.293}
\emlineto{34.864}{68.258}
\emmoveto{34.864}{68.248}
\emlineto{35.198}{68.212}
\emmoveto{35.198}{68.202}
\emlineto{35.530}{68.166}
\emmoveto{35.530}{68.156}
\emlineto{35.862}{68.119}
\emmoveto{35.862}{68.109}
\emlineto{36.193}{68.072}
\emmoveto{36.193}{68.062}
\emlineto{36.524}{68.025}
\emmoveto{36.524}{68.015}
\emlineto{36.855}{67.976}
\emmoveto{36.855}{67.966}
\emlineto{37.184}{67.927}
\emmoveto{37.184}{67.917}
\emlineto{37.513}{67.878}
\emmoveto{37.513}{67.868}
\emlineto{37.842}{67.828}
\emmoveto{37.842}{67.818}
\emlineto{38.170}{67.777}
\emmoveto{38.170}{67.767}
\emlineto{38.497}{67.726}
\emmoveto{38.497}{67.716}
\emlineto{38.824}{67.674}
\emmoveto{38.824}{67.664}
\emlineto{39.150}{67.622}
\emmoveto{39.150}{67.612}
\emlineto{39.476}{67.569}
\emmoveto{39.476}{67.559}
\emlineto{39.800}{67.515}
\emmoveto{39.800}{67.505}
\emlineto{40.125}{67.461}
\emmoveto{40.125}{67.451}
\emlineto{40.448}{67.407}
\emmoveto{40.448}{67.397}
\emlineto{40.771}{67.351}
\emmoveto{40.771}{67.341}
\emlineto{41.094}{67.296}
\emmoveto{41.094}{67.286}
\emlineto{41.415}{67.239}
\emmoveto{41.415}{67.229}
\emlineto{41.736}{67.182}
\emmoveto{41.736}{67.172}
\emlineto{42.056}{67.125}
\emmoveto{42.056}{67.115}
\emlineto{42.376}{67.067}
\emmoveto{42.376}{67.057}
\emlineto{42.695}{67.008}
\emmoveto{42.695}{66.998}
\emlineto{43.013}{66.949}
\emmoveto{43.013}{66.939}
\emlineto{43.331}{66.889}
\emmoveto{43.331}{66.879}
\emlineto{43.648}{66.829}
\emmoveto{43.648}{66.819}
\emlineto{43.964}{66.768}
\emmoveto{43.964}{66.758}
\emlineto{44.279}{66.707}
\emmoveto{44.279}{66.697}
\emlineto{44.594}{66.645}
\emmoveto{44.594}{66.635}
\emlineto{44.908}{66.582}
\emmoveto{44.908}{66.572}
\emlineto{45.221}{66.519}
\emmoveto{45.221}{66.509}
\emlineto{45.533}{66.456}
\emmoveto{45.533}{66.446}
\emlineto{45.845}{66.392}
\emmoveto{45.845}{66.382}
\emlineto{46.156}{66.327}
\emmoveto{46.156}{66.317}
\emlineto{46.466}{66.262}
\emmoveto{46.466}{66.252}
\emlineto{46.775}{66.197}
\emmoveto{46.775}{66.187}
\emlineto{47.084}{66.131}
\emmoveto{47.084}{66.121}
\emlineto{47.392}{66.064}
\emmoveto{47.392}{66.054}
\emlineto{47.699}{65.997}
\emmoveto{47.699}{65.987}
\emlineto{48.005}{65.930}
\emmoveto{48.005}{65.920}
\emlineto{48.311}{65.862}
\emmoveto{48.311}{65.852}
\emlineto{48.615}{65.793}
\emmoveto{48.615}{65.783}
\emlineto{48.919}{65.724}
\emmoveto{48.919}{65.714}
\emlineto{49.222}{65.655}
\emmoveto{49.222}{65.645}
\emlineto{49.524}{65.585}
\emmoveto{49.524}{65.575}
\emlineto{49.826}{65.514}
\emmoveto{49.826}{65.504}
\emlineto{50.126}{65.443}
\emmoveto{50.126}{65.433}
\emlineto{50.426}{65.372}
\emmoveto{50.426}{65.362}
\emlineto{50.725}{65.300}
\emmoveto{50.725}{65.290}
\emlineto{51.023}{65.228}
\emmoveto{51.023}{65.218}
\emlineto{51.320}{65.155}
\emmoveto{51.320}{65.145}
\emlineto{51.616}{65.082}
\emmoveto{51.616}{65.072}
\emlineto{51.912}{65.008}
\emmoveto{51.912}{64.998}
\emlineto{52.206}{64.933}
\emmoveto{52.206}{64.923}
\emlineto{52.500}{64.859}
\emmoveto{52.500}{64.849}
\emlineto{52.793}{64.784}
\emmoveto{52.793}{64.774}
\emlineto{53.084}{64.708}
\emmoveto{53.084}{64.698}
\emlineto{53.375}{64.632}
\emmoveto{53.375}{64.622}
\emlineto{53.666}{64.555}
\emmoveto{53.666}{64.545}
\emlineto{53.955}{64.478}
\emmoveto{53.955}{64.468}
\emlineto{54.243}{64.401}
\emmoveto{54.243}{64.391}
\emlineto{54.530}{64.323}
\emmoveto{54.530}{64.313}
\emlineto{54.817}{64.244}
\emmoveto{54.817}{64.234}
\emlineto{55.102}{64.166}
\emmoveto{55.102}{64.156}
\emlineto{55.387}{64.086}
\emmoveto{55.387}{64.076}
\emlineto{55.670}{64.006}
\emmoveto{55.670}{63.996}
\emlineto{55.953}{63.926}
\emmoveto{55.953}{63.916}
\emlineto{56.235}{63.845}
\emmoveto{56.235}{63.835}
\emlineto{56.516}{63.764}
\emmoveto{56.516}{63.754}
\emlineto{56.795}{63.683}
\emmoveto{56.795}{63.673}
\emlineto{57.074}{63.601}
\emmoveto{57.074}{63.591}
\emlineto{57.352}{63.518}
\emmoveto{57.352}{63.508}
\emlineto{57.629}{63.435}
\emmoveto{57.629}{63.425}
\emlineto{57.905}{63.352}
\emmoveto{57.905}{63.342}
\emlineto{58.180}{63.268}
\emmoveto{58.180}{63.258}
\emlineto{58.454}{63.183}
\emmoveto{58.454}{63.173}
\emlineto{58.727}{63.099}
\emmoveto{58.727}{63.089}
\emlineto{58.999}{63.013}
\emmoveto{58.999}{63.003}
\emlineto{59.270}{62.928}
\emmoveto{59.270}{62.918}
\emlineto{59.540}{62.842}
\emmoveto{59.540}{62.832}
\emlineto{59.809}{62.755}
\emmoveto{59.809}{62.745}
\emlineto{60.076}{62.668}
\emmoveto{60.076}{62.658}
\emlineto{60.343}{62.581}
\emmoveto{60.343}{62.571}
\emlineto{60.609}{62.493}
\emmoveto{60.609}{62.483}
\emlineto{60.874}{62.404}
\emmoveto{60.874}{62.394}
\emlineto{61.138}{62.316}
\emmoveto{61.138}{62.306}
\emlineto{61.400}{62.227}
\emmoveto{61.400}{62.217}
\emlineto{61.662}{62.137}
\emmoveto{61.662}{62.127}
\emlineto{61.922}{62.047}
\emmoveto{61.922}{62.037}
\emlineto{62.182}{61.956}
\emmoveto{62.182}{61.946}
\emlineto{62.440}{61.866}
\emmoveto{62.440}{61.856}
\emlineto{62.698}{61.774}
\emmoveto{62.698}{61.764}
\emlineto{62.954}{61.683}
\emmoveto{62.954}{61.673}
\emlineto{63.209}{61.591}
\emmoveto{63.209}{61.581}
\emlineto{63.463}{61.498}
\emmoveto{63.463}{61.488}
\emlineto{63.716}{61.405}
\emmoveto{63.716}{61.395}
\emlineto{63.968}{61.312}
\emmoveto{63.968}{61.302}
\emlineto{64.219}{61.218}
\emmoveto{64.219}{61.208}
\emlineto{64.469}{61.124}
\emmoveto{64.469}{61.114}
\emlineto{64.717}{61.029}
\emmoveto{64.717}{61.019}
\emlineto{64.965}{60.934}
\emmoveto{64.965}{60.924}
\emlineto{65.211}{60.839}
\emmoveto{65.211}{60.829}
\emlineto{65.456}{60.743}
\emmoveto{65.456}{60.733}
\emlineto{65.700}{60.647}
\emmoveto{65.700}{60.637}
\emlineto{65.943}{60.551}
\emmoveto{65.943}{60.541}
\emlineto{66.185}{60.454}
\emmoveto{66.185}{60.444}
\emlineto{66.426}{60.357}
\emmoveto{66.426}{60.347}
\emlineto{66.665}{60.259}
\emmoveto{66.665}{60.249}
\emlineto{66.904}{60.161}
\emmoveto{66.904}{60.151}
\emlineto{67.141}{60.062}
\emmoveto{67.141}{60.052}
\emlineto{67.377}{59.964}
\emmoveto{67.377}{59.954}
\emlineto{67.612}{59.865}
\emmoveto{67.612}{59.855}
\emlineto{67.846}{59.765}
\emmoveto{67.846}{59.755}
\emlineto{68.078}{59.665}
\emmoveto{68.078}{59.655}
\emlineto{68.309}{59.565}
\emmoveto{68.309}{59.555}
\emlineto{68.540}{59.464}
\emmoveto{68.540}{59.454}
\emlineto{68.769}{59.363}
\emmoveto{68.769}{59.353}
\emlineto{68.996}{59.262}
\emmoveto{68.996}{59.252}
\emlineto{69.223}{59.160}
\emmoveto{69.223}{59.150}
\emlineto{69.448}{59.058}
\emmoveto{69.448}{59.048}
\emlineto{69.672}{58.956}
\emmoveto{69.672}{58.946}
\emlineto{69.895}{58.853}
\emmoveto{69.895}{58.843}
\emlineto{70.117}{58.750}
\emmoveto{70.117}{58.740}
\emlineto{70.338}{58.646}
\emmoveto{70.338}{58.636}
\emlineto{70.557}{58.542}
\emmoveto{70.557}{58.532}
\emlineto{70.775}{58.438}
\emmoveto{70.775}{58.428}
\emlineto{70.992}{58.334}
\emmoveto{70.992}{58.324}
\emlineto{71.207}{58.230}
\emmoveto{71.207}{58.220}
\emlineto{71.422}{58.126}
\emmoveto{71.422}{58.116}
\emlineto{71.635}{58.024}
\emmoveto{71.635}{58.014}
\emlineto{71.847}{57.923}
\emmoveto{71.847}{57.913}
\emlineto{72.058}{57.823}
\emmoveto{72.058}{57.813}
\emlineto{72.267}{57.724}
\emmoveto{72.267}{57.714}
\emlineto{72.476}{57.626}
\emmoveto{72.476}{57.616}
\emlineto{72.683}{57.529}
\emmoveto{72.683}{57.519}
\emlineto{72.889}{57.433}
\emmoveto{72.889}{57.423}
\emlineto{73.094}{57.337}
\emmoveto{73.094}{57.327}
\emlineto{73.298}{57.242}
\emmoveto{73.298}{57.232}
\emlineto{73.501}{57.149}
\emmoveto{73.501}{57.139}
\emlineto{73.703}{57.055}
\emmoveto{73.703}{57.045}
\emlineto{73.903}{56.963}
\emmoveto{73.903}{56.953}
\emlineto{74.103}{56.871}
\emmoveto{74.103}{56.861}
\emlineto{74.301}{56.780}
\emmoveto{74.301}{56.770}
\emlineto{74.499}{56.690}
\emmoveto{74.499}{56.680}
\emlineto{74.695}{56.600}
\emmoveto{74.695}{56.590}
\emlineto{74.890}{56.511}
\emmoveto{74.890}{56.501}
\emlineto{75.084}{56.422}
\emmoveto{75.084}{56.412}
\emlineto{75.277}{56.334}
\emmoveto{75.277}{56.324}
\emlineto{75.469}{56.246}
\emmoveto{75.469}{56.236}
\emlineto{75.661}{56.159}
\emmoveto{75.661}{56.149}
\emlineto{75.851}{56.073}
\emmoveto{75.851}{56.063}
\emlineto{76.040}{55.987}
\emmoveto{76.040}{55.977}
\emlineto{76.228}{55.901}
\emmoveto{76.228}{55.891}
\emlineto{76.415}{55.816}
\emmoveto{76.415}{55.806}
\emlineto{76.601}{55.731}
\emmoveto{76.601}{55.721}
\emlineto{76.786}{55.646}
\emmoveto{76.786}{55.636}
\emlineto{76.970}{55.562}
\emmoveto{76.970}{55.552}
\emlineto{77.153}{55.479}
\emmoveto{77.153}{55.469}
\emlineto{77.335}{55.395}
\emmoveto{77.335}{55.385}
\emlineto{77.516}{55.312}
\emmoveto{77.516}{55.302}
\emlineto{77.696}{55.230}
\emmoveto{77.696}{55.220}
\emlineto{77.875}{55.147}
\emmoveto{77.875}{55.137}
\emlineto{78.053}{55.065}
\emmoveto{78.053}{55.055}
\emlineto{78.230}{54.983}
\emmoveto{78.230}{54.973}
\emlineto{78.407}{54.901}
\emmoveto{78.407}{54.891}
\emlineto{78.582}{54.820}
\emmoveto{78.582}{54.810}
\emlineto{78.756}{54.738}
\emmoveto{78.756}{54.728}
\emlineto{78.929}{54.657}
\emmoveto{78.929}{54.647}
\emlineto{79.102}{54.576}
\emmoveto{79.102}{54.566}
\emlineto{79.273}{54.495}
\emmoveto{79.273}{54.485}
\emlineto{79.444}{54.415}
\emmoveto{79.444}{54.405}
\emlineto{79.613}{54.334}
\emmoveto{79.613}{54.324}
\emlineto{79.782}{54.254}
\emmoveto{79.782}{54.244}
\emlineto{79.949}{54.174}
\emmoveto{79.949}{54.164}
\emlineto{80.116}{54.094}
\emmoveto{80.116}{54.084}
\emlineto{80.282}{54.013}
\emmoveto{80.282}{54.003}
\emlineto{80.446}{53.933}
\emmoveto{80.446}{53.923}
\emlineto{80.610}{53.854}
\emmoveto{80.610}{53.844}
\emlineto{80.773}{53.774}
\emmoveto{80.773}{53.764}
\emlineto{80.935}{53.694}
\emmoveto{80.935}{53.684}
\emlineto{81.096}{53.614}
\emmoveto{81.096}{53.604}
\emlineto{81.256}{53.535}
\emmoveto{81.256}{53.525}
\emlineto{81.415}{53.455}
\emmoveto{81.415}{53.445}
\emlineto{81.573}{53.376}
\emmoveto{81.573}{53.366}
\emlineto{81.731}{53.297}
\emmoveto{81.731}{53.287}
\emlineto{81.887}{53.218}
\emmoveto{81.887}{53.208}
\emlineto{82.042}{53.139}
\emmoveto{82.042}{53.129}
\emlineto{82.197}{53.061}
\emmoveto{82.197}{53.051}
\emlineto{82.350}{52.983}
\emmoveto{82.350}{52.973}
\emlineto{82.503}{52.904}
\emmoveto{82.503}{52.894}
\emlineto{82.655}{52.827}
\emmoveto{82.655}{52.817}
\emlineto{82.805}{52.749}
\emmoveto{82.805}{52.739}
\emlineto{82.955}{52.672}
\emmoveto{82.955}{52.662}
\emlineto{83.104}{52.594}
\emmoveto{83.104}{52.584}
\emlineto{83.252}{52.517}
\emmoveto{83.252}{52.507}
\emlineto{83.399}{52.441}
\emmoveto{83.399}{52.431}
\emlineto{83.546}{52.364}
\emmoveto{83.546}{52.354}
\emlineto{83.691}{52.288}
\emmoveto{83.691}{52.278}
\emlineto{83.835}{52.212}
\emmoveto{83.835}{52.202}
\emlineto{83.979}{52.136}
\emmoveto{83.979}{52.126}
\emlineto{84.122}{52.061}
\emmoveto{84.122}{52.051}
\emlineto{84.263}{51.986}
\emmoveto{84.263}{51.976}
\emlineto{84.404}{51.911}
\emmoveto{84.404}{51.901}
\emlineto{84.544}{51.836}
\emmoveto{84.544}{51.826}
\emlineto{84.683}{51.762}
\emmoveto{84.683}{51.752}
\emlineto{84.822}{51.688}
\emmoveto{84.822}{51.678}
\emlineto{84.959}{51.614}
\emmoveto{84.959}{51.604}
\emlineto{85.095}{51.540}
\emmoveto{85.095}{51.530}
\emlineto{85.231}{51.467}
\emmoveto{85.231}{51.457}
\emlineto{85.366}{51.394}
\emmoveto{85.366}{51.384}
\emlineto{85.500}{51.321}
\emmoveto{85.500}{51.311}
\emlineto{85.633}{51.249}
\emmoveto{85.633}{51.239}
\emlineto{85.765}{51.176}
\emmoveto{85.765}{51.166}
\emlineto{85.896}{51.104}
\emmoveto{85.896}{51.094}
\emlineto{86.027}{51.033}
\emmoveto{86.027}{51.023}
\emlineto{86.156}{50.961}
\emmoveto{86.156}{50.951}
\emlineto{86.285}{50.890}
\emmoveto{86.285}{50.880}
\emlineto{86.413}{50.819}
\emmoveto{86.413}{50.809}
\emlineto{86.540}{50.748}
\emmoveto{86.540}{50.738}
\emlineto{86.667}{50.677}
\emmoveto{86.667}{50.667}
\emlineto{86.792}{50.607}
\emmoveto{86.792}{50.597}
\emlineto{86.917}{50.536}
\emmoveto{86.917}{50.526}
\emlineto{87.040}{50.466}
\emmoveto{87.040}{50.456}
\emlineto{87.163}{50.397}
\emmoveto{87.163}{50.387}
\emlineto{87.286}{50.327}
\emmoveto{87.286}{50.317}
\emlineto{87.407}{50.258}
\emmoveto{87.407}{50.248}
\emlineto{87.527}{50.189}
\emmoveto{87.527}{50.179}
\emlineto{87.647}{50.120}
\emmoveto{87.647}{50.110}
\emlineto{87.766}{50.051}
\emmoveto{87.884}{49.972}
\emlineto{88.118}{49.845}
\emmoveto{88.233}{49.767}
\emlineto{88.462}{49.642}
\emmoveto{88.576}{49.564}
\emlineto{88.800}{49.439}
\emmoveto{88.911}{49.362}
\emlineto{89.130}{49.238}
\emmoveto{89.238}{49.162}
\emlineto{89.453}{49.039}
\emmoveto{89.559}{48.962}
\emlineto{89.769}{48.840}
\emmoveto{89.873}{48.764}
\emlineto{90.078}{48.643}
\emmoveto{90.180}{48.568}
\emlineto{90.380}{48.447}
\emmoveto{90.480}{48.373}
\emlineto{90.676}{48.253}
\emmoveto{90.773}{48.178}
\emlineto{90.964}{48.060}
\emmoveto{91.059}{47.986}
\emlineto{91.246}{47.868}
\emmoveto{91.338}{47.794}
\emlineto{91.520}{47.677}
\emmoveto{91.611}{47.604}
\emlineto{91.789}{47.488}
\emmoveto{91.876}{47.415}
\emlineto{92.050}{47.300}
\emmoveto{92.136}{47.227}
\emlineto{92.305}{47.113}
\emmoveto{92.388}{47.041}
\emlineto{92.553}{46.927}
\emmoveto{92.634}{46.856}
\emlineto{92.794}{46.743}
\emmoveto{92.873}{46.671}
\emlineto{93.029}{46.559}
\emmoveto{93.106}{46.488}
\emlineto{93.258}{46.377}
\emmoveto{93.333}{46.306}
\emlineto{93.480}{46.196}
\emmoveto{93.553}{46.125}
\emlineto{93.696}{46.015}
\emmoveto{93.767}{45.945}
\emlineto{93.905}{45.836}
\emmoveto{93.974}{45.766}
\emlineto{94.109}{45.657}
\emmoveto{94.175}{45.588}
\emlineto{94.305}{45.480}
\emmoveto{94.369}{45.411}
\emlineto{94.496}{45.303}
\emmoveto{94.558}{45.234}
\emlineto{94.680}{45.127}
\emmoveto{94.740}{45.059}
\emlineto{94.858}{44.952}
\emmoveto{94.916}{44.884}
\emlineto{95.030}{44.778}
\emmoveto{95.086}{44.710}
\emlineto{95.196}{44.604}
\emmoveto{95.250}{44.536}
\emlineto{95.355}{44.431}
\emmoveto{95.407}{44.364}
\emlineto{95.509}{44.259}
\emmoveto{95.559}{44.191}
\emlineto{95.656}{44.087}
\emmoveto{95.704}{44.020}
\emlineto{95.797}{43.916}
\emmoveto{95.843}{43.849}
\emlineto{95.933}{43.745}
\emmoveto{95.976}{43.679}
\emlineto{96.062}{43.575}
\emmoveto{96.104}{43.509}
\emlineto{96.185}{43.406}
\emmoveto{96.225}{43.339}
\emlineto{96.340}{43.181}
\emmoveto{96.413}{43.058}
\emlineto{96.519}{42.900}
\emmoveto{96.586}{42.778}
\emlineto{96.681}{42.621}
\emmoveto{96.741}{42.499}
\emlineto{96.827}{42.342}
\emmoveto{96.881}{42.221}
\emlineto{96.956}{42.065}
\emmoveto{97.003}{41.944}
\emlineto{97.069}{41.788}
\emmoveto{97.110}{41.667}
\emlineto{97.166}{41.512}
\emmoveto{97.200}{41.391}
\emlineto{97.247}{41.236}
\emmoveto{97.274}{41.116}
\emlineto{97.311}{40.960}
\emmoveto{97.332}{40.840}
\emlineto{97.359}{40.685}
\emmoveto{97.373}{40.565}
\emlineto{97.390}{40.409}
\emmoveto{97.398}{40.289}
\emlineto{97.406}{40.134}
\emmoveto{97.407}{40.013}
\emlineto{97.405}{39.858}
\emmoveto{97.400}{39.737}
\emlineto{97.388}{39.581}
\emmoveto{97.376}{39.460}
\emlineto{97.354}{39.304}
\emmoveto{97.336}{39.183}
\emlineto{97.305}{39.027}
\emmoveto{97.280}{38.905}
\emlineto{97.238}{38.748}
\emmoveto{97.207}{38.627}
\emlineto{97.156}{38.469}
\emmoveto{97.118}{38.347}
\emlineto{97.056}{38.188}
\emmoveto{97.012}{38.066}
\emlineto{96.941}{37.906}
\emmoveto{96.890}{37.783}
\emlineto{96.808}{37.623}
\emmoveto{96.750}{37.499}
\emlineto{96.659}{37.338}
\emmoveto{96.594}{37.214}
\emlineto{96.493}{37.052}
\emmoveto{96.422}{36.927}
\emlineto{96.310}{36.764}
\emmoveto{96.271}{36.696}
\emlineto{96.192}{36.590}
\emmoveto{96.151}{36.522}
\emlineto{96.068}{36.415}
\emmoveto{96.025}{36.347}
\emlineto{95.937}{36.240}
\emmoveto{95.892}{36.171}
\emlineto{95.801}{36.063}
\emmoveto{95.754}{35.994}
\emlineto{95.658}{35.886}
\emmoveto{95.609}{35.817}
\emlineto{95.509}{35.708}
\emmoveto{95.458}{35.639}
\emlineto{95.353}{35.529}
\emmoveto{95.300}{35.460}
\emlineto{95.191}{35.350}
\emmoveto{95.136}{35.279}
\emlineto{95.023}{35.169}
\emmoveto{94.966}{35.098}
\emlineto{94.849}{34.987}
\emmoveto{94.789}{34.916}
\emlineto{94.668}{34.804}
\emmoveto{94.606}{34.733}
\emlineto{94.480}{34.620}
\emmoveto{94.416}{34.549}
\emlineto{94.286}{34.435}
\emmoveto{94.220}{34.363}
\emlineto{94.085}{34.249}
\emmoveto{94.017}{34.177}
\emlineto{93.878}{34.062}
\emmoveto{93.808}{33.989}
\emlineto{93.664}{33.873}
\emmoveto{93.592}{33.800}
\emlineto{93.444}{33.683}
\emmoveto{93.369}{33.610}
\emlineto{93.216}{33.492}
\emmoveto{93.139}{33.418}
\emlineto{92.982}{33.300}
\emmoveto{92.903}{33.226}
\emlineto{92.741}{33.106}
\emmoveto{92.659}{33.031}
\emlineto{92.493}{32.911}
\emmoveto{92.409}{32.836}
\emlineto{92.239}{32.715}
\emmoveto{92.152}{32.639}
\emlineto{91.977}{32.517}
\emmoveto{91.888}{32.440}
\emlineto{91.708}{32.317}
\emmoveto{91.617}{32.240}
\emlineto{91.432}{32.116}
\emmoveto{91.339}{32.039}
\emlineto{91.149}{31.913}
\emmoveto{91.053}{31.836}
\emlineto{90.859}{31.709}
\emmoveto{90.761}{31.631}
\emlineto{90.562}{31.503}
\emmoveto{90.461}{31.424}
\emlineto{90.257}{31.296}
\emmoveto{90.153}{31.216}
\emlineto{89.944}{31.087}
\emmoveto{89.839}{31.006}
\emlineto{89.625}{30.876}
\emmoveto{89.517}{30.795}
\emlineto{89.298}{30.663}
\emmoveto{89.187}{30.581}
\emlineto{88.963}{30.448}
\emmoveto{88.850}{30.366}
\emlineto{88.621}{30.232}
\emmoveto{88.505}{30.149}
\emlineto{88.388}{30.086}
\emmoveto{88.388}{30.076}
\emlineto{88.271}{30.013}
\emmoveto{88.271}{30.003}
\emlineto{88.152}{29.940}
\emmoveto{88.152}{29.930}
\emlineto{88.033}{29.866}
\emmoveto{88.033}{29.856}
\emlineto{87.913}{29.793}
\emmoveto{87.913}{29.783}
\emlineto{87.792}{29.719}
\emmoveto{87.792}{29.709}
\emlineto{87.670}{29.645}
\emmoveto{87.670}{29.635}
\emlineto{87.547}{29.570}
\emmoveto{87.547}{29.560}
\emlineto{87.424}{29.496}
\emmoveto{87.424}{29.486}
\emlineto{87.299}{29.421}
\emmoveto{87.299}{29.411}
\emlineto{87.174}{29.346}
\emmoveto{87.174}{29.336}
\emlineto{87.047}{29.270}
\emmoveto{87.047}{29.260}
\emlineto{86.920}{29.195}
\emmoveto{86.920}{29.185}
\emlineto{86.792}{29.119}
\emmoveto{86.792}{29.109}
\emlineto{86.663}{29.043}
\emmoveto{86.663}{29.033}
\emlineto{86.533}{28.967}
\emmoveto{86.533}{28.957}
\emlineto{86.403}{28.890}
\emmoveto{86.403}{28.880}
\emlineto{86.271}{28.814}
\emmoveto{86.271}{28.804}
\emlineto{86.138}{28.737}
\emmoveto{86.138}{28.727}
\emlineto{86.005}{28.660}
\emmoveto{86.005}{28.650}
\emlineto{85.871}{28.582}
\emmoveto{85.871}{28.572}
\emlineto{85.735}{28.504}
\emmoveto{85.735}{28.494}
\emlineto{85.599}{28.426}
\emmoveto{85.599}{28.416}
\emlineto{85.462}{28.348}
\emmoveto{85.462}{28.338}
\emlineto{85.324}{28.270}
\emmoveto{85.324}{28.260}
\emlineto{85.185}{28.191}
\emmoveto{85.185}{28.181}
\emlineto{85.045}{28.112}
\emmoveto{85.045}{28.102}
\emlineto{84.904}{28.033}
\emmoveto{84.904}{28.023}
\emlineto{84.762}{27.954}
\emmoveto{84.762}{27.944}
\emlineto{84.620}{27.874}
\emmoveto{84.620}{27.864}
\emlineto{84.476}{27.794}
\emmoveto{84.476}{27.784}
\emlineto{84.331}{27.714}
\emmoveto{84.331}{27.704}
\emlineto{84.186}{27.633}
\emmoveto{84.186}{27.623}
\emlineto{84.039}{27.552}
\emmoveto{84.039}{27.542}
\emlineto{83.892}{27.471}
\emmoveto{83.892}{27.461}
\emlineto{83.743}{27.390}
\emmoveto{83.743}{27.380}
\emlineto{83.594}{27.308}
\emmoveto{83.594}{27.298}
\emlineto{83.443}{27.226}
\emmoveto{83.443}{27.216}
\emlineto{83.292}{27.144}
\emmoveto{83.292}{27.134}
\emlineto{83.140}{27.062}
\emmoveto{83.140}{27.052}
\emlineto{82.986}{26.979}
\emmoveto{82.986}{26.969}
\emlineto{82.832}{26.896}
\emmoveto{82.832}{26.886}
\emlineto{82.677}{26.813}
\emmoveto{82.677}{26.803}
\emlineto{82.521}{26.729}
\emmoveto{82.521}{26.719}
\emlineto{82.364}{26.645}
\emmoveto{82.364}{26.635}
\emlineto{82.205}{26.561}
\emmoveto{82.205}{26.551}
\emlineto{82.046}{26.477}
\emmoveto{82.046}{26.467}
\emlineto{81.886}{26.392}
\emmoveto{81.886}{26.382}
\emlineto{81.725}{26.307}
\emmoveto{81.725}{26.297}
\emlineto{81.563}{26.222}
\emmoveto{81.563}{26.212}
\emlineto{81.399}{26.136}
\emmoveto{81.399}{26.126}
\emlineto{81.235}{26.050}
\emmoveto{81.235}{26.040}
\emlineto{81.070}{25.964}
\emmoveto{81.070}{25.954}
\emlineto{80.904}{25.878}
\emmoveto{80.904}{25.868}
\emlineto{80.736}{25.791}
\emmoveto{80.736}{25.781}
\emlineto{80.568}{25.704}
\emmoveto{80.568}{25.694}
\emlineto{80.399}{25.616}
\emmoveto{80.399}{25.606}
\emlineto{80.228}{25.528}
\emmoveto{80.228}{25.518}
\emlineto{80.057}{25.440}
\emmoveto{80.057}{25.430}
\emlineto{79.885}{25.352}
\emmoveto{79.885}{25.342}
\emlineto{79.711}{25.263}
\emmoveto{79.711}{25.253}
\emlineto{79.537}{25.174}
\emmoveto{79.537}{25.164}
\emlineto{79.361}{25.085}
\emmoveto{79.361}{25.075}
\emlineto{79.184}{24.996}
\emmoveto{79.184}{24.986}
\emlineto{79.007}{24.906}
\emmoveto{79.007}{24.896}
\emlineto{78.828}{24.815}
\emmoveto{78.828}{24.805}
\emlineto{78.648}{24.725}
\emmoveto{78.648}{24.715}
\emlineto{78.467}{24.634}
\emmoveto{78.467}{24.624}
\emlineto{78.285}{24.543}
\emmoveto{78.285}{24.533}
\emlineto{78.102}{24.451}
\emmoveto{78.102}{24.441}
\emlineto{77.918}{24.359}
\emmoveto{77.918}{24.349}
\emlineto{77.733}{24.267}
\emmoveto{77.733}{24.257}
\emlineto{77.546}{24.174}
\emmoveto{77.546}{24.164}
\emlineto{77.359}{24.081}
\emmoveto{77.359}{24.071}
\emlineto{77.171}{23.988}
\emmoveto{77.171}{23.978}
\emlineto{76.981}{23.894}
\emmoveto{76.981}{23.884}
\emlineto{76.790}{23.800}
\emmoveto{76.790}{23.790}
\emlineto{76.598}{23.706}
\emmoveto{76.598}{23.696}
\emlineto{76.405}{23.611}
\emmoveto{76.405}{23.601}
\emlineto{76.211}{23.516}
\emmoveto{76.211}{23.506}
\emlineto{76.016}{23.421}
\emmoveto{76.016}{23.411}
\emlineto{75.820}{23.325}
\emmoveto{75.820}{23.315}
\emlineto{75.622}{23.229}
\emmoveto{75.622}{23.219}
\emlineto{75.424}{23.133}
\emmoveto{75.424}{23.123}
\emlineto{75.224}{23.036}
\emmoveto{75.224}{23.026}
\emlineto{75.023}{22.939}
\emmoveto{75.023}{22.929}
\emlineto{74.821}{22.841}
\emmoveto{74.821}{22.831}
\emlineto{74.618}{22.743}
\emmoveto{74.618}{22.733}
\emlineto{74.414}{22.645}
\emmoveto{74.414}{22.635}
\emlineto{74.208}{22.547}
\emmoveto{74.208}{22.537}
\emlineto{74.002}{22.448}
\emmoveto{74.002}{22.438}
\emlineto{73.794}{22.348}
\emmoveto{73.794}{22.338}
\emlineto{73.585}{22.248}
\emmoveto{73.585}{22.238}
\emlineto{73.375}{22.148}
\emmoveto{73.375}{22.138}
\emlineto{73.163}{22.048}
\emmoveto{73.163}{22.038}
\emlineto{72.951}{21.947}
\emmoveto{72.951}{21.937}
\emlineto{72.737}{21.846}
\emmoveto{72.737}{21.836}
\emlineto{72.522}{21.744}
\emmoveto{72.522}{21.734}
\emlineto{72.306}{21.642}
\emmoveto{72.306}{21.632}
\emlineto{72.088}{21.539}
\emmoveto{72.088}{21.529}
\emlineto{71.870}{21.436}
\emmoveto{71.870}{21.426}
\emlineto{71.650}{21.333}
\emmoveto{71.650}{21.323}
\emlineto{71.429}{21.230}
\emmoveto{71.429}{21.220}
\emlineto{71.207}{21.125}
\emmoveto{71.207}{21.115}
\emlineto{70.983}{21.021}
\emmoveto{70.983}{21.011}
\emlineto{70.759}{20.917}
\emmoveto{70.759}{20.907}
\emlineto{70.533}{20.814}
\emmoveto{70.533}{20.804}
\emlineto{70.306}{20.712}
\emmoveto{70.306}{20.702}
\emlineto{70.077}{20.611}
\emmoveto{70.077}{20.601}
\emlineto{69.848}{20.512}
\emmoveto{69.848}{20.502}
\emlineto{69.617}{20.413}
\emmoveto{69.617}{20.403}
\emlineto{69.385}{20.315}
\emmoveto{69.385}{20.305}
\emlineto{69.153}{20.218}
\emmoveto{69.153}{20.208}
\emlineto{68.918}{20.122}
\emmoveto{68.918}{20.112}
\emlineto{68.683}{20.027}
\emmoveto{68.683}{20.017}
\emlineto{68.447}{19.933}
\emmoveto{68.447}{19.923}
\emlineto{68.209}{19.840}
\emmoveto{68.209}{19.830}
\emlineto{67.971}{19.747}
\emmoveto{67.971}{19.737}
\emlineto{67.731}{19.656}
\emmoveto{67.731}{19.646}
\emlineto{67.490}{19.565}
\emmoveto{67.490}{19.555}
\emlineto{67.249}{19.475}
\emmoveto{67.249}{19.465}
\emlineto{67.006}{19.386}
\emmoveto{67.006}{19.376}
\emlineto{66.762}{19.297}
\emmoveto{66.762}{19.287}
\emlineto{66.517}{19.210}
\emmoveto{66.517}{19.200}
\emlineto{66.271}{19.123}
\emmoveto{66.271}{19.113}
\emlineto{66.024}{19.036}
\emmoveto{66.024}{19.026}
\emlineto{65.776}{18.951}
\emmoveto{65.776}{18.941}
\emlineto{65.527}{18.866}
\emmoveto{65.527}{18.856}
\emlineto{65.277}{18.781}
\emmoveto{65.277}{18.771}
\emlineto{65.026}{18.698}
\emmoveto{65.026}{18.688}
\emlineto{64.774}{18.615}
\emmoveto{64.774}{18.605}
\emlineto{64.521}{18.532}
\emmoveto{64.521}{18.522}
\emlineto{64.267}{18.450}
\emmoveto{64.267}{18.440}
\emlineto{64.012}{18.369}
\emmoveto{64.012}{18.359}
\emlineto{63.757}{18.288}
\emmoveto{63.757}{18.278}
\emlineto{63.500}{18.208}
\emmoveto{63.500}{18.198}
\emlineto{63.242}{18.128}
\emmoveto{63.242}{18.118}
\emlineto{62.983}{18.049}
\emmoveto{62.983}{18.039}
\emlineto{62.724}{17.970}
\emmoveto{62.724}{17.960}
\emlineto{62.463}{17.892}
\emmoveto{62.463}{17.882}
\emlineto{62.202}{17.814}
\emmoveto{62.202}{17.804}
\emlineto{61.939}{17.737}
\emmoveto{61.939}{17.727}
\emlineto{61.676}{17.660}
\emmoveto{61.676}{17.650}
\emlineto{61.412}{17.584}
\emmoveto{61.412}{17.574}
\emlineto{61.147}{17.508}
\emmoveto{61.147}{17.498}
\emlineto{60.881}{17.432}
\emmoveto{60.881}{17.422}
\emlineto{60.614}{17.357}
\emmoveto{60.614}{17.347}
\emlineto{60.346}{17.282}
\emmoveto{60.346}{17.272}
\emlineto{60.078}{17.208}
\emmoveto{60.078}{17.198}
\emlineto{59.808}{17.134}
\emmoveto{59.808}{17.124}
\emlineto{59.538}{17.060}
\emmoveto{59.538}{17.050}
\emlineto{59.266}{16.986}
\emmoveto{59.266}{16.976}
\emlineto{58.994}{16.913}
\emmoveto{58.994}{16.903}
\emlineto{58.721}{16.840}
\emmoveto{58.721}{16.830}
\emlineto{58.448}{16.768}
\emmoveto{58.448}{16.758}
\emlineto{58.173}{16.696}
\emmoveto{58.173}{16.686}
\emlineto{57.897}{16.624}
\emmoveto{57.897}{16.614}
\emlineto{57.621}{16.552}
\emmoveto{57.621}{16.542}
\emlineto{57.344}{16.481}
\emmoveto{57.344}{16.471}
\emlineto{57.066}{16.410}
\emmoveto{57.066}{16.400}
\emlineto{56.787}{16.339}
\emmoveto{56.787}{16.329}
\emlineto{56.507}{16.269}
\emmoveto{56.507}{16.259}
\emlineto{56.227}{16.199}
\emmoveto{56.227}{16.189}
\emlineto{55.945}{16.130}
\emmoveto{55.945}{16.120}
\emlineto{55.663}{16.061}
\emmoveto{55.663}{16.051}
\emlineto{55.380}{15.993}
\emmoveto{55.380}{15.983}
\emlineto{55.096}{15.925}
\emmoveto{55.096}{15.915}
\emlineto{54.812}{15.857}
\emmoveto{54.812}{15.847}
\emlineto{54.526}{15.790}
\emmoveto{54.526}{15.780}
\emlineto{54.240}{15.724}
\emmoveto{54.240}{15.714}
\emlineto{53.953}{15.657}
\emmoveto{53.953}{15.647}
\emlineto{53.665}{15.592}
\emmoveto{53.665}{15.582}
\emlineto{53.377}{15.527}
\emmoveto{53.377}{15.517}
\emlineto{53.087}{15.462}
\emmoveto{53.087}{15.452}
\emlineto{52.797}{15.398}
\emmoveto{52.797}{15.388}
\emlineto{52.507}{15.334}
\emmoveto{52.507}{15.324}
\emlineto{52.215}{15.271}
\emmoveto{52.215}{15.261}
\emlineto{51.923}{15.209}
\emmoveto{51.923}{15.199}
\emlineto{51.630}{15.147}
\emmoveto{51.630}{15.137}
\emlineto{51.336}{15.086}
\emmoveto{51.336}{15.076}
\emlineto{51.042}{15.025}
\emmoveto{51.042}{15.015}
\emlineto{50.746}{14.965}
\emmoveto{50.746}{14.955}
\emlineto{50.450}{14.905}
\emmoveto{50.450}{14.895}
\emlineto{50.154}{14.846}
\emmoveto{50.154}{14.836}
\emlineto{49.857}{14.787}
\emmoveto{49.857}{14.777}
\emlineto{49.559}{14.729}
\emmoveto{49.559}{14.719}
\emlineto{49.260}{14.672}
\emmoveto{49.260}{14.662}
\emlineto{48.961}{14.615}
\emmoveto{48.961}{14.605}
\emlineto{48.661}{14.558}
\emmoveto{48.661}{14.548}
\emlineto{48.360}{14.503}
\emmoveto{48.360}{14.493}
\emlineto{48.059}{14.448}
\emmoveto{48.059}{14.438}
\emlineto{47.757}{14.393}
\emmoveto{47.757}{14.383}
\emlineto{47.454}{14.339}
\emmoveto{47.454}{14.329}
\emlineto{47.151}{14.285}
\emmoveto{47.151}{14.275}
\emlineto{46.847}{14.233}
\emmoveto{46.847}{14.223}
\emlineto{46.543}{14.180}
\emmoveto{46.543}{14.170}
\emlineto{46.237}{14.128}
\emmoveto{46.237}{14.118}
\emlineto{45.932}{14.077}
\emmoveto{45.932}{14.067}
\emlineto{45.625}{14.027}
\emmoveto{45.625}{14.017}
\emlineto{45.319}{13.976}
\emmoveto{45.319}{13.966}
\emlineto{45.011}{13.927}
\emmoveto{45.011}{13.917}
\emlineto{44.703}{13.878}
\emmoveto{44.703}{13.868}
\emlineto{44.394}{13.830}
\emmoveto{44.394}{13.820}
\emlineto{44.085}{13.782}
\emmoveto{44.085}{13.772}
\emlineto{43.775}{13.734}
\emmoveto{43.775}{13.724}
\emlineto{43.465}{13.688}
\emmoveto{43.465}{13.678}
\emlineto{43.154}{13.641}
\emmoveto{43.154}{13.631}
\emlineto{42.843}{13.596}
\emmoveto{42.843}{13.586}
\emlineto{42.531}{13.551}
\emmoveto{42.531}{13.541}
\emlineto{42.218}{13.506}
\emmoveto{42.218}{13.496}
\emlineto{41.905}{13.462}
\emmoveto{41.905}{13.452}
\emlineto{41.592}{13.418}
\emmoveto{41.592}{13.408}
\emlineto{41.278}{13.375}
\emmoveto{41.278}{13.365}
\emlineto{40.963}{13.333}
\emmoveto{40.963}{13.323}
\emlineto{40.648}{13.291}
\emmoveto{40.648}{13.281}
\emlineto{40.333}{13.249}
\emmoveto{40.333}{13.239}
\emlineto{40.017}{13.208}
\emmoveto{40.017}{13.198}
\emlineto{39.700}{13.168}
\emmoveto{39.700}{13.158}
\emlineto{39.383}{13.128}
\emmoveto{39.383}{13.118}
\emlineto{39.066}{13.089}
\emmoveto{39.066}{13.079}
\emlineto{38.748}{13.050}
\emmoveto{38.748}{13.040}
\emlineto{38.429}{13.012}
\emmoveto{38.429}{13.002}
\emlineto{38.111}{12.974}
\emmoveto{38.111}{12.964}
\emlineto{37.791}{12.937}
\emmoveto{37.791}{12.927}
\emlineto{37.472}{12.900}
\emmoveto{37.472}{12.890}
\emlineto{37.152}{12.864}
\emmoveto{37.152}{12.854}
\emlineto{36.831}{12.829}
\emmoveto{36.831}{12.819}
\emlineto{36.510}{12.794}
\emmoveto{36.510}{12.784}
\emlineto{36.189}{12.759}
\emmoveto{36.189}{12.749}
\emlineto{35.867}{12.725}
\emmoveto{35.867}{12.715}
\emlineto{35.545}{12.692}
\emmoveto{35.545}{12.682}
\emlineto{35.222}{12.659}
\emmoveto{35.222}{12.649}
\emlineto{34.899}{12.626}
\emmoveto{34.899}{12.616}
\emlineto{34.576}{12.595}
\emmoveto{34.576}{12.585}
\emlineto{34.252}{12.563}
\emmoveto{34.252}{12.553}
\emlineto{33.928}{12.533}
\emmoveto{33.928}{12.523}
\emlineto{33.604}{12.503}
\emmoveto{33.604}{12.493}
\emlineto{33.279}{12.473}
\emmoveto{33.279}{12.463}
\emlineto{32.954}{12.444}
\emmoveto{32.954}{12.434}
\emlineto{32.629}{12.416}
\emmoveto{32.629}{12.406}
\emlineto{32.303}{12.388}
\emmoveto{32.303}{12.378}
\emlineto{31.977}{12.360}
\emmoveto{31.977}{12.350}
\emlineto{31.650}{12.334}
\emmoveto{31.650}{12.324}
\emlineto{31.324}{12.307}
\emmoveto{31.324}{12.297}
\emlineto{30.996}{12.282}
\emmoveto{30.996}{12.272}
\emlineto{30.669}{12.257}
\emmoveto{30.669}{12.247}
\emlineto{30.342}{12.232}
\emmoveto{30.342}{12.222}
\emlineto{30.014}{12.208}
\emmoveto{30.014}{12.198}
\emlineto{29.685}{12.185}
\emmoveto{29.685}{12.175}
\emlineto{29.357}{12.162}
\emmoveto{29.357}{12.152}
\emlineto{29.028}{12.140}
\emmoveto{29.028}{12.130}
\emlineto{28.699}{12.118}
\emmoveto{28.699}{12.108}
\emlineto{28.370}{12.097}
\emmoveto{28.370}{12.087}
\emlineto{28.040}{12.076}
\emmoveto{28.040}{12.066}
\emlineto{27.711}{12.056}
\emmoveto{27.711}{12.046}
\emlineto{27.381}{12.037}
\emmoveto{27.381}{12.027}
\emlineto{27.050}{12.018}
\emmoveto{27.050}{12.008}
\emlineto{26.720}{12.000}
\emmoveto{26.720}{11.990}
\emlineto{26.389}{11.982}
\emmoveto{26.389}{11.972}
\emlineto{26.059}{11.965}
\emmoveto{26.059}{11.955}
\emlineto{25.728}{11.948}
\emmoveto{25.728}{11.938}
\emlineto{25.396}{11.932}
\emmoveto{25.396}{11.922}
\emlineto{25.065}{11.917}
\emmoveto{25.065}{11.907}
\emlineto{24.733}{11.902}
\emmoveto{24.733}{11.892}
\emlineto{24.402}{11.888}
\emmoveto{24.402}{11.878}
\emlineto{24.070}{11.874}
\emmoveto{24.070}{11.864}
\emlineto{23.738}{11.861}
\emmoveto{23.738}{11.851}
\emlineto{23.405}{11.848}
\emmoveto{23.405}{11.838}
\emlineto{23.073}{11.837}
\emmoveto{23.073}{11.827}
\emlineto{22.740}{11.825}
\emmoveto{22.740}{11.815}
\emlineto{22.408}{11.814}
\emmoveto{22.408}{11.804}
\emlineto{22.075}{11.804}
\emmoveto{22.075}{11.794}
\emlineto{21.742}{11.794}
\emmoveto{21.742}{11.784}
\emlineto{21.409}{11.785}
\emmoveto{21.409}{11.775}
\emlineto{21.076}{11.777}
\emmoveto{21.076}{11.767}
\emlineto{20.743}{11.769}
\emmoveto{20.743}{11.759}
\emlineto{20.410}{11.761}
\emmoveto{20.410}{11.751}
\emlineto{20.076}{11.754}
\emmoveto{20.076}{11.744}
\emlineto{19.743}{11.748}
\emmoveto{19.743}{11.738}
\emlineto{19.409}{11.742}
\emmoveto{19.409}{11.732}
\emlineto{19.076}{11.737}
\emmoveto{19.076}{11.727}
\emlineto{18.742}{11.732}
\emmoveto{18.742}{11.722}
\emlineto{18.408}{11.728}
\emmoveto{18.408}{11.718}
\emlineto{18.075}{11.725}
\emmoveto{18.075}{11.715}
\emlineto{17.741}{11.722}
\emmoveto{17.741}{11.712}
\emlineto{17.407}{11.720}
\emmoveto{17.407}{11.710}
\emlineto{17.073}{11.718}
\emmoveto{17.073}{11.708}
\emlineto{16.739}{11.717}
\emmoveto{16.739}{11.707}
\emlineto{16.405}{11.716}
\emmoveto{16.405}{11.706}
\emlineto{16.072}{11.706}
\emlineto{16.072}{11.716}
\emmoveto{16.072}{11.706}
\emlineto{15.738}{11.716}
\emmoveto{15.738}{11.706}
\emlineto{15.404}{11.717}
\emmoveto{15.404}{11.707}
\emlineto{15.070}{11.719}
\emmoveto{15.070}{11.709}
\emlineto{14.736}{11.721}
\emmoveto{14.736}{11.711}
\emlineto{14.402}{11.724}
\emmoveto{14.402}{11.714}
\emlineto{14.069}{11.727}
\emmoveto{14.069}{11.717}
\emlineto{13.735}{11.731}
\emmoveto{13.735}{11.721}
\emlineto{13.401}{11.735}
\emmoveto{13.401}{11.725}
\emlineto{13.068}{11.740}
\emmoveto{13.068}{11.730}
\emlineto{12.734}{11.746}
\emmoveto{12.734}{11.736}
\emlineto{12.401}{11.752}
\emshow{95.780}{45.700}{k=1,v=0.3}
\emmoveto{12.000}{80.000}
\emlineto{12.472}{80.009}
\emmoveto{12.472}{79.999}
\emlineto{12.944}{80.007}
\emmoveto{12.944}{79.997}
\emlineto{13.416}{80.003}
\emmoveto{13.416}{79.993}
\emlineto{13.888}{79.998}
\emmoveto{13.888}{79.988}
\emlineto{14.360}{79.991}
\emmoveto{14.360}{79.981}
\emlineto{14.831}{79.982}
\emmoveto{14.831}{79.972}
\emlineto{15.303}{79.973}
\emmoveto{15.303}{79.963}
\emlineto{15.774}{79.961}
\emmoveto{15.774}{79.951}
\emlineto{16.246}{79.949}
\emmoveto{16.246}{79.939}
\emlineto{16.717}{79.935}
\emmoveto{16.717}{79.925}
\emlineto{17.188}{79.920}
\emmoveto{17.188}{79.910}
\emlineto{17.659}{79.904}
\emmoveto{17.659}{79.894}
\emlineto{18.130}{79.886}
\emmoveto{18.130}{79.876}
\emlineto{18.600}{79.867}
\emmoveto{18.600}{79.857}
\emlineto{19.070}{79.847}
\emmoveto{19.070}{79.837}
\emlineto{19.540}{79.826}
\emmoveto{19.540}{79.816}
\emlineto{20.010}{79.803}
\emmoveto{20.010}{79.793}
\emlineto{20.479}{79.780}
\emmoveto{20.479}{79.770}
\emlineto{20.948}{79.755}
\emmoveto{20.948}{79.745}
\emlineto{21.417}{79.729}
\emmoveto{21.417}{79.719}
\emlineto{21.886}{79.702}
\emmoveto{21.886}{79.692}
\emlineto{22.354}{79.674}
\emmoveto{22.354}{79.664}
\emlineto{22.822}{79.645}
\emmoveto{22.822}{79.635}
\emlineto{23.289}{79.616}
\emmoveto{23.289}{79.606}
\emlineto{23.756}{79.585}
\emmoveto{23.756}{79.575}
\emlineto{24.223}{79.553}
\emmoveto{24.223}{79.543}
\emlineto{24.690}{79.520}
\emmoveto{24.690}{79.510}
\emlineto{25.156}{79.486}
\emmoveto{25.156}{79.476}
\emlineto{25.621}{79.452}
\emmoveto{25.621}{79.442}
\emlineto{26.087}{79.416}
\emmoveto{26.087}{79.406}
\emlineto{26.551}{79.380}
\emmoveto{26.551}{79.370}
\emlineto{27.016}{79.343}
\emmoveto{27.016}{79.333}
\emlineto{27.480}{79.305}
\emmoveto{27.480}{79.295}
\emlineto{27.943}{79.266}
\emmoveto{27.943}{79.256}
\emlineto{28.406}{79.227}
\emmoveto{28.406}{79.217}
\emlineto{28.869}{79.186}
\emmoveto{28.869}{79.176}
\emlineto{29.331}{79.145}
\emmoveto{29.331}{79.135}
\emlineto{29.792}{79.103}
\emmoveto{29.792}{79.093}
\emlineto{30.253}{79.061}
\emmoveto{30.253}{79.051}
\emlineto{30.714}{79.018}
\emmoveto{30.714}{79.008}
\emlineto{31.174}{78.974}
\emmoveto{31.174}{78.964}
\emlineto{31.633}{78.929}
\emmoveto{31.633}{78.919}
\emlineto{32.092}{78.884}
\emmoveto{32.092}{78.874}
\emlineto{32.551}{78.838}
\emmoveto{32.551}{78.828}
\emlineto{33.009}{78.791}
\emmoveto{33.009}{78.781}
\emlineto{33.466}{78.744}
\emmoveto{33.466}{78.734}
\emlineto{33.923}{78.697}
\emmoveto{33.923}{78.687}
\emlineto{34.379}{78.648}
\emmoveto{34.379}{78.638}
\emlineto{34.835}{78.599}
\emmoveto{34.835}{78.589}
\emlineto{35.290}{78.550}
\emmoveto{35.290}{78.540}
\emlineto{35.744}{78.500}
\emmoveto{35.744}{78.490}
\emlineto{36.198}{78.449}
\emmoveto{36.198}{78.439}
\emlineto{36.651}{78.398}
\emmoveto{36.651}{78.388}
\emlineto{37.104}{78.346}
\emmoveto{37.104}{78.336}
\emlineto{37.556}{78.294}
\emmoveto{37.556}{78.284}
\emlineto{38.007}{78.241}
\emmoveto{38.007}{78.231}
\emlineto{38.458}{78.187}
\emmoveto{38.458}{78.177}
\emlineto{38.908}{78.133}
\emmoveto{38.908}{78.123}
\emlineto{39.358}{78.078}
\emmoveto{39.358}{78.068}
\emlineto{39.807}{78.022}
\emmoveto{39.807}{78.012}
\emlineto{40.255}{77.966}
\emmoveto{40.255}{77.956}
\emlineto{40.703}{77.909}
\emmoveto{40.703}{77.899}
\emlineto{41.149}{77.851}
\emmoveto{41.149}{77.841}
\emlineto{41.596}{77.793}
\emmoveto{41.596}{77.783}
\emlineto{42.041}{77.733}
\emmoveto{42.041}{77.723}
\emlineto{42.486}{77.674}
\emmoveto{42.486}{77.664}
\emlineto{42.930}{77.613}
\emmoveto{42.930}{77.603}
\emlineto{43.373}{77.551}
\emmoveto{43.373}{77.541}
\emlineto{43.816}{77.489}
\emmoveto{43.816}{77.479}
\emlineto{44.258}{77.426}
\emmoveto{44.258}{77.416}
\emlineto{44.699}{77.363}
\emmoveto{44.699}{77.353}
\emlineto{45.139}{77.298}
\emmoveto{45.139}{77.288}
\emlineto{45.579}{77.233}
\emmoveto{45.579}{77.223}
\emlineto{46.018}{77.167}
\emmoveto{46.018}{77.157}
\emlineto{46.456}{77.100}
\emmoveto{46.456}{77.090}
\emlineto{46.893}{77.033}
\emmoveto{46.893}{77.023}
\emlineto{47.330}{76.965}
\emmoveto{47.330}{76.955}
\emlineto{47.765}{76.896}
\emmoveto{47.765}{76.886}
\emlineto{48.200}{76.826}
\emmoveto{48.200}{76.816}
\emlineto{48.634}{76.755}
\emmoveto{48.634}{76.745}
\emlineto{49.067}{76.684}
\emmoveto{49.067}{76.674}
\emlineto{49.500}{76.612}
\emmoveto{49.500}{76.602}
\emlineto{49.931}{76.539}
\emmoveto{49.931}{76.529}
\emlineto{50.362}{76.465}
\emmoveto{50.362}{76.455}
\emlineto{50.791}{76.391}
\emmoveto{50.791}{76.381}
\emlineto{51.220}{76.316}
\emmoveto{51.220}{76.306}
\emlineto{51.648}{76.240}
\emmoveto{51.648}{76.230}
\emlineto{52.075}{76.163}
\emmoveto{52.075}{76.153}
\emlineto{52.501}{76.086}
\emmoveto{52.501}{76.076}
\emlineto{52.927}{76.008}
\emmoveto{52.927}{75.998}
\emlineto{53.351}{75.929}
\emmoveto{53.351}{75.919}
\emlineto{53.774}{75.849}
\emmoveto{53.774}{75.839}
\emlineto{54.197}{75.769}
\emmoveto{54.197}{75.759}
\emlineto{54.618}{75.688}
\emmoveto{54.618}{75.678}
\emlineto{55.039}{75.606}
\emmoveto{55.039}{75.596}
\emlineto{55.458}{75.523}
\emmoveto{55.458}{75.513}
\emlineto{55.877}{75.440}
\emmoveto{55.877}{75.430}
\emlineto{56.294}{75.356}
\emmoveto{56.294}{75.346}
\emlineto{56.711}{75.271}
\emmoveto{56.711}{75.261}
\emlineto{57.127}{75.186}
\emmoveto{57.127}{75.176}
\emlineto{57.541}{75.100}
\emmoveto{57.541}{75.090}
\emlineto{57.955}{75.013}
\emmoveto{57.955}{75.003}
\emlineto{58.367}{74.926}
\emmoveto{58.367}{74.916}
\emlineto{58.779}{74.838}
\emmoveto{58.779}{74.828}
\emlineto{59.189}{74.749}
\emmoveto{59.189}{74.739}
\emlineto{59.599}{74.660}
\emmoveto{59.599}{74.650}
\emlineto{60.007}{74.570}
\emmoveto{60.007}{74.560}
\emlineto{60.414}{74.479}
\emmoveto{60.414}{74.469}
\emlineto{60.820}{74.387}
\emmoveto{60.820}{74.377}
\emlineto{61.225}{74.295}
\emmoveto{61.225}{74.285}
\emlineto{61.630}{74.203}
\emmoveto{61.630}{74.193}
\emlineto{62.032}{74.110}
\emmoveto{62.032}{74.100}
\emlineto{62.434}{74.016}
\emmoveto{62.434}{74.006}
\emlineto{62.835}{73.921}
\emmoveto{62.835}{73.911}
\emlineto{63.235}{73.826}
\emmoveto{63.235}{73.816}
\emlineto{63.633}{73.730}
\emmoveto{63.633}{73.720}
\emlineto{64.030}{73.634}
\emmoveto{64.030}{73.624}
\emlineto{64.427}{73.536}
\emmoveto{64.427}{73.526}
\emlineto{64.822}{73.439}
\emmoveto{64.822}{73.429}
\emlineto{65.215}{73.340}
\emmoveto{65.215}{73.330}
\emlineto{65.608}{73.241}
\emmoveto{65.608}{73.231}
\emlineto{66.000}{73.142}
\emmoveto{66.000}{73.132}
\emlineto{66.390}{73.041}
\emmoveto{66.390}{73.031}
\emlineto{66.779}{72.941}
\emmoveto{66.779}{72.931}
\emlineto{67.167}{72.839}
\emmoveto{67.167}{72.829}
\emlineto{67.554}{72.737}
\emmoveto{67.554}{72.727}
\emlineto{67.940}{72.634}
\emmoveto{67.940}{72.624}
\emlineto{68.324}{72.531}
\emmoveto{68.324}{72.521}
\emlineto{68.707}{72.427}
\emmoveto{68.707}{72.417}
\emlineto{69.089}{72.323}
\emmoveto{69.089}{72.313}
\emlineto{69.470}{72.217}
\emmoveto{69.470}{72.207}
\emlineto{69.849}{72.112}
\emmoveto{69.849}{72.102}
\emlineto{70.227}{72.005}
\emmoveto{70.227}{71.995}
\emlineto{70.604}{71.898}
\emmoveto{70.604}{71.888}
\emlineto{70.980}{71.791}
\emmoveto{70.980}{71.781}
\emlineto{71.354}{71.683}
\emmoveto{71.354}{71.673}
\emlineto{71.727}{71.578}
\emmoveto{71.727}{71.568}
\emlineto{72.099}{71.475}
\emmoveto{72.099}{71.465}
\emlineto{72.470}{71.373}
\emmoveto{72.470}{71.363}
\emlineto{72.839}{71.273}
\emmoveto{72.839}{71.263}
\emlineto{73.208}{71.175}
\emmoveto{73.208}{71.165}
\emlineto{73.575}{71.078}
\emmoveto{73.575}{71.068}
\emlineto{73.941}{70.983}
\emmoveto{73.941}{70.973}
\emlineto{74.306}{70.890}
\emmoveto{74.306}{70.880}
\emlineto{74.670}{70.798}
\emmoveto{74.670}{70.788}
\emlineto{75.032}{70.707}
\emmoveto{75.032}{70.697}
\emlineto{75.394}{70.618}
\emmoveto{75.394}{70.608}
\emlineto{75.755}{70.530}
\emmoveto{75.755}{70.520}
\emlineto{76.114}{70.443}
\emmoveto{76.114}{70.433}
\emlineto{76.473}{70.357}
\emmoveto{76.473}{70.347}
\emlineto{76.831}{70.273}
\emmoveto{76.831}{70.263}
\emlineto{77.187}{70.190}
\emmoveto{77.187}{70.180}
\emlineto{77.543}{70.108}
\emmoveto{77.543}{70.098}
\emlineto{77.898}{70.027}
\emmoveto{77.898}{70.017}
\emlineto{78.251}{69.947}
\emmoveto{78.251}{69.937}
\emlineto{78.604}{69.868}
\emmoveto{78.604}{69.858}
\emlineto{78.956}{69.790}
\emmoveto{78.956}{69.780}
\emlineto{79.307}{69.713}
\emmoveto{79.307}{69.703}
\emlineto{79.657}{69.637}
\emmoveto{79.657}{69.627}
\emlineto{80.006}{69.562}
\emmoveto{80.006}{69.552}
\emlineto{80.354}{69.488}
\emmoveto{80.354}{69.478}
\emlineto{80.702}{69.414}
\emmoveto{80.702}{69.404}
\emlineto{81.048}{69.341}
\emmoveto{81.048}{69.331}
\emlineto{81.394}{69.269}
\emmoveto{81.394}{69.259}
\emlineto{81.739}{69.198}
\emmoveto{81.739}{69.188}
\emlineto{82.083}{69.127}
\emmoveto{82.083}{69.117}
\emlineto{82.426}{69.057}
\emmoveto{82.426}{69.047}
\emlineto{82.768}{68.988}
\emmoveto{82.768}{68.978}
\emlineto{83.110}{68.919}
\emmoveto{83.110}{68.909}
\emlineto{83.451}{68.850}
\emmoveto{83.451}{68.840}
\emlineto{83.790}{68.783}
\emmoveto{83.790}{68.773}
\emlineto{84.130}{68.715}
\emmoveto{84.130}{68.705}
\emlineto{84.468}{68.649}
\emmoveto{84.468}{68.639}
\emlineto{84.805}{68.582}
\emmoveto{84.805}{68.572}
\emlineto{85.142}{68.517}
\emmoveto{85.142}{68.507}
\emlineto{85.478}{68.451}
\emmoveto{85.478}{68.441}
\emlineto{85.813}{68.386}
\emmoveto{85.813}{68.376}
\emlineto{86.148}{68.322}
\emmoveto{86.148}{68.312}
\emlineto{86.482}{68.257}
\emmoveto{86.482}{68.247}
\emlineto{86.815}{68.193}
\emmoveto{86.815}{68.183}
\emlineto{87.147}{68.130}
\emmoveto{87.147}{68.120}
\emlineto{87.478}{68.066}
\emmoveto{87.478}{68.056}
\emlineto{87.809}{68.003}
\emmoveto{87.809}{67.993}
\emlineto{88.139}{67.940}
\emmoveto{88.139}{67.930}
\emlineto{88.468}{67.878}
\emmoveto{88.468}{67.868}
\emlineto{88.796}{67.815}
\emmoveto{88.796}{67.805}
\emlineto{89.124}{67.753}
\emmoveto{89.124}{67.743}
\emlineto{89.451}{67.691}
\emmoveto{89.451}{67.681}
\emlineto{89.778}{67.630}
\emmoveto{89.778}{67.620}
\emlineto{90.103}{67.568}
\emmoveto{90.103}{67.558}
\emlineto{90.428}{67.507}
\emmoveto{90.428}{67.497}
\emlineto{90.752}{67.447}
\emmoveto{90.752}{67.437}
\emlineto{91.075}{67.386}
\emmoveto{91.075}{67.376}
\emlineto{91.398}{67.327}
\emmoveto{91.398}{67.317}
\emlineto{91.720}{67.267}
\emmoveto{91.720}{67.257}
\emlineto{92.041}{67.208}
\emmoveto{92.041}{67.198}
\emlineto{92.362}{67.150}
\emmoveto{92.362}{67.140}
\emlineto{92.682}{67.092}
\emmoveto{92.682}{67.082}
\emlineto{93.001}{67.034}
\emmoveto{93.001}{67.024}
\emlineto{93.320}{66.977}
\emmoveto{93.320}{66.967}
\emlineto{93.638}{66.921}
\emmoveto{93.638}{66.911}
\emlineto{93.955}{66.865}
\emmoveto{93.955}{66.855}
\emlineto{94.271}{66.810}
\emmoveto{94.271}{66.800}
\emlineto{94.587}{66.755}
\emmoveto{94.587}{66.745}
\emlineto{94.902}{66.700}
\emmoveto{94.902}{66.690}
\emlineto{95.217}{66.647}
\emmoveto{95.217}{66.637}
\emlineto{95.531}{66.593}
\emmoveto{95.531}{66.583}
\emlineto{95.844}{66.541}
\emmoveto{95.844}{66.531}
\emlineto{96.157}{66.489}
\emmoveto{96.157}{66.479}
\emlineto{96.469}{66.437}
\emmoveto{96.469}{66.427}
\emlineto{96.781}{66.386}
\emmoveto{96.781}{66.376}
\emlineto{97.092}{66.336}
\emmoveto{97.092}{66.326}
\emlineto{97.402}{66.286}
\emmoveto{97.402}{66.276}
\emlineto{97.712}{66.237}
\emmoveto{97.712}{66.227}
\emlineto{98.021}{66.189}
\emmoveto{98.021}{66.179}
\emlineto{98.330}{66.141}
\emmoveto{98.330}{66.131}
\emlineto{98.638}{66.094}
\emmoveto{98.638}{66.084}
\emlineto{98.945}{66.047}
\emmoveto{98.945}{66.037}
\emlineto{99.252}{66.001}
\emmoveto{99.252}{65.991}
\emlineto{99.559}{65.955}
\emmoveto{99.559}{65.945}
\emlineto{99.865}{65.910}
\emmoveto{99.865}{65.900}
\emlineto{100.170}{65.866}
\emmoveto{100.170}{65.856}
\emlineto{100.475}{65.822}
\emmoveto{100.475}{65.812}
\emlineto{100.779}{65.779}
\emmoveto{100.779}{65.769}
\emlineto{101.083}{65.736}
\emmoveto{101.083}{65.726}
\emlineto{101.386}{65.694}
\emmoveto{101.386}{65.684}
\emlineto{101.689}{65.652}
\emmoveto{101.689}{65.642}
\emlineto{101.991}{65.611}
\emmoveto{101.991}{65.601}
\emlineto{102.293}{65.571}
\emmoveto{102.293}{65.561}
\emlineto{102.595}{65.531}
\emmoveto{102.595}{65.521}
\emlineto{102.896}{65.491}
\emmoveto{102.896}{65.481}
\emlineto{103.196}{65.453}
\emmoveto{103.196}{65.443}
\emlineto{103.496}{65.414}
\emmoveto{103.496}{65.404}
\emlineto{103.796}{65.376}
\emmoveto{103.796}{65.366}
\emlineto{104.095}{65.339}
\emmoveto{104.095}{65.329}
\emlineto{104.393}{65.302}
\emmoveto{104.393}{65.292}
\emlineto{104.692}{65.266}
\emmoveto{104.692}{65.256}
\emlineto{104.989}{65.230}
\emmoveto{104.989}{65.220}
\emlineto{105.287}{65.195}
\emmoveto{105.287}{65.185}
\emlineto{105.584}{65.160}
\emmoveto{105.584}{65.150}
\emlineto{105.880}{65.126}
\emmoveto{105.880}{65.116}
\emlineto{106.177}{65.092}
\emmoveto{106.177}{65.082}
\emlineto{106.472}{65.058}
\emmoveto{106.472}{65.048}
\emlineto{106.768}{65.025}
\emmoveto{106.768}{65.015}
\emlineto{107.063}{64.993}
\emmoveto{107.063}{64.983}
\emlineto{107.357}{64.961}
\emmoveto{107.357}{64.951}
\emlineto{107.651}{64.929}
\emmoveto{107.651}{64.919}
\emlineto{107.945}{64.898}
\emmoveto{107.945}{64.888}
\emlineto{108.239}{64.868}
\emmoveto{108.239}{64.858}
\emlineto{108.532}{64.837}
\emmoveto{108.532}{64.827}
\emlineto{108.825}{64.808}
\emmoveto{108.825}{64.798}
\emlineto{109.117}{64.778}
\emmoveto{109.117}{64.768}
\emlineto{109.409}{64.749}
\emmoveto{109.409}{64.739}
\emlineto{109.701}{64.721}
\emmoveto{109.701}{64.711}
\emlineto{109.992}{64.693}
\emmoveto{109.992}{64.683}
\emlineto{110.284}{64.666}
\emmoveto{110.284}{64.656}
\emlineto{110.574}{64.639}
\emmoveto{110.574}{64.629}
\emlineto{110.865}{64.612}
\emmoveto{110.865}{64.602}
\emlineto{111.155}{64.586}
\emmoveto{111.155}{64.576}
\emlineto{111.445}{64.561}
\emmoveto{111.445}{64.551}
\emlineto{111.734}{64.535}
\emmoveto{111.734}{64.525}
\emlineto{112.024}{64.511}
\emmoveto{112.024}{64.501}
\emlineto{112.313}{64.487}
\emmoveto{112.313}{64.477}
\emlineto{112.601}{64.463}
\emmoveto{112.601}{64.453}
\emlineto{112.890}{64.440}
\emmoveto{112.890}{64.430}
\emlineto{113.178}{64.417}
\emmoveto{113.178}{64.407}
\emlineto{113.466}{64.395}
\emmoveto{113.466}{64.385}
\emlineto{113.753}{64.373}
\emmoveto{113.753}{64.363}
\emlineto{114.041}{64.351}
\emmoveto{114.041}{64.341}
\emlineto{114.328}{64.330}
\emmoveto{114.328}{64.320}
\emlineto{114.615}{64.310}
\emmoveto{114.615}{64.300}
\emlineto{114.901}{64.290}
\emmoveto{114.901}{64.280}
\emlineto{115.188}{64.271}
\emmoveto{115.188}{64.261}
\emlineto{115.474}{64.252}
\emmoveto{115.474}{64.242}
\emlineto{115.760}{64.233}
\emmoveto{115.760}{64.223}
\emlineto{116.045}{64.215}
\emmoveto{116.045}{64.205}
\emlineto{116.331}{64.198}
\emmoveto{116.331}{64.188}
\emlineto{116.616}{64.181}
\emmoveto{116.616}{64.171}
\emlineto{116.901}{64.164}
\emmoveto{116.901}{64.154}
\emlineto{117.186}{64.148}
\emmoveto{117.186}{64.138}
\emlineto{117.471}{64.133}
\emmoveto{117.471}{64.123}
\emlineto{117.756}{64.118}
\emmoveto{117.756}{64.108}
\emlineto{118.040}{64.103}
\emmoveto{118.040}{64.093}
\emlineto{118.324}{64.089}
\emmoveto{118.324}{64.079}
\emlineto{118.608}{64.075}
\emmoveto{118.608}{64.065}
\emlineto{118.892}{64.062}
\emmoveto{118.892}{64.052}
\emlineto{119.176}{64.050}
\emmoveto{119.176}{64.040}
\emlineto{119.460}{64.038}
\emmoveto{119.460}{64.028}
\emlineto{119.743}{64.026}
\emmoveto{119.743}{64.016}
\emlineto{120.026}{64.015}
\emmoveto{120.026}{64.005}
\emlineto{120.309}{64.004}
\emmoveto{120.309}{63.994}
\emlineto{120.593}{63.994}
\emmoveto{120.593}{63.984}
\emlineto{120.876}{63.984}
\emmoveto{120.876}{63.974}
\emlineto{121.158}{63.975}
\emmoveto{121.158}{63.965}
\emlineto{121.441}{63.966}
\emmoveto{121.441}{63.956}
\emlineto{121.724}{63.958}
\emmoveto{121.724}{63.948}
\emlineto{122.006}{63.950}
\emmoveto{122.006}{63.940}
\emlineto{122.289}{63.943}
\emmoveto{122.289}{63.933}
\emlineto{122.571}{63.936}
\emmoveto{122.571}{63.926}
\emlineto{122.853}{63.930}
\emmoveto{122.853}{63.920}
\emlineto{123.136}{63.924}
\emmoveto{123.136}{63.914}
\emlineto{123.418}{63.919}
\emmoveto{123.418}{63.909}
\emlineto{123.700}{63.914}
\emmoveto{123.700}{63.904}
\emlineto{123.982}{63.909}
\emmoveto{123.982}{63.899}
\emlineto{124.264}{63.906}
\emmoveto{124.264}{63.896}
\emlineto{124.546}{63.902}
\emmoveto{124.546}{63.892}
\emlineto{124.828}{63.899}
\emmoveto{124.828}{63.889}
\emlineto{125.110}{63.897}
\emmoveto{125.110}{63.887}
\emlineto{125.391}{63.895}
\emmoveto{125.391}{63.885}
\emlineto{125.673}{63.893}
\emmoveto{125.673}{63.883}
\emlineto{125.955}{63.892}
\emmoveto{125.955}{63.882}
\emlineto{126.237}{63.891}
\emmoveto{126.237}{63.881}
\emlineto{126.519}{63.891}
\emmoveto{126.519}{63.881}
\emlineto{126.800}{63.892}
\emmoveto{126.800}{63.882}
\emlineto{127.082}{63.892}
\emmoveto{127.082}{63.882}
\emlineto{127.364}{63.894}
\emmoveto{127.364}{63.884}
\emlineto{127.646}{63.895}
\emmoveto{127.646}{63.885}
\emlineto{127.928}{63.898}
\emmoveto{127.928}{63.888}
\emlineto{128.210}{63.900}
\emmoveto{128.210}{63.890}
\emlineto{128.492}{63.904}
\emmoveto{128.492}{63.894}
\emlineto{128.774}{63.907}
\emmoveto{128.774}{63.897}
\emlineto{129.056}{63.911}
\emmoveto{129.056}{63.901}
\emlineto{129.338}{63.916}
\emmoveto{129.338}{63.906}
\emlineto{129.620}{63.921}
\emshow{95.780}{66.700}{k=1,v=0.4}
\emshow{1.000}{10.000}{-3.00e-1}
\emshow{1.000}{17.000}{-2.30e-1}
\emshow{1.000}{24.000}{-1.60e-1}
\emshow{1.000}{31.000}{-9.00e-2}
\emshow{1.000}{38.000}{-2.00e-2}
\emshow{1.000}{45.000}{5.00e-2}
\emshow{1.000}{52.000}{1.20e-1}
\emshow{1.000}{59.000}{1.90e-1}
\emshow{1.000}{66.000}{2.60e-1}
\emshow{1.000}{73.000}{3.30e-1}
\emshow{1.000}{80.000}{4.00e-1}
\emshow{12.000}{5.000}{-5.00e-1}
\emshow{23.800}{5.000}{-3.00e-1}
\emshow{35.600}{5.000}{-1.00e-1}
\emshow{47.400}{5.000}{1.00e-1}
\emshow{59.200}{5.000}{3.00e-1}
\emshow{71.000}{5.000}{5.00e-1}
\emshow{82.800}{5.000}{7.00e-1}
\emshow{94.600}{5.000}{9.00e-1}
\emshow{106.400}{5.000}{1.10e0}
\emshow{118.200}{5.000}{1.30e0}
\emshow{130.000}{5.000}{1.50e0}

\centerline{\bf{Fig.A.2}}
\eject
\newcount\numpoint
\newcount\numpointo
\numpoint=1 \numpointo=1
\def\emmoveto#1#2{\offinterlineskip
\hbox to 0 true cm{\vbox to 0
true cm{\vskip - #2 true mm
\hskip #1 true mm \special{em:point
\the\numpoint}\vss}\hss}
\numpointo=\numpoint
\global\advance \numpoint by 1}
\def\emlineto#1#2{\offinterlineskip
\hbox to 0 true cm{\vbox to 0
true cm{\vskip - #2 true mm
\hskip #1 true mm \special{em:point
\the\numpoint}\vss}\hss}
\special{em:line
\the\numpointo,\the\numpoint}
\numpointo=\numpoint
\global\advance \numpoint by 1}
\def\emshow#1#2#3{\offinterlineskip
\hbox to 0 true cm{\vbox to 0
true cm{\vskip - #2 true mm
\hskip #1 true mm \vbox to 0
true cm{\vss\hbox{#3\hss
}}\vss}\hss}}
\special{em:linewidth 0.8pt}

\vrule width 0 mm height                0 mm depth 90.000 true mm

\special{em:linewidth 0.8pt}
\emmoveto{130.000}{10.000}
\emlineto{12.000}{10.000}
\emlineto{12.000}{80.000}
\emmoveto{71.000}{10.000}
\emlineto{71.000}{80.000}
\emmoveto{12.000}{45.000}
\emlineto{130.000}{45.000}
\emmoveto{130.000}{10.000}
\emlineto{130.000}{80.000}
\emlineto{12.000}{80.000}
\emlineto{12.000}{10.000}
\emlineto{130.000}{10.000}
\special{em:linewidth 0.4pt}
\emmoveto{12.000}{17.000}
\emlineto{130.000}{17.000}
\emmoveto{12.000}{24.000}
\emlineto{130.000}{24.000}
\emmoveto{12.000}{31.000}
\emlineto{130.000}{31.000}
\emmoveto{12.000}{38.000}
\emlineto{130.000}{38.000}
\emmoveto{12.000}{45.000}
\emlineto{130.000}{45.000}
\emmoveto{12.000}{52.000}
\emlineto{130.000}{52.000}
\emmoveto{12.000}{59.000}
\emlineto{130.000}{59.000}
\emmoveto{12.000}{66.000}
\emlineto{130.000}{66.000}
\emmoveto{12.000}{73.000}
\emlineto{130.000}{73.000}
\emmoveto{23.800}{10.000}
\emlineto{23.800}{80.000}
\emmoveto{35.600}{10.000}
\emlineto{35.600}{80.000}
\emmoveto{47.400}{10.000}
\emlineto{47.400}{80.000}
\emmoveto{59.200}{10.000}
\emlineto{59.200}{80.000}
\emmoveto{71.000}{10.000}
\emlineto{71.000}{80.000}
\emmoveto{82.800}{10.000}
\emlineto{82.800}{80.000}
\emmoveto{94.600}{10.000}
\emlineto{94.600}{80.000}
\emmoveto{106.400}{10.000}
\emlineto{106.400}{80.000}
\emmoveto{118.200}{10.000}
\emlineto{118.200}{80.000}
\special{em:linewidth 0.8pt}
\emmoveto{12.000}{42.308}
\emlineto{12.142}{42.317}
\emmoveto{12.142}{42.307}
\emlineto{12.283}{42.317}
\emmoveto{12.283}{42.307}
\emlineto{12.425}{42.315}
\emmoveto{12.425}{42.305}
\emlineto{12.566}{42.314}
\emmoveto{12.566}{42.304}
\emlineto{12.708}{42.312}
\emmoveto{12.708}{42.302}
\emlineto{12.849}{42.310}
\emmoveto{12.849}{42.300}
\emlineto{12.991}{42.307}
\emmoveto{12.991}{42.297}
\emlineto{13.132}{42.305}
\emmoveto{13.132}{42.295}
\emlineto{13.274}{42.302}
\emmoveto{13.274}{42.292}
\emlineto{13.415}{42.299}
\emmoveto{13.415}{42.289}
\emlineto{13.557}{42.297}
\emmoveto{13.557}{42.287}
\emlineto{13.698}{42.294}
\emmoveto{13.698}{42.284}
\emlineto{13.840}{42.291}
\emmoveto{13.840}{42.281}
\emlineto{13.981}{42.288}
\emmoveto{13.981}{42.278}
\emlineto{14.122}{42.285}
\emmoveto{14.122}{42.275}
\emlineto{14.264}{42.282}
\emmoveto{14.264}{42.272}
\emlineto{14.405}{42.278}
\emmoveto{14.405}{42.268}
\emlineto{14.546}{42.275}
\emmoveto{14.546}{42.265}
\emlineto{14.687}{42.272}
\emmoveto{14.687}{42.262}
\emlineto{14.828}{42.269}
\emmoveto{14.828}{42.259}
\emlineto{14.970}{42.266}
\emmoveto{14.970}{42.256}
\emlineto{15.111}{42.263}
\emmoveto{15.111}{42.253}
\emlineto{15.252}{42.259}
\emmoveto{15.252}{42.249}
\emlineto{15.393}{42.256}
\emmoveto{15.393}{42.246}
\emlineto{15.534}{42.253}
\emmoveto{15.534}{42.243}
\emlineto{15.675}{42.250}
\emmoveto{15.675}{42.240}
\emlineto{15.816}{42.247}
\emmoveto{15.816}{42.237}
\emlineto{15.957}{42.243}
\emmoveto{15.957}{42.233}
\emlineto{16.098}{42.240}
\emmoveto{16.098}{42.230}
\emlineto{16.239}{42.237}
\emmoveto{16.239}{42.227}
\emlineto{16.380}{42.234}
\emmoveto{16.380}{42.224}
\emlineto{16.521}{42.231}
\emmoveto{16.521}{42.221}
\emlineto{16.661}{42.227}
\emmoveto{16.661}{42.217}
\emlineto{16.802}{42.224}
\emmoveto{16.802}{42.214}
\emlineto{16.943}{42.221}
\emmoveto{16.943}{42.211}
\emlineto{17.084}{42.218}
\emmoveto{17.084}{42.208}
\emlineto{17.224}{42.215}
\emmoveto{17.224}{42.205}
\emlineto{17.365}{42.211}
\emmoveto{17.365}{42.201}
\emlineto{17.506}{42.208}
\emmoveto{17.506}{42.198}
\emlineto{17.646}{42.205}
\emmoveto{17.646}{42.195}
\emlineto{17.787}{42.202}
\emmoveto{17.787}{42.192}
\emlineto{17.927}{42.198}
\emmoveto{17.927}{42.188}
\emlineto{18.068}{42.195}
\emmoveto{18.068}{42.185}
\emlineto{18.209}{42.192}
\emmoveto{18.209}{42.182}
\emlineto{18.349}{42.189}
\emmoveto{18.349}{42.179}
\emlineto{18.489}{42.186}
\emmoveto{18.489}{42.176}
\emlineto{18.630}{42.182}
\emmoveto{18.630}{42.172}
\emlineto{18.770}{42.179}
\emmoveto{18.770}{42.169}
\emlineto{18.911}{42.176}
\emmoveto{18.911}{42.166}
\emlineto{19.051}{42.173}
\emmoveto{19.051}{42.163}
\emlineto{19.191}{42.170}
\emmoveto{19.191}{42.160}
\emlineto{19.332}{42.166}
\emmoveto{19.332}{42.156}
\emlineto{19.472}{42.163}
\emmoveto{19.472}{42.153}
\emlineto{19.612}{42.159}
\emmoveto{19.612}{42.149}
\emlineto{19.752}{42.155}
\emmoveto{19.752}{42.145}
\emlineto{19.892}{42.151}
\emmoveto{19.892}{42.141}
\emlineto{20.033}{42.147}
\emmoveto{20.033}{42.137}
\emlineto{20.173}{42.142}
\emmoveto{20.173}{42.132}
\emlineto{20.313}{42.137}
\emmoveto{20.313}{42.127}
\emlineto{20.453}{42.132}
\emmoveto{20.453}{42.122}
\emlineto{20.593}{42.127}
\emmoveto{20.593}{42.117}
\emlineto{20.733}{42.122}
\emmoveto{20.733}{42.112}
\emlineto{20.872}{42.116}
\emmoveto{20.872}{42.106}
\emlineto{21.012}{42.110}
\emmoveto{21.012}{42.100}
\emlineto{21.152}{42.105}
\emmoveto{21.152}{42.095}
\emlineto{21.292}{42.099}
\emmoveto{21.292}{42.089}
\emlineto{21.431}{42.093}
\emmoveto{21.431}{42.083}
\emlineto{21.571}{42.087}
\emmoveto{21.571}{42.077}
\emlineto{21.711}{42.081}
\emmoveto{21.711}{42.071}
\emlineto{21.850}{42.075}
\emmoveto{21.850}{42.065}
\emlineto{21.989}{42.069}
\emmoveto{21.989}{42.059}
\emlineto{22.129}{42.063}
\emmoveto{22.129}{42.053}
\emlineto{22.268}{42.056}
\emmoveto{22.268}{42.046}
\emlineto{22.407}{42.050}
\emmoveto{22.407}{42.040}
\emlineto{22.547}{42.044}
\emmoveto{22.547}{42.034}
\emlineto{22.686}{42.038}
\emmoveto{22.686}{42.028}
\emlineto{22.825}{42.031}
\emmoveto{22.825}{42.021}
\emlineto{22.964}{42.025}
\emmoveto{22.964}{42.015}
\emlineto{23.103}{42.019}
\emmoveto{23.103}{42.009}
\emlineto{23.242}{42.012}
\emmoveto{23.242}{42.002}
\emlineto{23.381}{42.006}
\emmoveto{23.381}{41.996}
\emlineto{23.520}{42.000}
\emmoveto{23.520}{41.990}
\emlineto{23.659}{41.993}
\emmoveto{23.659}{41.983}
\emlineto{23.797}{41.987}
\emmoveto{23.797}{41.977}
\emlineto{23.936}{41.980}
\emmoveto{23.936}{41.970}
\emlineto{24.075}{41.974}
\emmoveto{24.075}{41.964}
\emlineto{24.213}{41.968}
\emmoveto{24.213}{41.958}
\emlineto{24.352}{41.961}
\emmoveto{24.352}{41.951}
\emlineto{24.490}{41.955}
\emmoveto{24.490}{41.945}
\emlineto{24.628}{41.949}
\emmoveto{24.628}{41.939}
\emlineto{24.767}{41.942}
\emmoveto{24.767}{41.932}
\emlineto{24.905}{41.936}
\emmoveto{24.905}{41.926}
\emlineto{25.043}{41.930}
\emmoveto{25.043}{41.920}
\emlineto{25.181}{41.923}
\emmoveto{25.181}{41.913}
\emlineto{25.320}{41.917}
\emmoveto{25.320}{41.907}
\emlineto{25.458}{41.910}
\emmoveto{25.458}{41.900}
\emlineto{25.596}{41.904}
\emmoveto{25.596}{41.894}
\emlineto{25.734}{41.898}
\emmoveto{25.734}{41.888}
\emlineto{25.871}{41.891}
\emmoveto{25.871}{41.881}
\emlineto{26.009}{41.885}
\emmoveto{26.009}{41.875}
\emlineto{26.147}{41.879}
\emmoveto{26.147}{41.869}
\emlineto{26.285}{41.872}
\emmoveto{26.285}{41.862}
\emlineto{26.422}{41.866}
\emmoveto{26.422}{41.856}
\emlineto{26.560}{41.859}
\emmoveto{26.560}{41.849}
\emlineto{26.698}{41.853}
\emmoveto{26.698}{41.843}
\emlineto{26.835}{41.846}
\emmoveto{26.835}{41.836}
\emlineto{26.973}{41.839}
\emmoveto{26.973}{41.829}
\emlineto{27.110}{41.832}
\emmoveto{27.110}{41.822}
\emlineto{27.247}{41.825}
\emmoveto{27.247}{41.815}
\emlineto{27.384}{41.818}
\emmoveto{27.384}{41.808}
\emlineto{27.522}{41.811}
\emmoveto{27.522}{41.801}
\emlineto{27.659}{41.803}
\emmoveto{27.659}{41.793}
\emlineto{27.796}{41.795}
\emmoveto{27.796}{41.785}
\emlineto{27.933}{41.787}
\emmoveto{28.070}{41.769}
\emlineto{28.343}{41.762}
\emmoveto{28.480}{41.744}
\emlineto{28.753}{41.736}
\emmoveto{28.890}{41.718}
\emlineto{29.162}{41.710}
\emmoveto{29.299}{41.691}
\emlineto{29.571}{41.682}
\emmoveto{29.707}{41.663}
\emlineto{29.979}{41.655}
\emmoveto{30.114}{41.635}
\emlineto{30.386}{41.627}
\emmoveto{30.521}{41.607}
\emlineto{30.792}{41.599}
\emmoveto{30.927}{41.579}
\emlineto{31.197}{41.570}
\emmoveto{31.332}{41.551}
\emlineto{31.602}{41.542}
\emmoveto{31.737}{41.523}
\emlineto{32.006}{41.514}
\emmoveto{32.141}{41.494}
\emlineto{32.409}{41.486}
\emmoveto{32.544}{41.466}
\emlineto{32.812}{41.457}
\emmoveto{32.946}{41.438}
\emlineto{33.214}{41.429}
\emmoveto{33.348}{41.410}
\emlineto{33.615}{41.401}
\emmoveto{33.748}{41.381}
\emlineto{34.015}{41.372}
\emmoveto{34.148}{41.352}
\emlineto{34.415}{41.342}
\emmoveto{34.548}{41.322}
\emlineto{34.813}{41.311}
\emmoveto{34.946}{41.290}
\emlineto{35.211}{41.279}
\emmoveto{35.344}{41.258}
\emlineto{35.608}{41.246}
\emmoveto{35.741}{41.224}
\emlineto{36.005}{41.211}
\emmoveto{36.136}{41.189}
\emlineto{36.400}{41.176}
\emmoveto{36.531}{41.154}
\emlineto{36.794}{41.140}
\emmoveto{36.925}{41.118}
\emlineto{37.187}{41.104}
\emmoveto{37.318}{41.081}
\emlineto{37.580}{41.067}
\emmoveto{37.710}{41.045}
\emlineto{37.971}{41.030}
\emmoveto{38.102}{41.008}
\emlineto{38.362}{40.993}
\emmoveto{38.492}{40.971}
\emlineto{38.751}{40.956}
\emmoveto{38.881}{40.934}
\emlineto{39.140}{40.919}
\emmoveto{39.269}{40.896}
\emlineto{39.527}{40.882}
\emmoveto{39.656}{40.859}
\emlineto{39.914}{40.845}
\emmoveto{40.042}{40.822}
\emlineto{40.299}{40.807}
\emmoveto{40.428}{40.785}
\emlineto{40.684}{40.770}
\emmoveto{40.812}{40.747}
\emlineto{41.068}{40.732}
\emmoveto{41.195}{40.709}
\emlineto{41.450}{40.694}
\emmoveto{41.577}{40.671}
\emlineto{41.832}{40.654}
\emmoveto{41.959}{40.631}
\emlineto{42.212}{40.613}
\emmoveto{42.339}{40.590}
\emlineto{42.592}{40.572}
\emmoveto{42.718}{40.548}
\emlineto{42.970}{40.529}
\emmoveto{43.096}{40.505}
\emlineto{43.347}{40.486}
\emmoveto{43.473}{40.461}
\emlineto{43.723}{40.442}
\emmoveto{43.848}{40.417}
\emlineto{44.098}{40.397}
\emmoveto{44.223}{40.372}
\emlineto{44.472}{40.352}
\emmoveto{44.596}{40.327}
\emlineto{44.844}{40.307}
\emmoveto{44.968}{40.282}
\emlineto{45.216}{40.261}
\emmoveto{45.339}{40.236}
\emlineto{45.586}{40.216}
\emmoveto{45.709}{40.191}
\emlineto{45.955}{40.170}
\emmoveto{46.077}{40.145}
\emlineto{46.322}{40.125}
\emmoveto{46.445}{40.100}
\emlineto{46.689}{40.079}
\emmoveto{46.811}{40.054}
\emlineto{47.054}{40.034}
\emmoveto{47.176}{40.009}
\emlineto{47.418}{39.988}
\emmoveto{47.540}{39.963}
\emlineto{47.781}{39.942}
\emmoveto{47.902}{39.916}
\emlineto{48.143}{39.895}
\emmoveto{48.263}{39.869}
\emlineto{48.504}{39.847}
\emmoveto{48.623}{39.821}
\emlineto{48.863}{39.798}
\emmoveto{48.982}{39.772}
\emlineto{49.221}{39.749}
\emmoveto{49.340}{39.722}
\emlineto{49.577}{39.698}
\emmoveto{49.696}{39.671}
\emlineto{49.933}{39.647}
\emmoveto{50.051}{39.620}
\emlineto{50.286}{39.595}
\emmoveto{50.404}{39.568}
\emlineto{50.639}{39.543}
\emmoveto{50.756}{39.515}
\emlineto{50.990}{39.490}
\emmoveto{51.107}{39.463}
\emlineto{51.340}{39.437}
\emmoveto{51.456}{39.410}
\emlineto{51.688}{39.384}
\emmoveto{51.804}{39.356}
\emlineto{52.035}{39.331}
\emmoveto{52.150}{39.303}
\emlineto{52.381}{39.277}
\emmoveto{52.496}{39.250}
\emlineto{52.725}{39.224}
\emmoveto{52.839}{39.196}
\emlineto{53.068}{39.171}
\emmoveto{53.181}{39.143}
\emlineto{53.409}{39.117}
\emmoveto{53.522}{39.089}
\emlineto{53.749}{39.063}
\emmoveto{53.862}{39.035}
\emlineto{54.087}{39.009}
\emmoveto{54.200}{38.981}
\emlineto{54.424}{38.955}
\emmoveto{54.536}{38.926}
\emlineto{54.760}{38.899}
\emmoveto{54.872}{38.871}
\emlineto{55.094}{38.843}
\emmoveto{55.205}{38.814}
\emlineto{55.427}{38.786}
\emmoveto{55.537}{38.757}
\emlineto{55.758}{38.728}
\emmoveto{55.868}{38.699}
\emlineto{56.088}{38.670}
\emmoveto{56.197}{38.640}
\emlineto{56.416}{38.611}
\emmoveto{56.525}{38.581}
\emlineto{56.742}{38.551}
\emmoveto{56.851}{38.521}
\emlineto{57.067}{38.491}
\emmoveto{57.175}{38.461}
\emlineto{57.391}{38.431}
\emmoveto{57.498}{38.401}
\emlineto{57.712}{38.371}
\emmoveto{57.819}{38.341}
\emlineto{58.033}{38.310}
\emmoveto{58.139}{38.280}
\emlineto{58.351}{38.250}
\emmoveto{58.457}{38.219}
\emlineto{58.668}{38.189}
\emmoveto{58.774}{38.159}
\emlineto{58.984}{38.128}
\emmoveto{59.089}{38.098}
\emlineto{59.298}{38.068}
\emmoveto{59.402}{38.037}
\emlineto{59.610}{38.007}
\emmoveto{59.714}{37.976}
\emlineto{59.921}{37.945}
\emmoveto{60.024}{37.915}
\emlineto{60.229}{37.884}
\emmoveto{60.332}{37.853}
\emlineto{60.537}{37.821}
\emmoveto{60.639}{37.790}
\emlineto{60.843}{37.758}
\emmoveto{60.944}{37.727}
\emlineto{61.147}{37.695}
\emmoveto{61.248}{37.663}
\emlineto{61.449}{37.630}
\emmoveto{61.549}{37.598}
\emlineto{61.750}{37.565}
\emmoveto{61.850}{37.533}
\emlineto{62.049}{37.499}
\emmoveto{62.148}{37.467}
\emlineto{62.346}{37.433}
\emmoveto{62.445}{37.401}
\emlineto{62.641}{37.367}
\emmoveto{62.740}{37.335}
\emlineto{62.935}{37.300}
\emmoveto{63.033}{37.268}
\emlineto{63.227}{37.233}
\emmoveto{63.324}{37.201}
\emlineto{63.517}{37.166}
\emmoveto{63.614}{37.134}
\emlineto{63.806}{37.099}
\emmoveto{63.901}{37.066}
\emlineto{64.092}{37.032}
\emmoveto{64.188}{36.999}
\emlineto{64.377}{36.964}
\emmoveto{64.472}{36.932}
\emlineto{64.660}{36.897}
\emmoveto{64.754}{36.865}
\emlineto{64.942}{36.830}
\emmoveto{65.035}{36.797}
\emlineto{65.221}{36.762}
\emmoveto{65.314}{36.729}
\emlineto{65.499}{36.694}
\emmoveto{65.591}{36.661}
\emlineto{65.775}{36.625}
\emmoveto{65.867}{36.592}
\emlineto{66.049}{36.556}
\emmoveto{66.140}{36.523}
\emlineto{66.322}{36.486}
\emmoveto{66.412}{36.453}
\emlineto{66.592}{36.416}
\emmoveto{66.682}{36.382}
\emlineto{66.861}{36.344}
\emmoveto{66.950}{36.311}
\emlineto{67.128}{36.273}
\emmoveto{67.216}{36.239}
\emlineto{67.393}{36.201}
\emmoveto{67.481}{36.167}
\emlineto{67.656}{36.129}
\emmoveto{67.743}{36.094}
\emlineto{67.917}{36.056}
\emmoveto{68.003}{36.022}
\emlineto{68.176}{35.983}
\emmoveto{68.262}{35.949}
\emlineto{68.433}{35.910}
\emmoveto{68.519}{35.876}
\emlineto{68.688}{35.837}
\emmoveto{68.773}{35.803}
\emlineto{68.942}{35.764}
\emmoveto{69.026}{35.730}
\emlineto{69.193}{35.691}
\emmoveto{69.277}{35.657}
\emlineto{69.443}{35.618}
\emmoveto{69.526}{35.584}
\emlineto{69.691}{35.545}
\emmoveto{69.773}{35.511}
\emlineto{69.936}{35.472}
\emmoveto{70.018}{35.437}
\emlineto{70.180}{35.398}
\emmoveto{70.261}{35.363}
\emlineto{70.422}{35.324}
\emmoveto{70.502}{35.289}
\emlineto{70.662}{35.249}
\emmoveto{70.742}{35.214}
\emlineto{70.900}{35.174}
\emmoveto{70.979}{35.139}
\emlineto{71.136}{35.099}
\emmoveto{71.214}{35.065}
\emlineto{71.370}{35.028}
\emmoveto{71.447}{34.995}
\emlineto{71.602}{34.959}
\emmoveto{71.679}{34.926}
\emlineto{71.832}{34.891}
\emmoveto{71.909}{34.858}
\emlineto{72.061}{34.823}
\emmoveto{72.137}{34.790}
\emlineto{72.288}{34.755}
\emmoveto{72.363}{34.722}
\emlineto{72.513}{34.687}
\emmoveto{72.588}{34.655}
\emlineto{72.736}{34.620}
\emmoveto{72.810}{34.587}
\emlineto{72.958}{34.552}
\emmoveto{73.031}{34.519}
\emlineto{73.177}{34.484}
\emmoveto{73.250}{34.451}
\emlineto{73.395}{34.416}
\emmoveto{73.467}{34.383}
\emlineto{73.611}{34.348}
\emmoveto{73.683}{34.315}
\emlineto{73.826}{34.279}
\emmoveto{73.897}{34.247}
\emlineto{74.038}{34.211}
\emmoveto{74.109}{34.178}
\emlineto{74.249}{34.142}
\emmoveto{74.319}{34.109}
\emlineto{74.458}{34.073}
\emmoveto{74.527}{34.039}
\emlineto{74.665}{34.003}
\emmoveto{74.733}{33.969}
\emlineto{74.870}{33.933}
\emmoveto{74.938}{33.900}
\emlineto{75.073}{33.864}
\emmoveto{75.141}{33.832}
\emlineto{75.275}{33.797}
\emmoveto{75.342}{33.765}
\emlineto{75.475}{33.732}
\emmoveto{75.541}{33.700}
\emlineto{75.673}{33.667}
\emmoveto{75.739}{33.636}
\emlineto{75.870}{33.604}
\emmoveto{76.000}{33.551}
\emlineto{76.193}{33.498}
\emmoveto{76.322}{33.445}
\emlineto{76.512}{33.392}
\emmoveto{76.639}{33.340}
\emlineto{76.827}{33.286}
\emmoveto{76.951}{33.234}
\emlineto{77.137}{33.180}
\emmoveto{77.259}{33.127}
\emlineto{77.442}{33.073}
\emmoveto{77.563}{33.020}
\emlineto{77.742}{32.965}
\emmoveto{77.861}{32.912}
\emlineto{78.038}{32.857}
\emmoveto{78.155}{32.803}
\emlineto{78.329}{32.748}
\emmoveto{78.444}{32.695}
\emlineto{78.615}{32.642}
\emmoveto{78.728}{32.591}
\emlineto{78.897}{32.539}
\emmoveto{79.008}{32.489}
\emlineto{79.174}{32.438}
\emmoveto{79.284}{32.388}
\emlineto{79.447}{32.338}
\emmoveto{79.555}{32.288}
\emlineto{79.715}{32.238}
\emmoveto{79.822}{32.189}
\emlineto{79.979}{32.139}
\emmoveto{80.084}{32.088}
\emlineto{80.239}{32.038}
\emmoveto{80.342}{31.988}
\emlineto{80.494}{31.937}
\emmoveto{80.595}{31.887}
\emlineto{80.745}{31.835}
\emmoveto{80.844}{31.784}
\emlineto{80.991}{31.732}
\emmoveto{81.089}{31.682}
\emlineto{81.233}{31.631}
\emmoveto{81.329}{31.580}
\emlineto{81.471}{31.531}
\emmoveto{81.564}{31.481}
\emlineto{81.704}{31.433}
\emmoveto{81.796}{31.385}
\emlineto{81.933}{31.337}
\emmoveto{82.023}{31.289}
\emlineto{82.157}{31.242}
\emmoveto{82.246}{31.194}
\emlineto{82.378}{31.147}
\emmoveto{82.465}{31.099}
\emlineto{82.594}{31.052}
\emmoveto{82.679}{31.004}
\emlineto{82.806}{30.956}
\emmoveto{82.890}{30.908}
\emlineto{83.055}{30.841}
\emmoveto{83.177}{30.772}
\emlineto{83.338}{30.704}
\emmoveto{83.457}{30.636}
\emlineto{83.612}{30.568}
\emmoveto{83.727}{30.501}
\emlineto{83.879}{30.436}
\emmoveto{83.990}{30.370}
\emlineto{84.137}{30.306}
\emmoveto{84.245}{30.241}
\emlineto{84.387}{30.178}
\emmoveto{84.492}{30.114}
\emlineto{84.629}{30.051}
\emmoveto{84.731}{29.986}
\emlineto{84.864}{29.922}
\emmoveto{84.962}{29.857}
\emlineto{85.091}{29.792}
\emmoveto{85.185}{29.726}
\emlineto{85.309}{29.661}
\emmoveto{85.400}{29.595}
\emlineto{85.520}{29.531}
\emmoveto{85.608}{29.467}
\emlineto{85.751}{29.386}
\emmoveto{85.863}{29.305}
\emlineto{85.999}{29.226}
\emmoveto{86.105}{29.146}
\emlineto{86.234}{29.067}
\emmoveto{86.335}{28.987}
\emlineto{86.457}{28.907}
\emmoveto{86.552}{28.825}
\emlineto{86.667}{28.745}
\emmoveto{86.756}{28.662}
\emlineto{86.864}{28.582}
\emmoveto{86.968}{28.483}
\emlineto{87.088}{28.388}
\emmoveto{87.183}{28.291}
\emlineto{87.293}{28.197}
\emmoveto{87.380}{28.101}
\emlineto{87.480}{28.006}
\emmoveto{87.559}{27.909}
\emlineto{87.648}{27.813}
\emmoveto{87.732}{27.696}
\emlineto{87.823}{27.583}
\emmoveto{87.895}{27.468}
\emlineto{87.972}{27.358}
\emmoveto{88.032}{27.246}
\emlineto{88.104}{27.119}
\emmoveto{88.159}{26.988}
\emlineto{88.212}{26.858}
\emmoveto{88.251}{26.725}
\emlineto{88.286}{26.596}
\emmoveto{88.309}{26.466}
\emlineto{88.326}{26.339}
\emmoveto{88.334}{26.192}
\emlineto{88.332}{26.064}
\emmoveto{88.322}{25.932}
\emlineto{88.301}{25.802}
\emmoveto{88.275}{25.669}
\emlineto{88.236}{25.541}
\emmoveto{88.194}{25.410}
\emlineto{88.137}{25.282}
\emmoveto{88.080}{25.150}
\emlineto{88.014}{25.038}
\emmoveto{87.952}{24.921}
\emlineto{87.873}{24.807}
\emmoveto{87.799}{24.691}
\emlineto{87.705}{24.577}
\emmoveto{87.619}{24.462}
\emlineto{87.527}{24.366}
\emmoveto{87.446}{24.268}
\emlineto{87.344}{24.171}
\emmoveto{87.255}{24.072}
\emlineto{87.143}{23.973}
\emmoveto{87.045}{23.872}
\emlineto{86.922}{23.773}
\emmoveto{86.837}{23.691}
\emlineto{86.727}{23.610}
\emmoveto{86.636}{23.528}
\emlineto{86.519}{23.447}
\emmoveto{86.422}{23.365}
\emlineto{86.298}{23.284}
\emmoveto{86.196}{23.200}
\emlineto{86.064}{23.118}
\emmoveto{85.956}{23.033}
\emlineto{85.818}{22.950}
\emmoveto{85.704}{22.865}
\emlineto{85.587}{22.799}
\emmoveto{85.498}{22.733}
\emlineto{85.376}{22.668}
\emmoveto{85.284}{22.602}
\emlineto{85.158}{22.536}
\emmoveto{85.062}{22.470}
\emlineto{84.931}{22.404}
\emmoveto{84.832}{22.338}
\emlineto{84.696}{22.272}
\emmoveto{84.593}{22.204}
\emlineto{84.454}{22.138}
\emmoveto{84.347}{22.070}
\emlineto{84.202}{22.002}
\emmoveto{84.092}{21.934}
\emlineto{83.943}{21.865}
\emmoveto{83.829}{21.797}
\emlineto{83.675}{21.728}
\emmoveto{83.557}{21.659}
\emlineto{83.398}{21.590}
\emmoveto{83.277}{21.521}
\emlineto{83.114}{21.452}
\emmoveto{83.031}{21.403}
\emlineto{82.905}{21.353}
\emmoveto{82.820}{21.304}
\emlineto{82.692}{21.254}
\emmoveto{82.605}{21.204}
\emlineto{82.475}{21.154}
\emmoveto{82.386}{21.104}
\emlineto{82.253}{21.053}
\emmoveto{82.163}{21.002}
\emlineto{82.027}{20.951}
\emmoveto{81.935}{20.900}
\emlineto{81.796}{20.849}
\emmoveto{81.702}{20.798}
\emlineto{81.561}{20.746}
\emmoveto{81.465}{20.694}
\emlineto{81.321}{20.642}
\emmoveto{81.224}{20.591}
\emlineto{81.077}{20.539}
\emmoveto{80.978}{20.487}
\emlineto{80.828}{20.435}
\emmoveto{80.727}{20.383}
\emlineto{80.574}{20.330}
\emmoveto{80.472}{20.278}
\emlineto{80.316}{20.225}
\emmoveto{80.212}{20.173}
\emlineto{80.054}{20.120}
\emmoveto{79.947}{20.068}
\emlineto{79.787}{20.014}
\emmoveto{79.678}{19.961}
\emlineto{79.515}{19.907}
\emmoveto{79.405}{19.854}
\emlineto{79.238}{19.799}
\emmoveto{79.126}{19.745}
\emlineto{78.957}{19.690}
\emmoveto{78.843}{19.636}
\emlineto{78.671}{19.581}
\emmoveto{78.555}{19.527}
\emlineto{78.380}{19.471}
\emmoveto{78.262}{19.417}
\emlineto{78.084}{19.361}
\emmoveto{77.964}{19.306}
\emlineto{77.783}{19.250}
\emmoveto{77.662}{19.195}
\emlineto{77.478}{19.139}
\emmoveto{77.354}{19.084}
\emlineto{77.168}{19.027}
\emmoveto{77.042}{18.972}
\emlineto{76.852}{18.914}
\emmoveto{76.725}{18.859}
\emlineto{76.532}{18.801}
\emmoveto{76.467}{18.768}
\emlineto{76.337}{18.732}
\emmoveto{76.272}{18.699}
\emlineto{76.141}{18.663}
\emmoveto{76.075}{18.630}
\emlineto{75.943}{18.594}
\emmoveto{75.877}{18.561}
\emlineto{75.743}{18.525}
\emmoveto{75.676}{18.491}
\emlineto{75.541}{18.455}
\emmoveto{75.474}{18.422}
\emlineto{75.338}{18.385}
\emmoveto{75.269}{18.351}
\emlineto{75.132}{18.314}
\emmoveto{75.063}{18.281}
\emlineto{74.925}{18.244}
\emmoveto{74.855}{18.210}
\emlineto{74.716}{18.173}
\emmoveto{74.645}{18.139}
\emlineto{74.504}{18.102}
\emmoveto{74.434}{18.068}
\emlineto{74.292}{18.031}
\emmoveto{74.220}{17.997}
\emlineto{74.077}{17.959}
\emmoveto{74.005}{17.926}
\emlineto{73.860}{17.888}
\emmoveto{73.787}{17.854}
\emlineto{73.641}{17.816}
\emmoveto{73.568}{17.782}
\emlineto{73.421}{17.744}
\emmoveto{73.347}{17.710}
\emlineto{73.199}{17.671}
\emmoveto{73.124}{17.637}
\emlineto{72.974}{17.599}
\emmoveto{72.899}{17.564}
\emlineto{72.748}{17.525}
\emmoveto{72.672}{17.491}
\emlineto{72.520}{17.452}
\emmoveto{72.443}{17.417}
\emlineto{72.290}{17.378}
\emmoveto{72.213}{17.343}
\emlineto{72.058}{17.304}
\emmoveto{71.980}{17.269}
\emlineto{71.824}{17.229}
\emmoveto{71.745}{17.194}
\emlineto{71.588}{17.154}
\emmoveto{71.509}{17.119}
\emlineto{71.350}{17.079}
\emmoveto{71.270}{17.043}
\emlineto{71.110}{17.003}
\emmoveto{71.030}{16.968}
\emlineto{70.868}{16.928}
\emmoveto{70.787}{16.894}
\emlineto{70.624}{16.857}
\emmoveto{70.543}{16.824}
\emlineto{70.379}{16.788}
\emmoveto{70.296}{16.756}
\emlineto{70.131}{16.721}
\emmoveto{70.048}{16.689}
\emlineto{69.882}{16.654}
\emmoveto{69.799}{16.622}
\emlineto{69.631}{16.588}
\emmoveto{69.547}{16.556}
\emlineto{69.378}{16.521}
\emmoveto{69.294}{16.489}
\emlineto{69.124}{16.455}
\emmoveto{69.039}{16.423}
\emlineto{68.868}{16.388}
\emmoveto{68.782}{16.356}
\emlineto{68.610}{16.321}
\emmoveto{68.523}{16.289}
\emlineto{68.350}{16.254}
\emmoveto{68.263}{16.221}
\emlineto{68.089}{16.186}
\emmoveto{68.001}{16.154}
\emlineto{67.825}{16.118}
\emmoveto{67.737}{16.086}
\emlineto{67.560}{16.050}
\emmoveto{67.472}{16.017}
\emlineto{67.293}{15.982}
\emmoveto{67.204}{15.949}
\emlineto{67.025}{15.913}
\emmoveto{66.935}{15.880}
\emlineto{66.754}{15.844}
\emmoveto{66.664}{15.811}
\emlineto{66.482}{15.775}
\emmoveto{66.391}{15.742}
\emlineto{66.208}{15.707}
\emmoveto{66.116}{15.675}
\emlineto{65.932}{15.642}
\emmoveto{65.840}{15.611}
\emlineto{65.655}{15.579}
\emmoveto{65.562}{15.548}
\emlineto{65.375}{15.517}
\emmoveto{65.282}{15.486}
\emlineto{65.095}{15.456}
\emmoveto{65.001}{15.425}
\emlineto{64.812}{15.395}
\emmoveto{64.718}{15.365}
\emlineto{64.528}{15.335}
\emmoveto{64.433}{15.304}
\emlineto{64.243}{15.274}
\emmoveto{64.147}{15.244}
\emlineto{63.955}{15.214}
\emmoveto{63.859}{15.184}
\emlineto{63.667}{15.153}
\emmoveto{63.570}{15.123}
\emlineto{63.376}{15.093}
\emmoveto{63.279}{15.063}
\emlineto{63.084}{15.032}
\emmoveto{62.987}{15.002}
\emlineto{62.791}{14.971}
\emmoveto{62.693}{14.940}
\emlineto{62.496}{14.909}
\emmoveto{62.397}{14.879}
\emlineto{62.199}{14.848}
\emmoveto{62.100}{14.817}
\emlineto{61.901}{14.786}
\emmoveto{61.801}{14.755}
\emlineto{61.601}{14.724}
\emmoveto{61.500}{14.693}
\emlineto{61.299}{14.662}
\emmoveto{61.198}{14.632}
\emlineto{60.996}{14.602}
\emmoveto{60.894}{14.572}
\emlineto{60.691}{14.543}
\emmoveto{60.589}{14.514}
\emlineto{60.384}{14.486}
\emmoveto{60.282}{14.458}
\emlineto{60.077}{14.430}
\emmoveto{59.974}{14.402}
\emlineto{59.767}{14.376}
\emmoveto{59.664}{14.348}
\emlineto{59.457}{14.322}
\emmoveto{59.353}{14.294}
\emlineto{59.144}{14.268}
\emmoveto{59.040}{14.240}
\emlineto{58.831}{14.214}
\emmoveto{58.726}{14.187}
\emlineto{58.516}{14.161}
\emmoveto{58.411}{14.133}
\emlineto{58.200}{14.108}
\emmoveto{58.094}{14.080}
\emlineto{57.882}{14.054}
\emmoveto{57.776}{14.026}
\emlineto{57.563}{14.000}
\emmoveto{57.456}{13.972}
\emlineto{57.242}{13.946}
\emmoveto{57.135}{13.918}
\emlineto{56.920}{13.892}
\emmoveto{56.813}{13.864}
\emlineto{56.597}{13.838}
\emmoveto{56.489}{13.810}
\emlineto{56.272}{13.784}
\emmoveto{56.163}{13.755}
\emlineto{55.946}{13.729}
\emmoveto{55.837}{13.701}
\emlineto{55.618}{13.676}
\emmoveto{55.508}{13.648}
\emlineto{55.289}{13.623}
\emmoveto{55.179}{13.596}
\emlineto{54.958}{13.572}
\emmoveto{54.848}{13.545}
\emlineto{54.627}{13.522}
\emmoveto{54.516}{13.495}
\emlineto{54.294}{13.473}
\emmoveto{54.182}{13.447}
\emlineto{53.959}{13.425}
\emmoveto{53.847}{13.399}
\emlineto{53.624}{13.378}
\emmoveto{53.511}{13.353}
\emlineto{53.287}{13.332}
\emmoveto{53.174}{13.306}
\emlineto{52.949}{13.286}
\emmoveto{52.836}{13.260}
\emlineto{52.609}{13.240}
\emmoveto{52.496}{13.214}
\emlineto{52.269}{13.194}
\emmoveto{52.155}{13.168}
\emlineto{51.927}{13.148}
\emmoveto{51.813}{13.123}
\emlineto{51.584}{13.102}
\emmoveto{51.470}{13.077}
\emlineto{51.240}{13.056}
\emmoveto{51.125}{13.030}
\emlineto{50.895}{13.010}
\emmoveto{50.780}{12.984}
\emlineto{50.548}{12.963}
\emmoveto{50.433}{12.938}
\emlineto{50.201}{12.917}
\emmoveto{50.084}{12.891}
\emlineto{49.852}{12.871}
\emmoveto{49.735}{12.846}
\emlineto{49.502}{12.825}
\emmoveto{49.385}{12.800}
\emlineto{49.150}{12.781}
\emmoveto{49.033}{12.756}
\emlineto{48.798}{12.737}
\emmoveto{48.680}{12.713}
\emlineto{48.444}{12.695}
\emmoveto{48.326}{12.671}
\emlineto{48.089}{12.654}
\emmoveto{47.971}{12.630}
\emlineto{47.733}{12.614}
\emmoveto{47.615}{12.590}
\emlineto{47.377}{12.574}
\emmoveto{47.257}{12.551}
\emlineto{47.019}{12.535}
\emmoveto{46.899}{12.513}
\emlineto{46.660}{12.497}
\emmoveto{46.540}{12.474}
\emlineto{46.300}{12.459}
\emmoveto{46.180}{12.436}
\emlineto{45.939}{12.421}
\emmoveto{45.819}{12.399}
\emlineto{45.577}{12.383}
\emmoveto{45.456}{12.361}
\emlineto{45.214}{12.346}
\emmoveto{45.093}{12.323}
\emlineto{44.850}{12.308}
\emmoveto{44.729}{12.285}
\emlineto{44.486}{12.270}
\emmoveto{44.364}{12.247}
\emlineto{44.120}{12.232}
\emmoveto{43.997}{12.209}
\emlineto{43.753}{12.194}
\emmoveto{43.630}{12.171}
\emlineto{43.385}{12.156}
\emmoveto{43.262}{12.134}
\emlineto{43.016}{12.119}
\emmoveto{42.893}{12.097}
\emlineto{42.646}{12.083}
\emmoveto{42.523}{12.061}
\emlineto{42.276}{12.048}
\emmoveto{42.152}{12.027}
\emlineto{41.904}{12.014}
\emmoveto{41.780}{11.993}
\emlineto{41.531}{11.981}
\emmoveto{41.407}{11.960}
\emlineto{41.158}{11.949}
\emmoveto{41.033}{11.928}
\emlineto{40.784}{11.917}
\emmoveto{40.659}{11.897}
\emlineto{40.409}{11.887}
\emmoveto{40.283}{11.867}
\emlineto{40.033}{11.857}
\emmoveto{39.907}{11.837}
\emlineto{39.656}{11.827}
\emmoveto{39.530}{11.807}
\emlineto{39.279}{11.797}
\emmoveto{39.153}{11.778}
\emlineto{38.901}{11.768}
\emmoveto{38.774}{11.748}
\emlineto{38.522}{11.739}
\emmoveto{38.395}{11.719}
\emlineto{38.142}{11.710}
\emmoveto{38.015}{11.690}
\emlineto{37.761}{11.680}
\emmoveto{37.634}{11.661}
\emlineto{37.380}{11.651}
\emmoveto{37.253}{11.631}
\emlineto{36.998}{11.622}
\emmoveto{36.871}{11.602}
\emlineto{36.615}{11.593}
\emmoveto{36.488}{11.574}
\emlineto{36.232}{11.565}
\emmoveto{36.104}{11.546}
\emlineto{35.848}{11.538}
\emmoveto{35.719}{11.519}
\emlineto{35.463}{11.511}
\emmoveto{35.334}{11.492}
\emlineto{35.077}{11.485}
\emmoveto{34.948}{11.467}
\emlineto{34.691}{11.461}
\emmoveto{34.562}{11.443}
\emlineto{34.304}{11.437}
\emmoveto{34.175}{11.419}
\emlineto{33.916}{11.414}
\emmoveto{33.787}{11.397}
\emlineto{33.528}{11.392}
\emmoveto{33.398}{11.375}
\emlineto{33.139}{11.370}
\emmoveto{33.009}{11.353}
\emlineto{32.750}{11.349}
\emmoveto{32.620}{11.332}
\emlineto{32.360}{11.328}
\emmoveto{32.230}{11.311}
\emlineto{31.970}{11.308}
\emmoveto{31.839}{11.291}
\emlineto{31.579}{11.287}
\emmoveto{31.448}{11.270}
\emlineto{31.187}{11.267}
\emmoveto{31.057}{11.250}
\emlineto{30.795}{11.246}
\emmoveto{30.664}{11.230}
\emlineto{30.403}{11.226}
\emmoveto{30.272}{11.209}
\emlineto{30.010}{11.206}
\emmoveto{29.878}{11.189}
\emlineto{29.616}{11.186}
\emmoveto{29.485}{11.170}
\emlineto{29.222}{11.167}
\emmoveto{29.090}{11.151}
\emlineto{28.827}{11.149}
\emmoveto{28.696}{11.133}
\emlineto{28.432}{11.131}
\emmoveto{28.300}{11.115}
\emlineto{28.036}{11.114}
\emmoveto{27.905}{11.099}
\emlineto{27.641}{11.098}
\emmoveto{27.508}{11.083}
\emlineto{27.244}{11.083}
\emmoveto{27.112}{11.068}
\emlineto{26.847}{11.069}
\emmoveto{26.715}{11.054}
\emlineto{26.450}{11.056}
\emmoveto{26.318}{11.041}
\emlineto{26.053}{11.043}
\emmoveto{25.920}{11.028}
\emlineto{25.655}{11.030}
\emmoveto{25.522}{11.016}
\emlineto{25.257}{11.018}
\emmoveto{25.124}{11.004}
\emlineto{24.858}{11.006}
\emmoveto{24.726}{10.993}
\emlineto{24.460}{10.995}
\emmoveto{24.327}{10.981}
\emlineto{24.061}{10.983}
\emmoveto{23.927}{10.970}
\emlineto{23.661}{10.972}
\emmoveto{23.528}{10.958}
\emlineto{23.262}{10.961}
\emmoveto{23.128}{10.947}
\emlineto{22.862}{10.950}
\emmoveto{22.728}{10.936}
\emlineto{22.461}{10.939}
\emmoveto{22.328}{10.926}
\emlineto{22.061}{10.929}
\emmoveto{21.927}{10.916}
\emlineto{21.660}{10.920}
\emmoveto{21.527}{10.907}
\emlineto{21.259}{10.911}
\emmoveto{21.125}{10.899}
\emlineto{20.858}{10.903}
\emmoveto{20.724}{10.891}
\emlineto{20.457}{10.896}
\emmoveto{20.323}{10.884}
\emlineto{20.055}{10.890}
\emmoveto{19.921}{10.878}
\emlineto{19.653}{10.885}
\emmoveto{19.519}{10.873}
\emlineto{19.251}{10.880}
\emmoveto{19.117}{10.869}
\emlineto{18.849}{10.876}
\emmoveto{18.715}{10.865}
\emlineto{18.447}{10.873}
\emmoveto{18.313}{10.862}
\emlineto{18.045}{10.870}
\emmoveto{17.911}{10.859}
\emlineto{17.643}{10.867}
\emmoveto{17.509}{10.856}
\emlineto{17.241}{10.864}
\emmoveto{17.107}{10.853}
\emlineto{16.838}{10.862}
\emmoveto{16.704}{10.851}
\emlineto{16.436}{10.860}
\emmoveto{16.302}{10.849}
\emlineto{16.033}{10.858}
\emmoveto{15.899}{10.847}
\emlineto{15.631}{10.856}
\emmoveto{15.497}{10.845}
\emlineto{15.228}{10.854}
\emmoveto{15.094}{10.844}
\emlineto{14.826}{10.853}
\emmoveto{14.692}{10.843}
\emlineto{14.423}{10.843}
\emlineto{14.423}{10.853}
\emmoveto{14.289}{10.843}
\emlineto{14.020}{10.854}
\emmoveto{13.886}{10.844}
\emlineto{13.618}{10.855}
\emmoveto{13.484}{10.845}
\emlineto{13.215}{10.857}
\emmoveto{13.081}{10.848}
\emlineto{12.813}{10.859}
\emmoveto{12.679}{10.851}
\emlineto{12.410}{10.863}
\emmoveto{12.276}{10.854}
\emlineto{12.008}{10.867}
\emshow{36.780}{17.700}{k=10,v=0.12}
\emmoveto{12.000}{66.538}
\emlineto{12.354}{66.548}
\emmoveto{12.354}{66.538}
\emlineto{12.708}{66.546}
\emmoveto{12.708}{66.536}
\emlineto{13.062}{66.542}
\emmoveto{13.062}{66.532}
\emlineto{13.416}{66.538}
\emmoveto{13.416}{66.528}
\emlineto{13.770}{66.534}
\emmoveto{13.770}{66.524}
\emlineto{14.124}{66.528}
\emmoveto{14.124}{66.518}
\emlineto{14.477}{66.522}
\emmoveto{14.477}{66.512}
\emlineto{14.831}{66.516}
\emmoveto{14.831}{66.506}
\emlineto{15.185}{66.509}
\emmoveto{15.185}{66.499}
\emlineto{15.538}{66.503}
\emmoveto{15.538}{66.493}
\emlineto{15.892}{66.496}
\emmoveto{15.892}{66.486}
\emlineto{16.246}{66.488}
\emmoveto{16.246}{66.478}
\emlineto{16.599}{66.481}
\emmoveto{16.599}{66.471}
\emlineto{16.952}{66.473}
\emmoveto{16.952}{66.463}
\emlineto{17.306}{66.466}
\emmoveto{17.306}{66.456}
\emlineto{17.659}{66.458}
\emmoveto{17.659}{66.448}
\emlineto{18.012}{66.450}
\emmoveto{18.012}{66.440}
\emlineto{18.365}{66.442}
\emmoveto{18.365}{66.432}
\emlineto{18.718}{66.435}
\emmoveto{18.718}{66.425}
\emlineto{19.071}{66.427}
\emmoveto{19.071}{66.417}
\emlineto{19.424}{66.419}
\emmoveto{19.424}{66.409}
\emlineto{19.777}{66.411}
\emmoveto{19.777}{66.401}
\emlineto{20.130}{66.403}
\emmoveto{20.130}{66.393}
\emlineto{20.482}{66.395}
\emmoveto{20.482}{66.385}
\emlineto{20.835}{66.387}
\emmoveto{20.835}{66.377}
\emlineto{21.188}{66.379}
\emmoveto{21.188}{66.369}
\emlineto{21.540}{66.371}
\emmoveto{21.540}{66.361}
\emlineto{21.892}{66.363}
\emmoveto{21.892}{66.353}
\emlineto{22.245}{66.355}
\emmoveto{22.245}{66.345}
\emlineto{22.597}{66.347}
\emmoveto{22.597}{66.337}
\emlineto{22.949}{66.339}
\emmoveto{22.949}{66.329}
\emlineto{23.301}{66.331}
\emmoveto{23.301}{66.321}
\emlineto{23.653}{66.323}
\emmoveto{23.653}{66.313}
\emlineto{24.005}{66.315}
\emmoveto{24.005}{66.305}
\emlineto{24.357}{66.307}
\emmoveto{24.357}{66.297}
\emlineto{24.709}{66.299}
\emmoveto{24.709}{66.289}
\emlineto{25.061}{66.291}
\emmoveto{25.061}{66.281}
\emlineto{25.413}{66.283}
\emmoveto{25.413}{66.273}
\emlineto{25.764}{66.275}
\emmoveto{25.764}{66.265}
\emlineto{26.116}{66.267}
\emmoveto{26.116}{66.257}
\emlineto{26.467}{66.258}
\emmoveto{26.467}{66.248}
\emlineto{26.819}{66.250}
\emmoveto{26.819}{66.240}
\emlineto{27.170}{66.242}
\emmoveto{27.170}{66.232}
\emlineto{27.521}{66.234}
\emmoveto{27.521}{66.224}
\emlineto{27.873}{66.226}
\emmoveto{27.873}{66.216}
\emlineto{28.224}{66.218}
\emmoveto{28.224}{66.208}
\emlineto{28.575}{66.210}
\emmoveto{28.575}{66.200}
\emlineto{28.926}{66.202}
\emmoveto{28.926}{66.192}
\emlineto{29.277}{66.194}
\emmoveto{29.277}{66.184}
\emlineto{29.628}{66.186}
\emmoveto{29.628}{66.176}
\emlineto{29.978}{66.178}
\emmoveto{29.978}{66.168}
\emlineto{30.329}{66.170}
\emmoveto{30.329}{66.160}
\emlineto{30.680}{66.161}
\emmoveto{30.680}{66.151}
\emlineto{31.030}{66.152}
\emmoveto{31.030}{66.142}
\emlineto{31.381}{66.142}
\emmoveto{31.381}{66.132}
\emlineto{31.731}{66.132}
\emmoveto{31.731}{66.122}
\emlineto{32.081}{66.121}
\emmoveto{32.081}{66.111}
\emlineto{32.432}{66.109}
\emmoveto{32.432}{66.099}
\emlineto{32.782}{66.097}
\emmoveto{32.782}{66.087}
\emlineto{33.132}{66.084}
\emmoveto{33.132}{66.074}
\emlineto{33.482}{66.071}
\emmoveto{33.482}{66.061}
\emlineto{33.831}{66.058}
\emmoveto{33.831}{66.048}
\emlineto{34.181}{66.044}
\emmoveto{34.181}{66.034}
\emlineto{34.531}{66.030}
\emmoveto{34.531}{66.020}
\emlineto{34.880}{66.016}
\emmoveto{34.880}{66.006}
\emlineto{35.229}{66.001}
\emmoveto{35.229}{65.991}
\emlineto{35.578}{65.987}
\emmoveto{35.578}{65.977}
\emlineto{35.927}{65.972}
\emmoveto{35.927}{65.962}
\emlineto{36.276}{65.957}
\emmoveto{36.276}{65.947}
\emlineto{36.625}{65.941}
\emmoveto{36.625}{65.931}
\emlineto{36.974}{65.926}
\emmoveto{36.974}{65.916}
\emlineto{37.322}{65.911}
\emmoveto{37.322}{65.901}
\emlineto{37.670}{65.895}
\emmoveto{37.670}{65.885}
\emlineto{38.019}{65.879}
\emmoveto{38.019}{65.869}
\emlineto{38.367}{65.864}
\emmoveto{38.367}{65.854}
\emlineto{38.715}{65.848}
\emmoveto{38.715}{65.838}
\emlineto{39.062}{65.832}
\emmoveto{39.062}{65.822}
\emlineto{39.410}{65.816}
\emmoveto{39.410}{65.806}
\emlineto{39.758}{65.801}
\emmoveto{39.758}{65.791}
\emlineto{40.105}{65.785}
\emmoveto{40.105}{65.775}
\emlineto{40.452}{65.769}
\emmoveto{40.452}{65.759}
\emlineto{40.799}{65.753}
\emmoveto{40.799}{65.743}
\emlineto{41.146}{65.737}
\emmoveto{41.146}{65.727}
\emlineto{41.493}{65.721}
\emmoveto{41.493}{65.711}
\emlineto{41.840}{65.705}
\emmoveto{41.840}{65.695}
\emlineto{42.186}{65.689}
\emmoveto{42.186}{65.679}
\emlineto{42.533}{65.674}
\emmoveto{42.533}{65.664}
\emlineto{42.879}{65.658}
\emmoveto{42.879}{65.648}
\emlineto{43.225}{65.642}
\emmoveto{43.225}{65.632}
\emlineto{43.571}{65.626}
\emmoveto{43.571}{65.616}
\emlineto{43.917}{65.610}
\emmoveto{43.917}{65.600}
\emlineto{44.263}{65.594}
\emmoveto{44.263}{65.584}
\emlineto{44.608}{65.578}
\emmoveto{44.608}{65.568}
\emlineto{44.954}{65.562}
\emmoveto{44.954}{65.552}
\emlineto{45.299}{65.546}
\emmoveto{45.299}{65.536}
\emlineto{45.644}{65.530}
\emmoveto{45.644}{65.520}
\emlineto{45.989}{65.514}
\emmoveto{45.989}{65.504}
\emlineto{46.334}{65.498}
\emmoveto{46.334}{65.488}
\emlineto{46.679}{65.483}
\emmoveto{46.679}{65.473}
\emlineto{47.023}{65.467}
\emmoveto{47.023}{65.457}
\emlineto{47.368}{65.451}
\emmoveto{47.368}{65.441}
\emlineto{47.712}{65.435}
\emmoveto{47.712}{65.425}
\emlineto{48.056}{65.419}
\emmoveto{48.056}{65.409}
\emlineto{48.400}{65.403}
\emmoveto{48.400}{65.393}
\emlineto{48.744}{65.386}
\emmoveto{48.744}{65.376}
\emlineto{49.088}{65.370}
\emmoveto{49.088}{65.360}
\emlineto{49.431}{65.353}
\emmoveto{49.431}{65.343}
\emlineto{49.775}{65.335}
\emmoveto{49.775}{65.325}
\emlineto{50.118}{65.317}
\emmoveto{50.118}{65.307}
\emlineto{50.461}{65.299}
\emmoveto{50.461}{65.289}
\emlineto{50.804}{65.281}
\emmoveto{50.804}{65.271}
\emlineto{51.147}{65.261}
\emmoveto{51.147}{65.251}
\emlineto{51.490}{65.242}
\emmoveto{51.490}{65.232}
\emlineto{51.832}{65.222}
\emmoveto{51.832}{65.212}
\emlineto{52.174}{65.202}
\emmoveto{52.174}{65.192}
\emlineto{52.516}{65.181}
\emmoveto{52.516}{65.171}
\emlineto{52.858}{65.160}
\emmoveto{52.858}{65.150}
\emlineto{53.200}{65.139}
\emmoveto{53.200}{65.129}
\emlineto{53.542}{65.117}
\emmoveto{53.542}{65.107}
\emlineto{53.883}{65.095}
\emmoveto{53.883}{65.085}
\emlineto{54.224}{65.073}
\emmoveto{54.224}{65.063}
\emlineto{54.565}{65.051}
\emmoveto{54.565}{65.041}
\emlineto{54.906}{65.029}
\emmoveto{54.906}{65.019}
\emlineto{55.247}{65.006}
\emmoveto{55.247}{64.996}
\emlineto{55.587}{64.983}
\emmoveto{55.587}{64.973}
\emlineto{55.927}{64.960}
\emmoveto{55.927}{64.950}
\emlineto{56.267}{64.937}
\emmoveto{56.267}{64.927}
\emlineto{56.607}{64.914}
\emmoveto{56.607}{64.904}
\emlineto{56.946}{64.891}
\emmoveto{56.946}{64.881}
\emlineto{57.286}{64.868}
\emmoveto{57.286}{64.858}
\emlineto{57.625}{64.845}
\emmoveto{57.625}{64.835}
\emlineto{57.964}{64.821}
\emmoveto{57.964}{64.811}
\emlineto{58.303}{64.798}
\emmoveto{58.303}{64.788}
\emlineto{58.641}{64.774}
\emmoveto{58.641}{64.764}
\emlineto{58.980}{64.751}
\emmoveto{58.980}{64.741}
\emlineto{59.318}{64.727}
\emmoveto{59.318}{64.717}
\emlineto{59.656}{64.704}
\emmoveto{59.656}{64.694}
\emlineto{59.993}{64.680}
\emmoveto{59.993}{64.670}
\emlineto{60.331}{64.657}
\emmoveto{60.331}{64.647}
\emlineto{60.668}{64.633}
\emmoveto{60.668}{64.623}
\emlineto{61.005}{64.610}
\emmoveto{61.005}{64.600}
\emlineto{61.342}{64.586}
\emmoveto{61.342}{64.576}
\emlineto{61.679}{64.562}
\emmoveto{61.679}{64.552}
\emlineto{62.015}{64.539}
\emmoveto{62.015}{64.529}
\emlineto{62.352}{64.515}
\emmoveto{62.352}{64.505}
\emlineto{62.688}{64.492}
\emmoveto{62.688}{64.482}
\emlineto{63.024}{64.468}
\emmoveto{63.024}{64.458}
\emlineto{63.359}{64.445}
\emmoveto{63.359}{64.435}
\emlineto{63.695}{64.421}
\emmoveto{63.695}{64.411}
\emlineto{64.030}{64.397}
\emmoveto{64.030}{64.387}
\emlineto{64.365}{64.374}
\emmoveto{64.365}{64.364}
\emlineto{64.700}{64.350}
\emmoveto{64.700}{64.340}
\emlineto{65.035}{64.327}
\emmoveto{65.035}{64.317}
\emlineto{65.369}{64.303}
\emmoveto{65.369}{64.293}
\emlineto{65.703}{64.279}
\emmoveto{65.703}{64.269}
\emlineto{66.037}{64.256}
\emmoveto{66.037}{64.246}
\emlineto{66.371}{64.232}
\emmoveto{66.371}{64.222}
\emlineto{66.705}{64.208}
\emmoveto{66.705}{64.198}
\emlineto{67.038}{64.184}
\emmoveto{67.038}{64.174}
\emlineto{67.371}{64.159}
\emmoveto{67.371}{64.149}
\emlineto{67.704}{64.134}
\emmoveto{67.704}{64.124}
\emlineto{68.037}{64.109}
\emmoveto{68.037}{64.099}
\emlineto{68.369}{64.084}
\emmoveto{68.369}{64.074}
\emlineto{68.702}{64.058}
\emmoveto{68.702}{64.048}
\emlineto{69.034}{64.032}
\emmoveto{69.034}{64.022}
\emlineto{69.365}{64.005}
\emmoveto{69.365}{63.995}
\emlineto{69.697}{63.979}
\emmoveto{69.697}{63.969}
\emlineto{70.028}{63.951}
\emmoveto{70.028}{63.941}
\emlineto{70.359}{63.924}
\emmoveto{70.359}{63.914}
\emlineto{70.690}{63.896}
\emmoveto{70.690}{63.886}
\emlineto{71.021}{63.868}
\emmoveto{71.021}{63.858}
\emlineto{71.351}{63.841}
\emmoveto{71.351}{63.831}
\emlineto{71.682}{63.817}
\emmoveto{71.682}{63.807}
\emlineto{72.011}{63.794}
\emmoveto{72.011}{63.784}
\emlineto{72.341}{63.772}
\emmoveto{72.341}{63.762}
\emlineto{72.671}{63.752}
\emmoveto{72.671}{63.742}
\emlineto{73.000}{63.732}
\emmoveto{73.000}{63.722}
\emlineto{73.329}{63.714}
\emmoveto{73.329}{63.704}
\emlineto{73.659}{63.695}
\emmoveto{73.659}{63.685}
\emlineto{73.987}{63.678}
\emmoveto{73.987}{63.668}
\emlineto{74.316}{63.660}
\emmoveto{74.316}{63.650}
\emlineto{74.645}{63.643}
\emmoveto{74.645}{63.633}
\emlineto{74.973}{63.626}
\emmoveto{74.973}{63.616}
\emlineto{75.302}{63.610}
\emmoveto{75.302}{63.600}
\emlineto{75.630}{63.593}
\emmoveto{75.630}{63.583}
\emlineto{75.958}{63.577}
\emmoveto{75.958}{63.567}
\emlineto{76.286}{63.560}
\emmoveto{76.286}{63.550}
\emlineto{76.613}{63.544}
\emmoveto{76.613}{63.534}
\emlineto{76.941}{63.528}
\emmoveto{76.941}{63.518}
\emlineto{77.268}{63.511}
\emmoveto{77.268}{63.501}
\emlineto{77.596}{63.495}
\emmoveto{77.596}{63.485}
\emlineto{77.923}{63.479}
\emmoveto{77.923}{63.469}
\emlineto{78.250}{63.463}
\emmoveto{78.250}{63.453}
\emlineto{78.577}{63.446}
\emmoveto{78.577}{63.436}
\emlineto{78.904}{63.430}
\emmoveto{78.904}{63.420}
\emlineto{79.230}{63.414}
\emmoveto{79.230}{63.404}
\emlineto{79.557}{63.398}
\emmoveto{79.557}{63.388}
\emlineto{79.883}{63.382}
\emmoveto{79.883}{63.372}
\emlineto{80.209}{63.366}
\emmoveto{80.209}{63.356}
\emlineto{80.535}{63.350}
\emmoveto{80.535}{63.340}
\emlineto{80.861}{63.333}
\emmoveto{80.861}{63.323}
\emlineto{81.187}{63.317}
\emmoveto{81.187}{63.307}
\emlineto{81.512}{63.301}
\emmoveto{81.512}{63.291}
\emlineto{81.838}{63.285}
\emmoveto{81.838}{63.275}
\emlineto{82.163}{63.269}
\emmoveto{82.163}{63.259}
\emlineto{82.488}{63.253}
\emmoveto{82.488}{63.243}
\emlineto{82.813}{63.236}
\emmoveto{82.813}{63.226}
\emlineto{83.138}{63.220}
\emmoveto{83.138}{63.210}
\emlineto{83.463}{63.204}
\emmoveto{83.463}{63.194}
\emlineto{83.788}{63.187}
\emmoveto{83.788}{63.177}
\emlineto{84.112}{63.171}
\emmoveto{84.112}{63.161}
\emlineto{84.436}{63.154}
\emmoveto{84.436}{63.144}
\emlineto{84.761}{63.137}
\emmoveto{84.761}{63.127}
\emlineto{85.085}{63.120}
\emmoveto{85.085}{63.110}
\emlineto{85.408}{63.102}
\emmoveto{85.408}{63.092}
\emlineto{85.732}{63.085}
\emmoveto{85.732}{63.075}
\emlineto{86.056}{63.067}
\emmoveto{86.056}{63.057}
\emlineto{86.379}{63.048}
\emmoveto{86.379}{63.038}
\emlineto{86.702}{63.030}
\emmoveto{86.702}{63.020}
\emlineto{87.025}{63.011}
\emmoveto{87.025}{63.001}
\emlineto{87.348}{62.992}
\emmoveto{87.348}{62.982}
\emlineto{87.671}{62.972}
\emmoveto{87.671}{62.962}
\emlineto{87.994}{62.953}
\emmoveto{87.994}{62.943}
\emlineto{88.316}{62.934}
\emmoveto{88.316}{62.924}
\emlineto{88.638}{62.916}
\emmoveto{88.638}{62.906}
\emlineto{88.960}{62.899}
\emmoveto{88.960}{62.889}
\emlineto{89.282}{62.883}
\emmoveto{89.282}{62.873}
\emlineto{89.604}{62.867}
\emmoveto{89.604}{62.857}
\emlineto{89.926}{62.853}
\emmoveto{89.926}{62.843}
\emlineto{90.247}{62.839}
\emmoveto{90.247}{62.829}
\emlineto{90.569}{62.825}
\emmoveto{90.569}{62.815}
\emlineto{90.890}{62.813}
\emmoveto{90.890}{62.803}
\emlineto{91.211}{62.801}
\emmoveto{91.211}{62.791}
\emlineto{91.532}{62.789}
\emmoveto{91.532}{62.779}
\emlineto{91.853}{62.778}
\emmoveto{91.853}{62.768}
\emlineto{92.174}{62.767}
\emmoveto{92.174}{62.757}
\emlineto{92.495}{62.756}
\emmoveto{92.495}{62.746}
\emlineto{92.816}{62.746}
\emmoveto{92.816}{62.736}
\emlineto{93.136}{62.736}
\emmoveto{93.136}{62.726}
\emlineto{93.457}{62.726}
\emmoveto{93.457}{62.716}
\emlineto{93.777}{62.717}
\emmoveto{93.777}{62.707}
\emlineto{94.098}{62.707}
\emmoveto{94.098}{62.697}
\emlineto{94.418}{62.698}
\emmoveto{94.418}{62.688}
\emlineto{94.738}{62.688}
\emmoveto{94.738}{62.678}
\emlineto{95.058}{62.679}
\emmoveto{95.058}{62.669}
\emlineto{95.378}{62.670}
\emmoveto{95.378}{62.660}
\emlineto{95.698}{62.661}
\emmoveto{95.698}{62.651}
\emlineto{96.018}{62.652}
\emmoveto{96.018}{62.642}
\emlineto{96.338}{62.643}
\emmoveto{96.338}{62.633}
\emlineto{96.658}{62.634}
\emmoveto{96.658}{62.624}
\emlineto{96.977}{62.625}
\emmoveto{96.977}{62.615}
\emlineto{97.297}{62.616}
\emmoveto{97.297}{62.606}
\emlineto{97.616}{62.607}
\emmoveto{97.616}{62.597}
\emlineto{97.936}{62.598}
\emmoveto{97.936}{62.588}
\emlineto{98.255}{62.589}
\emmoveto{98.255}{62.579}
\emlineto{98.574}{62.580}
\emmoveto{98.574}{62.570}
\emlineto{98.894}{62.571}
\emmoveto{98.894}{62.561}
\emlineto{99.213}{62.562}
\emmoveto{99.213}{62.552}
\emlineto{99.532}{62.553}
\emmoveto{99.532}{62.543}
\emlineto{99.851}{62.544}
\emmoveto{99.851}{62.534}
\emlineto{100.170}{62.535}
\emmoveto{100.170}{62.525}
\emlineto{100.488}{62.526}
\emmoveto{100.488}{62.516}
\emlineto{100.807}{62.517}
\emmoveto{100.807}{62.507}
\emlineto{101.126}{62.508}
\emmoveto{101.126}{62.498}
\emlineto{101.444}{62.498}
\emmoveto{101.444}{62.488}
\emlineto{101.763}{62.489}
\emmoveto{101.763}{62.479}
\emlineto{102.081}{62.479}
\emmoveto{102.081}{62.469}
\emlineto{102.399}{62.469}
\emmoveto{102.399}{62.459}
\emlineto{102.718}{62.459}
\emmoveto{102.718}{62.449}
\emlineto{103.036}{62.448}
\emmoveto{103.036}{62.438}
\emlineto{103.354}{62.438}
\emmoveto{103.354}{62.428}
\emlineto{103.672}{62.427}
\emmoveto{103.672}{62.417}
\emlineto{103.989}{62.416}
\emmoveto{103.989}{62.406}
\emlineto{104.307}{62.404}
\emmoveto{104.307}{62.394}
\emlineto{104.625}{62.393}
\emmoveto{104.625}{62.383}
\emlineto{104.942}{62.382}
\emmoveto{104.942}{62.372}
\emlineto{105.260}{62.371}
\emmoveto{105.260}{62.361}
\emlineto{105.577}{62.361}
\emmoveto{105.577}{62.351}
\emlineto{105.894}{62.351}
\emmoveto{105.894}{62.341}
\emlineto{106.211}{62.342}
\emmoveto{106.211}{62.332}
\emlineto{106.529}{62.333}
\emmoveto{106.529}{62.323}
\emlineto{106.846}{62.324}
\emmoveto{106.846}{62.314}
\emlineto{107.162}{62.317}
\emmoveto{107.162}{62.307}
\emlineto{107.479}{62.310}
\emmoveto{107.479}{62.300}
\emlineto{107.796}{62.303}
\emmoveto{107.796}{62.293}
\emlineto{108.113}{62.297}
\emmoveto{108.113}{62.287}
\emlineto{108.430}{62.291}
\emmoveto{108.430}{62.281}
\emlineto{108.746}{62.286}
\emmoveto{108.746}{62.276}
\emlineto{109.063}{62.281}
\emmoveto{109.063}{62.271}
\emlineto{109.379}{62.277}
\emmoveto{109.379}{62.267}
\emlineto{109.696}{62.273}
\emmoveto{109.696}{62.263}
\emlineto{110.013}{62.269}
\emmoveto{110.013}{62.259}
\emlineto{110.329}{62.266}
\emmoveto{110.329}{62.256}
\emlineto{110.645}{62.262}
\emmoveto{110.645}{62.252}
\emlineto{110.962}{62.259}
\emmoveto{110.962}{62.249}
\emlineto{111.278}{62.257}
\emmoveto{111.278}{62.247}
\emlineto{111.595}{62.254}
\emmoveto{111.595}{62.244}
\emlineto{111.911}{62.251}
\emmoveto{111.911}{62.241}
\emlineto{112.227}{62.249}
\emmoveto{112.227}{62.239}
\emlineto{112.544}{62.247}
\emmoveto{112.544}{62.237}
\emlineto{112.860}{62.244}
\emmoveto{112.860}{62.234}
\emlineto{113.176}{62.242}
\emmoveto{113.176}{62.232}
\emlineto{113.492}{62.240}
\emmoveto{113.492}{62.230}
\emlineto{113.809}{62.238}
\emmoveto{113.809}{62.228}
\emlineto{114.125}{62.236}
\emmoveto{114.125}{62.226}
\emlineto{114.441}{62.234}
\emmoveto{114.441}{62.224}
\emlineto{114.757}{62.233}
\emmoveto{114.757}{62.223}
\emlineto{115.073}{62.231}
\emmoveto{115.073}{62.221}
\emlineto{115.389}{62.229}
\emmoveto{115.389}{62.219}
\emlineto{115.706}{62.227}
\emmoveto{115.706}{62.217}
\emlineto{116.022}{62.225}
\emmoveto{116.022}{62.215}
\emlineto{116.338}{62.224}
\emmoveto{116.338}{62.214}
\emlineto{116.654}{62.222}
\emmoveto{116.654}{62.212}
\emlineto{116.970}{62.220}
\emmoveto{116.970}{62.210}
\emlineto{117.286}{62.218}
\emmoveto{117.286}{62.208}
\emlineto{117.602}{62.216}
\emmoveto{117.602}{62.206}
\emlineto{117.918}{62.214}
\emmoveto{117.918}{62.204}
\emlineto{118.234}{62.212}
\emmoveto{118.234}{62.202}
\emlineto{118.550}{62.209}
\emmoveto{118.550}{62.199}
\emlineto{118.866}{62.207}
\emmoveto{118.866}{62.197}
\emlineto{119.182}{62.204}
\emmoveto{119.182}{62.194}
\emlineto{119.498}{62.202}
\emmoveto{119.498}{62.192}
\emlineto{119.814}{62.199}
\emmoveto{119.814}{62.189}
\emlineto{120.130}{62.195}
\emmoveto{120.130}{62.185}
\emlineto{120.445}{62.192}
\emmoveto{120.445}{62.182}
\emlineto{120.761}{62.189}
\emmoveto{120.761}{62.179}
\emlineto{121.077}{62.185}
\emmoveto{121.077}{62.175}
\emlineto{121.393}{62.182}
\emmoveto{121.393}{62.172}
\emlineto{121.708}{62.178}
\emmoveto{121.708}{62.168}
\emlineto{122.024}{62.175}
\emmoveto{122.024}{62.165}
\emlineto{122.340}{62.172}
\emmoveto{122.340}{62.162}
\emlineto{122.655}{62.170}
\emmoveto{122.655}{62.160}
\emlineto{122.971}{62.167}
\emmoveto{122.971}{62.157}
\emlineto{123.287}{62.165}
\emmoveto{123.287}{62.155}
\emlineto{123.602}{62.164}
\emmoveto{123.602}{62.154}
\emlineto{123.918}{62.163}
\emmoveto{123.918}{62.153}
\emlineto{124.233}{62.162}
\emmoveto{124.233}{62.152}
\emlineto{124.549}{62.162}
\emmoveto{124.549}{62.152}
\emlineto{124.864}{62.162}
\emmoveto{124.864}{62.152}
\emlineto{125.180}{62.163}
\emmoveto{125.180}{62.153}
\emlineto{125.495}{62.164}
\emmoveto{125.495}{62.154}
\emlineto{125.811}{62.165}
\emmoveto{125.811}{62.155}
\emlineto{126.127}{62.167}
\emmoveto{126.127}{62.157}
\emlineto{126.442}{62.169}
\emmoveto{126.442}{62.159}
\emlineto{126.758}{62.172}
\emmoveto{126.758}{62.162}
\emlineto{127.073}{62.175}
\emmoveto{127.073}{62.165}
\emlineto{127.389}{62.178}
\emmoveto{127.389}{62.168}
\emlineto{127.705}{62.181}
\emmoveto{127.705}{62.171}
\emlineto{128.021}{62.185}
\emmoveto{128.021}{62.175}
\emlineto{128.336}{62.189}
\emmoveto{128.336}{62.179}
\emlineto{128.652}{62.193}
\emmoveto{128.652}{62.183}
\emlineto{128.968}{62.197}
\emmoveto{128.968}{62.187}
\emlineto{129.284}{62.202}
\emmoveto{129.284}{62.192}
\emlineto{129.600}{62.206}
\emshow{48.580}{59.700}{k=10,v=0.3}
\emmoveto{12.000}{80.000}
\emlineto{12.472}{80.009}
\emmoveto{12.472}{79.999}
\emlineto{12.944}{80.006}
\emmoveto{12.944}{79.996}
\emlineto{13.416}{80.002}
\emmoveto{13.416}{79.992}
\emlineto{13.888}{79.997}
\emmoveto{13.888}{79.987}
\emlineto{14.360}{79.990}
\emmoveto{14.360}{79.980}
\emlineto{14.831}{79.983}
\emmoveto{14.831}{79.973}
\emlineto{15.303}{79.975}
\emmoveto{15.303}{79.965}
\emlineto{15.775}{79.967}
\emmoveto{15.775}{79.957}
\emlineto{16.246}{79.958}
\emmoveto{16.246}{79.948}
\emlineto{16.718}{79.949}
\emmoveto{16.718}{79.939}
\emlineto{17.189}{79.939}
\emmoveto{17.189}{79.929}
\emlineto{17.661}{79.930}
\emmoveto{17.661}{79.920}
\emlineto{18.132}{79.920}
\emmoveto{18.132}{79.910}
\emlineto{18.603}{79.910}
\emmoveto{18.603}{79.900}
\emlineto{19.074}{79.900}
\emmoveto{19.074}{79.890}
\emlineto{19.545}{79.889}
\emmoveto{19.545}{79.879}
\emlineto{20.016}{79.879}
\emmoveto{20.016}{79.869}
\emlineto{20.487}{79.869}
\emmoveto{20.487}{79.859}
\emlineto{20.958}{79.858}
\emmoveto{20.958}{79.848}
\emlineto{21.428}{79.848}
\emmoveto{21.428}{79.838}
\emlineto{21.899}{79.837}
\emmoveto{21.899}{79.827}
\emlineto{22.369}{79.826}
\emmoveto{22.369}{79.816}
\emlineto{22.840}{79.816}
\emmoveto{22.840}{79.806}
\emlineto{23.310}{79.805}
\emmoveto{23.310}{79.795}
\emlineto{23.780}{79.795}
\emmoveto{23.780}{79.785}
\emlineto{24.250}{79.784}
\emmoveto{24.250}{79.774}
\emlineto{24.720}{79.773}
\emmoveto{24.720}{79.763}
\emlineto{25.190}{79.763}
\emmoveto{25.190}{79.753}
\emlineto{25.660}{79.752}
\emmoveto{25.660}{79.742}
\emlineto{26.129}{79.741}
\emmoveto{26.129}{79.731}
\emlineto{26.599}{79.730}
\emmoveto{26.599}{79.720}
\emlineto{27.068}{79.720}
\emmoveto{27.068}{79.710}
\emlineto{27.538}{79.709}
\emmoveto{27.538}{79.699}
\emlineto{28.007}{79.698}
\emmoveto{28.007}{79.688}
\emlineto{28.476}{79.688}
\emmoveto{28.476}{79.678}
\emlineto{28.946}{79.677}
\emmoveto{28.946}{79.667}
\emlineto{29.415}{79.666}
\emmoveto{29.415}{79.656}
\emlineto{29.883}{79.655}
\emmoveto{29.883}{79.645}
\emlineto{30.352}{79.645}
\emmoveto{30.352}{79.635}
\emlineto{30.821}{79.634}
\emmoveto{30.821}{79.624}
\emlineto{31.290}{79.623}
\emmoveto{31.290}{79.613}
\emlineto{31.758}{79.613}
\emmoveto{31.758}{79.603}
\emlineto{32.227}{79.602}
\emmoveto{32.227}{79.592}
\emlineto{32.695}{79.591}
\emmoveto{32.695}{79.581}
\emlineto{33.163}{79.581}
\emmoveto{33.163}{79.571}
\emlineto{33.632}{79.570}
\emmoveto{33.632}{79.560}
\emlineto{34.100}{79.559}
\emmoveto{34.100}{79.549}
\emlineto{34.568}{79.549}
\emmoveto{34.568}{79.539}
\emlineto{35.036}{79.538}
\emmoveto{35.036}{79.528}
\emlineto{35.503}{79.527}
\emmoveto{35.503}{79.517}
\emlineto{35.971}{79.516}
\emmoveto{35.971}{79.506}
\emlineto{36.439}{79.505}
\emmoveto{36.439}{79.495}
\emlineto{36.906}{79.494}
\emmoveto{36.906}{79.484}
\emlineto{37.374}{79.481}
\emmoveto{37.374}{79.471}
\emlineto{37.841}{79.468}
\emmoveto{37.841}{79.458}
\emlineto{38.308}{79.454}
\emmoveto{38.308}{79.444}
\emlineto{38.775}{79.440}
\emmoveto{38.775}{79.430}
\emlineto{39.242}{79.424}
\emmoveto{39.242}{79.414}
\emlineto{39.709}{79.408}
\emmoveto{39.709}{79.398}
\emlineto{40.176}{79.391}
\emmoveto{40.176}{79.381}
\emlineto{40.642}{79.374}
\emmoveto{40.642}{79.364}
\emlineto{41.109}{79.356}
\emmoveto{41.109}{79.346}
\emlineto{41.575}{79.338}
\emmoveto{41.575}{79.328}
\emlineto{42.041}{79.319}
\emmoveto{42.041}{79.309}
\emlineto{42.507}{79.300}
\emmoveto{42.507}{79.290}
\emlineto{42.972}{79.281}
\emmoveto{42.972}{79.271}
\emlineto{43.438}{79.261}
\emmoveto{43.438}{79.251}
\emlineto{43.903}{79.241}
\emmoveto{43.903}{79.231}
\emlineto{44.368}{79.221}
\emmoveto{44.368}{79.211}
\emlineto{44.833}{79.200}
\emmoveto{44.833}{79.190}
\emlineto{45.298}{79.180}
\emmoveto{45.298}{79.170}
\emlineto{45.763}{79.159}
\emmoveto{45.763}{79.149}
\emlineto{46.227}{79.139}
\emmoveto{46.227}{79.129}
\emlineto{46.692}{79.118}
\emmoveto{46.692}{79.108}
\emlineto{47.156}{79.097}
\emmoveto{47.156}{79.087}
\emlineto{47.619}{79.076}
\emmoveto{47.619}{79.066}
\emlineto{48.083}{79.055}
\emmoveto{48.083}{79.045}
\emlineto{48.547}{79.034}
\emmoveto{48.547}{79.024}
\emlineto{49.010}{79.013}
\emmoveto{49.010}{79.003}
\emlineto{49.473}{78.992}
\emmoveto{49.473}{78.982}
\emlineto{49.936}{78.971}
\emmoveto{49.936}{78.961}
\emlineto{50.399}{78.949}
\emmoveto{50.399}{78.939}
\emlineto{50.862}{78.928}
\emmoveto{50.862}{78.918}
\emlineto{51.324}{78.907}
\emmoveto{51.324}{78.897}
\emlineto{51.786}{78.886}
\emmoveto{51.786}{78.876}
\emlineto{52.248}{78.865}
\emmoveto{52.248}{78.855}
\emlineto{52.710}{78.843}
\emmoveto{52.710}{78.833}
\emlineto{53.172}{78.822}
\emmoveto{53.172}{78.812}
\emlineto{53.633}{78.801}
\emmoveto{53.633}{78.791}
\emlineto{54.095}{78.780}
\emmoveto{54.095}{78.770}
\emlineto{54.556}{78.759}
\emmoveto{54.556}{78.749}
\emlineto{55.017}{78.737}
\emmoveto{55.017}{78.727}
\emlineto{55.478}{78.716}
\emmoveto{55.478}{78.706}
\emlineto{55.938}{78.695}
\emmoveto{55.938}{78.685}
\emlineto{56.399}{78.674}
\emmoveto{56.399}{78.664}
\emlineto{56.859}{78.652}
\emmoveto{56.859}{78.642}
\emlineto{57.319}{78.631}
\emmoveto{57.319}{78.621}
\emlineto{57.779}{78.610}
\emmoveto{57.779}{78.600}
\emlineto{58.238}{78.589}
\emmoveto{58.238}{78.579}
\emlineto{58.698}{78.568}
\emmoveto{58.698}{78.558}
\emlineto{59.157}{78.546}
\emmoveto{59.157}{78.536}
\emlineto{59.616}{78.525}
\emmoveto{59.616}{78.515}
\emlineto{60.075}{78.504}
\emmoveto{60.075}{78.494}
\emlineto{60.534}{78.482}
\emmoveto{60.534}{78.472}
\emlineto{60.992}{78.461}
\emmoveto{60.992}{78.451}
\emlineto{61.450}{78.438}
\emmoveto{61.450}{78.428}
\emlineto{61.909}{78.416}
\emmoveto{61.909}{78.406}
\emlineto{62.366}{78.392}
\emmoveto{62.366}{78.382}
\emlineto{62.824}{78.369}
\emmoveto{62.824}{78.359}
\emlineto{63.282}{78.344}
\emmoveto{63.282}{78.334}
\emlineto{63.739}{78.319}
\emmoveto{63.739}{78.309}
\emlineto{64.196}{78.294}
\emmoveto{64.196}{78.284}
\emlineto{64.653}{78.268}
\emmoveto{64.653}{78.258}
\emlineto{65.109}{78.241}
\emmoveto{65.109}{78.231}
\emlineto{65.566}{78.214}
\emmoveto{65.566}{78.204}
\emlineto{66.022}{78.187}
\emmoveto{66.022}{78.177}
\emlineto{66.478}{78.159}
\emmoveto{66.478}{78.149}
\emlineto{66.934}{78.130}
\emmoveto{66.934}{78.120}
\emlineto{67.389}{78.102}
\emmoveto{67.389}{78.092}
\emlineto{67.844}{78.073}
\emmoveto{67.844}{78.063}
\emlineto{68.299}{78.043}
\emmoveto{68.299}{78.033}
\emlineto{68.754}{78.014}
\emmoveto{68.754}{78.004}
\emlineto{69.208}{77.984}
\emmoveto{69.208}{77.974}
\emlineto{69.662}{77.953}
\emmoveto{69.662}{77.943}
\emlineto{70.116}{77.923}
\emmoveto{70.116}{77.913}
\emlineto{70.569}{77.893}
\emmoveto{70.569}{77.883}
\emlineto{71.023}{77.862}
\emmoveto{71.023}{77.852}
\emlineto{71.476}{77.833}
\emmoveto{71.476}{77.823}
\emlineto{71.929}{77.808}
\emmoveto{71.929}{77.798}
\emlineto{72.381}{77.785}
\emmoveto{72.381}{77.775}
\emlineto{72.834}{77.765}
\emmoveto{72.834}{77.755}
\emlineto{73.286}{77.746}
\emmoveto{73.286}{77.736}
\emlineto{73.738}{77.728}
\emmoveto{73.738}{77.718}
\emlineto{74.190}{77.712}
\emmoveto{74.190}{77.702}
\emlineto{74.642}{77.697}
\emmoveto{74.642}{77.687}
\emlineto{75.093}{77.682}
\emmoveto{75.093}{77.672}
\emlineto{75.545}{77.669}
\emmoveto{75.545}{77.659}
\emlineto{75.996}{77.655}
\emmoveto{75.996}{77.645}
\emlineto{76.448}{77.643}
\emmoveto{76.448}{77.633}
\emlineto{76.899}{77.630}
\emmoveto{76.899}{77.620}
\emlineto{77.350}{77.618}
\emmoveto{77.350}{77.608}
\emlineto{77.801}{77.606}
\emmoveto{77.801}{77.596}
\emlineto{78.252}{77.594}
\emmoveto{78.252}{77.584}
\emlineto{78.702}{77.582}
\emmoveto{78.702}{77.572}
\emlineto{79.153}{77.571}
\emmoveto{79.153}{77.561}
\emlineto{79.604}{77.559}
\emmoveto{79.604}{77.549}
\emlineto{80.054}{77.548}
\emmoveto{80.054}{77.538}
\emlineto{80.505}{77.537}
\emmoveto{80.505}{77.527}
\emlineto{80.955}{77.526}
\emmoveto{80.955}{77.516}
\emlineto{81.405}{77.515}
\emmoveto{81.405}{77.505}
\emlineto{81.855}{77.503}
\emmoveto{81.855}{77.493}
\emlineto{82.305}{77.492}
\emmoveto{82.305}{77.482}
\emlineto{82.755}{77.481}
\emmoveto{82.755}{77.471}
\emlineto{83.205}{77.470}
\emmoveto{83.205}{77.460}
\emlineto{83.654}{77.459}
\emmoveto{83.654}{77.449}
\emlineto{84.104}{77.448}
\emmoveto{84.104}{77.438}
\emlineto{84.553}{77.436}
\emmoveto{84.553}{77.426}
\emlineto{85.003}{77.425}
\emmoveto{85.003}{77.415}
\emlineto{85.452}{77.413}
\emmoveto{85.452}{77.403}
\emlineto{85.901}{77.400}
\emmoveto{85.901}{77.390}
\emlineto{86.350}{77.388}
\emmoveto{86.350}{77.378}
\emlineto{86.799}{77.375}
\emmoveto{86.799}{77.365}
\emlineto{87.248}{77.361}
\emmoveto{87.248}{77.351}
\emlineto{87.697}{77.347}
\emmoveto{87.697}{77.337}
\emlineto{88.146}{77.333}
\emmoveto{88.146}{77.323}
\emlineto{88.594}{77.318}
\emmoveto{88.594}{77.308}
\emlineto{89.042}{77.302}
\emmoveto{89.042}{77.292}
\emlineto{89.491}{77.287}
\emmoveto{89.491}{77.277}
\emlineto{89.939}{77.270}
\emmoveto{89.939}{77.260}
\emlineto{90.387}{77.254}
\emmoveto{90.387}{77.244}
\emlineto{90.834}{77.236}
\emmoveto{90.834}{77.226}
\emlineto{91.282}{77.219}
\emmoveto{91.282}{77.209}
\emlineto{91.729}{77.201}
\emmoveto{91.729}{77.191}
\emlineto{92.177}{77.183}
\emmoveto{92.177}{77.173}
\emlineto{92.624}{77.164}
\emmoveto{92.624}{77.154}
\emlineto{93.071}{77.145}
\emmoveto{93.071}{77.135}
\emlineto{93.518}{77.126}
\emmoveto{93.518}{77.116}
\emlineto{93.964}{77.107}
\emmoveto{93.964}{77.097}
\emlineto{94.411}{77.088}
\emmoveto{94.411}{77.078}
\emlineto{94.857}{77.070}
\emmoveto{94.857}{77.060}
\emlineto{95.303}{77.053}
\emmoveto{95.303}{77.043}
\emlineto{95.749}{77.038}
\emmoveto{95.749}{77.028}
\emlineto{96.195}{77.024}
\emmoveto{96.195}{77.014}
\emlineto{96.641}{77.011}
\emmoveto{96.641}{77.001}
\emlineto{97.086}{77.000}
\emmoveto{97.086}{76.990}
\emlineto{97.532}{76.990}
\emmoveto{97.532}{76.980}
\emlineto{97.977}{76.981}
\emmoveto{97.977}{76.971}
\emlineto{98.423}{76.973}
\emmoveto{98.423}{76.963}
\emlineto{98.868}{76.965}
\emmoveto{98.868}{76.955}
\emlineto{99.314}{76.959}
\emmoveto{99.314}{76.949}
\emlineto{99.759}{76.954}
\emmoveto{99.759}{76.944}
\emlineto{100.204}{76.949}
\emmoveto{100.204}{76.939}
\emlineto{100.649}{76.945}
\emmoveto{100.649}{76.935}
\emlineto{101.094}{76.941}
\emmoveto{101.094}{76.931}
\emlineto{101.539}{76.937}
\emmoveto{101.539}{76.927}
\emlineto{101.984}{76.934}
\emmoveto{101.984}{76.924}
\emlineto{102.429}{76.932}
\emmoveto{102.429}{76.922}
\emlineto{102.874}{76.929}
\emmoveto{102.874}{76.919}
\emlineto{103.319}{76.927}
\emmoveto{103.319}{76.917}
\emlineto{103.764}{76.925}
\emmoveto{103.764}{76.915}
\emlineto{104.209}{76.923}
\emmoveto{104.209}{76.913}
\emlineto{104.654}{76.922}
\emmoveto{104.654}{76.912}
\emlineto{105.099}{76.920}
\emmoveto{105.099}{76.910}
\emlineto{105.544}{76.919}
\emmoveto{105.544}{76.909}
\emlineto{105.989}{76.918}
\emmoveto{105.989}{76.908}
\emlineto{106.434}{76.916}
\emmoveto{106.434}{76.906}
\emlineto{106.879}{76.915}
\emmoveto{106.879}{76.905}
\emlineto{107.324}{76.914}
\emmoveto{107.324}{76.904}
\emlineto{107.768}{76.912}
\emmoveto{107.768}{76.902}
\emlineto{108.213}{76.911}
\emmoveto{108.213}{76.901}
\emlineto{108.658}{76.909}
\emmoveto{108.658}{76.899}
\emlineto{109.103}{76.908}
\emmoveto{109.103}{76.898}
\emlineto{109.548}{76.906}
\emmoveto{109.548}{76.896}
\emlineto{109.992}{76.903}
\emmoveto{109.992}{76.893}
\emlineto{110.437}{76.901}
\emmoveto{110.437}{76.891}
\emlineto{110.882}{76.898}
\emmoveto{110.882}{76.888}
\emlineto{111.327}{76.894}
\emmoveto{111.327}{76.884}
\emlineto{111.771}{76.890}
\emmoveto{111.771}{76.880}
\emlineto{112.216}{76.886}
\emmoveto{112.216}{76.876}
\emlineto{112.661}{76.882}
\emmoveto{112.661}{76.872}
\emlineto{113.105}{76.877}
\emmoveto{113.105}{76.867}
\emlineto{113.550}{76.871}
\emmoveto{113.550}{76.861}
\emlineto{113.994}{76.866}
\emmoveto{113.994}{76.856}
\emlineto{114.439}{76.859}
\emmoveto{114.439}{76.849}
\emlineto{114.883}{76.853}
\emmoveto{114.883}{76.843}
\emlineto{115.327}{76.846}
\emmoveto{115.327}{76.836}
\emlineto{115.771}{76.839}
\emmoveto{115.771}{76.829}
\emlineto{116.216}{76.831}
\emmoveto{116.216}{76.821}
\emlineto{116.660}{76.823}
\emmoveto{116.660}{76.813}
\emlineto{117.104}{76.815}
\emmoveto{117.104}{76.805}
\emlineto{117.548}{76.807}
\emmoveto{117.548}{76.797}
\emlineto{117.992}{76.800}
\emmoveto{117.992}{76.790}
\emlineto{118.435}{76.793}
\emmoveto{118.435}{76.783}
\emlineto{118.879}{76.787}
\emmoveto{118.879}{76.777}
\emlineto{119.323}{76.781}
\emmoveto{119.323}{76.771}
\emlineto{119.767}{76.777}
\emmoveto{119.767}{76.767}
\emlineto{120.210}{76.773}
\emmoveto{120.210}{76.763}
\emlineto{120.654}{76.770}
\emmoveto{120.654}{76.760}
\emlineto{121.097}{76.769}
\emmoveto{121.097}{76.759}
\emlineto{121.541}{76.768}
\emmoveto{121.541}{76.758}
\emlineto{121.985}{76.768}
\emmoveto{121.985}{76.758}
\emlineto{122.428}{76.769}
\emmoveto{122.428}{76.759}
\emlineto{122.872}{76.771}
\emmoveto{122.872}{76.761}
\emlineto{123.315}{76.773}
\emmoveto{123.315}{76.763}
\emlineto{123.759}{76.777}
\emmoveto{123.759}{76.767}
\emlineto{124.203}{76.781}
\emmoveto{124.203}{76.771}
\emlineto{124.646}{76.785}
\emmoveto{124.646}{76.775}
\emlineto{125.090}{76.790}
\emmoveto{125.090}{76.780}
\emlineto{125.534}{76.796}
\emmoveto{125.534}{76.786}
\emlineto{125.978}{76.802}
\emmoveto{125.978}{76.792}
\emlineto{126.422}{76.809}
\emmoveto{126.422}{76.799}
\emlineto{126.866}{76.815}
\emmoveto{126.866}{76.805}
\emlineto{127.310}{76.822}
\emmoveto{127.310}{76.812}
\emlineto{127.754}{76.830}
\emmoveto{127.754}{76.820}
\emlineto{128.198}{76.838}
\emmoveto{128.198}{76.828}
\emlineto{128.642}{76.845}
\emmoveto{128.642}{76.835}
\emlineto{129.086}{76.854}
\emmoveto{129.086}{76.844}
\emlineto{129.531}{76.862}
\emshow{60.380}{73.700}{k=10,v=0.4}
\emshow{1.000}{10.000}{-1.20e-1}
\emshow{1.000}{17.000}{-6.80e-2}
\emshow{1.000}{24.000}{-1.60e-2}
\emshow{1.000}{31.000}{3.60e-2}
\emshow{1.000}{38.000}{8.80e-2}
\emshow{1.000}{45.000}{1.40e-1}
\emshow{1.000}{52.000}{1.92e-1}
\emshow{1.000}{59.000}{2.44e-1}
\emshow{1.000}{66.000}{2.96e-1}
\emshow{1.000}{73.000}{3.48e-1}
\emshow{1.000}{80.000}{4.00e-1}
\emshow{12.000}{5.000}{-5.00e-1}
\emshow{23.800}{5.000}{-3.00e-1}
\emshow{35.600}{5.000}{-1.00e-1}
\emshow{47.400}{5.000}{1.00e-1}
\emshow{59.200}{5.000}{3.00e-1}
\emshow{71.000}{5.000}{5.00e-1}
\emshow{82.800}{5.000}{7.00e-1}
\emshow{94.600}{5.000}{9.00e-1}
\emshow{106.400}{5.000}{1.10e0}
\emshow{118.200}{5.000}{1.30e0}
\emshow{130.000}{5.000}{1.50e0}

\centerline{\bf{Fig.A.3}}
\eject

 
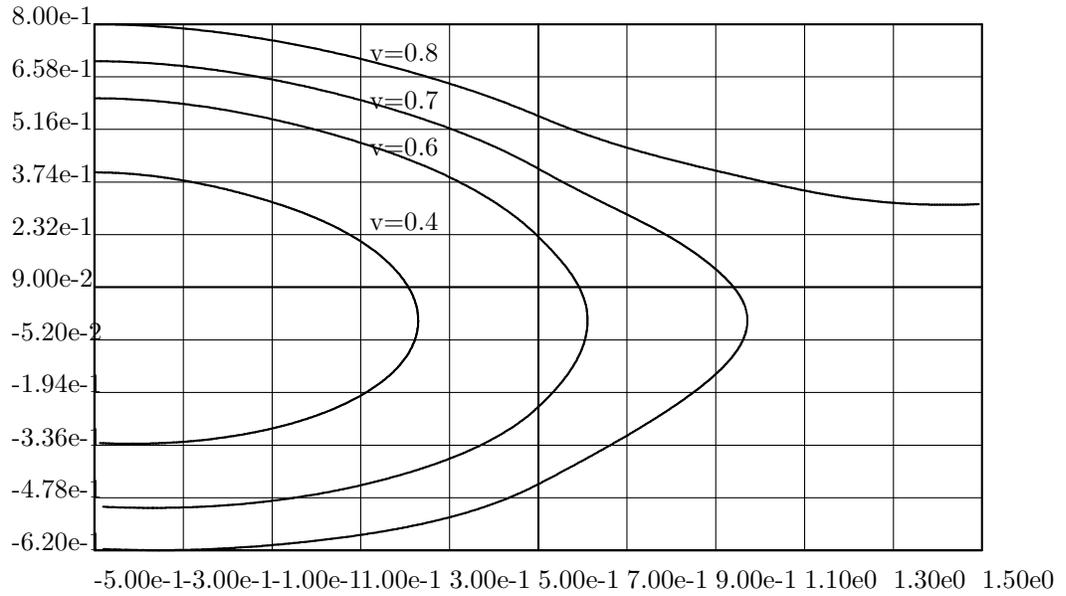
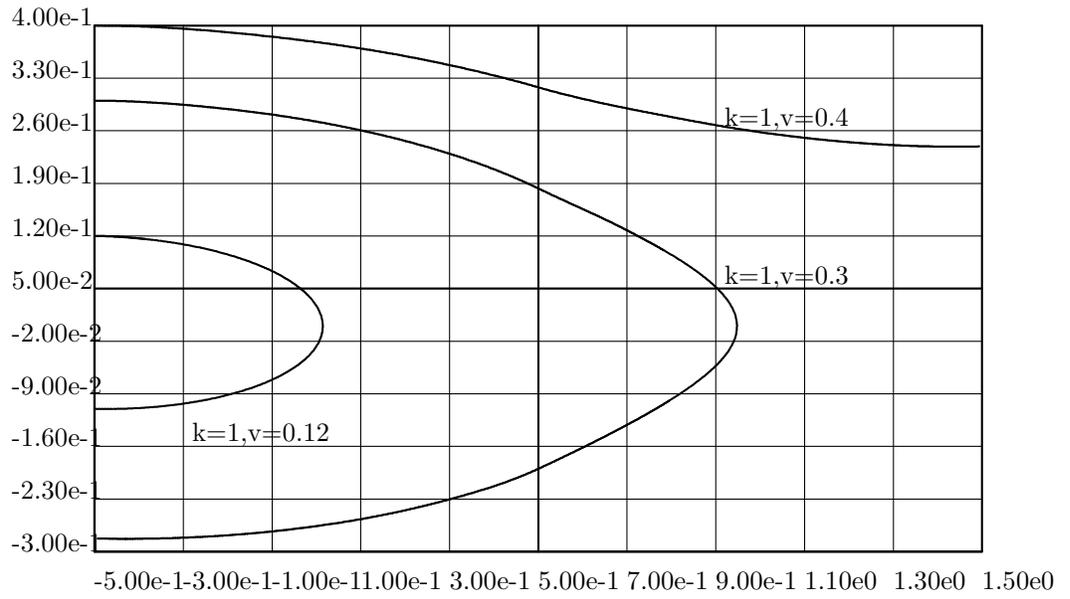
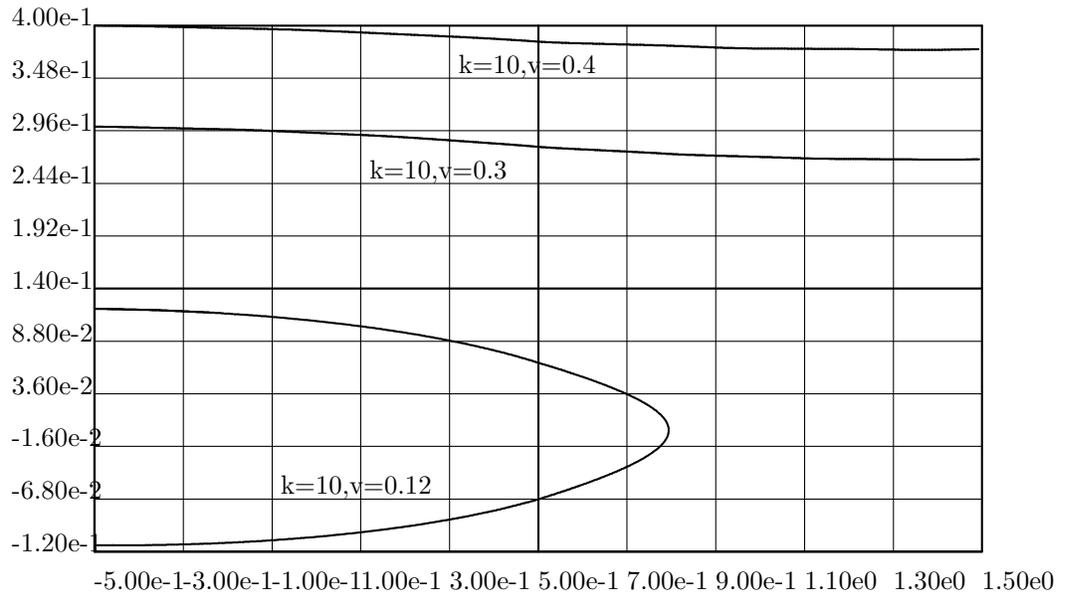
\end{document}